\documentclass[reprint,aps,prd,superscriptaddress,nofootinbib,floatfix,showpacs]{revtex4-1} 

\usepackage{hyperref}
\usepackage{bm}
\usepackage[usenames,dvipsnames]{color}
\usepackage{ulem}
\usepackage{amsmath}
\usepackage{amssymb}
\usepackage{graphicx}
\usepackage{dcolumn}
\usepackage{bm}
\usepackage{amssymb,amsmath}
\usepackage{latexsym}
\usepackage{color}
\usepackage{cancel}
\allowdisplaybreaks
\usepackage{amsfonts}
\usepackage{color}
\usepackage{subfigure}

\def\laq{\raise 0.4ex\hbox{$<$}\kern -0.8em\lower 0.62ex\hbox{$\sim$}}
\def\gaq{\raise 0.4ex\hbox{$>$}\kern -0.7em\lower 0.62ex\hbox{$\sim$}}

\allowdisplaybreaks

\newcommand{\Maryland}{\affiliation{Maryland Center for Fundamental
    Physics \& Joint Space-Science Institute,\\ Department of Physics,
University of Maryland, College
    Park, MD 20742, USA}}
\newcommand{\CIFAR}{\affiliation{Canadian Institute for Advanced Research, 180 Dundas St.~West, Toronto, Ontario M5G 1Z8, Canada}} %
\newcommand{\CITA}{\affiliation{Canadian Institute for Theoretical Astrophysics, University of Toronto, Toronto, Ontario M5S 3H8, Canada}}
\newcommand{\Cornell}{\affiliation{Center for Radiophysics and Space Research, Cornell University, Ithaca, NY 14853 USA}}
\newcommand{\CSUF}{\affiliation{Gravitational Wave Physics and Astronomy Center, California State University Fullerton, Fullerton, CA 92831, USA}}
\newcommand{\Caltech}{\affiliation{Theoretical Astrophysics 350-17, California Institute of Technology, Pasadena, CA 91125, USA}}

\begin{document}

\title{Periastron advance in spinning black hole binaries: \\ comparing effective-one-body and numerical relativity}
\author{Tanja Hinderer}\Maryland
\author{Alessandra Buonanno}\Maryland%
\author{Abdul~H. Mrou\'e}\CITA%
\author{Daniel~A. Hemberger}\Cornell\Caltech%
\author{Geoffrey Lovelace}\Caltech\CSUF%
\author{Harald~P. Pfeiffer}\CITA\CIFAR%
\author{Lawrence~E. Kidder}\Cornell%
\author{Mark~A. Scheel}\Caltech%
\author{Bela Szilagyi}\Caltech%
\author{Nicholas~W. Taylor}\Caltech%
\author{Saul~A. Teukolsky}\Cornell%
\date{\today}

\begin{abstract}
We compute the periastron advance using the effective-one-body formalism for
binary black holes moving on quasi-circular orbits and 
having spins collinear with the orbital angular momentum. We compare the
predictions with the periastron advance recently computed in accurate 
numerical-relativity simulations and find remarkable agreement for a wide 
range of spins and mass ratios. These results do not use any numerical-relativity 
calibration of the effective-one-body model, and stem from two key ingredients 
in the effective-one-body Hamiltonian: (i) the mapping of the two-body dynamics of 
spinning particles onto the dynamics of an effective  spinning particle in 
a (deformed) Kerr spacetime, fully symmetrized with respect to the two-body masses and spins, 
and (ii) the resummation, in the test-particle limit,  of all post-Newtonian (PN) corrections 
linear in the spin of the particle. In fact, even when only the leading spin PN corrections are 
included in the effective-one-body spinning Hamiltonian but all the test-particle corrections linear in the spin 
of the particle are resummed we find very good agreement with the numerical results 
(within the numerical error for equal-mass binaries and discrepancies of at most $1\%$ for larger mass ratios). 
Furthermore, we specialize to the extreme mass-ratio limit and derive, 
using the equations of motion in the gravitational skeleton approach, analytical expressions for the 
periastron advance, the meridional Lense-Thirring precession and spin 
precession frequency in the case of a spinning particle on a nearly circular equatorial orbit in Kerr spacetime, 
including also terms quadratic in the spin.
\end{abstract}


\maketitle

\def\be{\begin{equation}}
\def\ee{\end{equation}}
\def\bea{\begin{eqnarray}}
\def\eea{\end{eqnarray}}
\newcommand{\bes}{\begin{subequations}}
\newcommand{\ees}{\end{subequations}}

\section{Introduction}
\label{sec:intro}

The periastron precession in a two-body system describes the angular
advance of the line joining the points of closest and farthest
approach of an elliptic orbit. This secular effect occurs whenever the
ratio of frequencies of the radial and azimuthal motions is different
from unity, and it is caused by relativistic effects, the bodies'
multipole moments or other perturbations.  In the Solar System, the
periastron advance (PA) has been measured for several planets
\cite{PAsolar,standish1992orbital} and is mainly due to perturbations
from the presence of the other planets. The sun's oblateness also
contributes to the PA, but after initial controversies about a
potentially large effect \cite{1967PhRvL..18..313D}, subsequent
helioseismology measurements found that the contribution is
negligibly small \cite{2011EPJH...36..407R}. The residual rates of
precession are entirely accounted for by general relativity and
provide important constraints on possible deviations
\cite{1990grg..conf..313S}. The periastron shifts are also measured in
numerous binary systems \cite{1999AJ....117..587P}. These include
relativistic binary pulsars where spin-orbit effects in the PA could
constrain the neutron stars' moments of inertia and hence the nuclear
equation of state~\cite{DamourSchaferPA2PN,2005ApJ...629..979L}. The most extreme values
of the PA occur for zoom-whirl orbits in highly relativistic binaries
near the threshold of an instability in the radial motion
\cite{PhysRevD.66.044002,2007CQGra..24S..83P}.

In the limit that the binary's orbit is a small perturbation to a
strictly circular orbit, its epicyclic frequency becomes independent
of the eccentricity. The PA in this limit, when expressed as a
function of the azimuthal frequency, is a gauge invariant quantity. As
such it provides an important tool for comparing and connecting
different approaches to modeling the binary dynamics. Accurate
analytical models of coalescing binaries are the
foundation for computing templates for gravitational waves
that could be observed with detectors coming online within the next few years, 
such as advanced LIGO and Virgo. This has motivated
substantial recent interest in using the PA to assess
the performance of perturbative post-Newtonian (PN)~\cite{Blanchet2006}, gravitational
self-force~\cite{2011LRR....14....7P} and effective-one-body (EOB)~\cite{Buonanno99,Buonanno00,2000PhRvD..62h4011D,Damour01c} approaches. For binaries at
large orbital separations, the PN computations of Refs.~\cite{DamourSchaferPA2PN,DJS2000,Damour:2007nc,KG3PNandSO,1979Ap&SS..62..185A,1989PThPh..81..679O}
apply to binaries with arbitrary mass ratios and include the
spin-orbit effects.  Finite size effects such as stellar oscillations
and tidal interactions were considered in
Refs.~\cite{1938MNRAS..98..734C, 1980MNRAS.193..603P}. The radial
epicyclic frequency for test particles in Kerr spacetime is frequently
used when modeling phenomena in relativistic thin accretion disks
\cite{2013LRR....16....1A}. For generic geodesics in Kerr spacetime, the PA is
known in terms of elliptic integrals \cite{2002CQGra..19.2743S} and
its PN expansion has been calculated explicitly
\cite{2013PhRvL.111b1101H}. Postgeodesic effects in the PA for
nonspinning black holes in the small mass-ratio limit were obtained
from the conservative gravitational self-force in
Ref.~\cite{Barack:2011ed} and used to improve the EOB model
\cite{Damour:2009sm, Barack:2010ny,Barausse:2011dq}. In 2010, the computation of the
PA from numerical relativity (NR) simulations became possible
\cite{Mroue:2010re}, enabling tests of the veracity of the various
perturbative approaches \cite{LeTiec:2011bk}.

In this paper we extend previous calculations of the PA for
nonspinning binaries to include nonprecessional spin effects. For
comparable mass binaries, we compute the PA in the limit of circular
equatorial orbits within the EOB approach \cite{Buonanno99}.  In
the EOB model the conservative dynamics with spins is generated by a
Hamiltonian \cite{Damour01c,Damour:024009,Barausse:2009aa,Barausse:2009xi,Barausse:2011ys,
2011PhRvD..84h4028N,2013PhRvD..87l4036B}. The Hamiltonian used 
here~\cite{Barausse:2009aa,Barausse:2009xi,Barausse:2011ys} has the 
structure of the constrained Hamiltonian for a spinning 
particle in Kerr spacetime but the metric functions and the particle's
spin are augmented with mass-ratio dependent deformations chosen so as
to reproduce the PN Hamiltonian \cite{Damour:2007nc,Steinhoff:2008zr}
in the weak-field limit. The Hamiltonian for generic spinning binaries is usually expressed in a fixed source frame in terms of Cartesian coordinates. Here, we write this Hamiltonian in coordinates adapted to the binary geometry, which could be useful in future work on precessing binaries~\cite{Pan:2013rra}.
Then, specializing to the case of equatorial orbits, we apply a linear stability analysis to the
canonical equations of motion to obtain the epicyclic frequency in
terms of derivatives of the Hamiltonian. Combining the result with the
algebraic relations for determining the angular momentum and
radius-frequency relationship for circular orbits leads to the gauge
invariant expression for the PA as a function of the orbital
frequency.  We compare the results, which we evaluate numerically,
with data from the NR simulations of Refs.~\cite{Mroueinprep, 2012arXiv1210.2958M,Hemberger2013,2013arXiv1304.6077M} and assess the
performance of spinning EOB models. 

By construction, the results derived from the EOB Hamiltonian when
specialized to the extreme mass-ratio limit directly reduce to the dynamics of a spinning particle in a  Kerr spacetime, to linear order in the particle's spin. As an independent check and an extension to
quadratic order in the spin we also compute the PA from the equations 
of motion in the gravitational skeleton approach~\cite{Mathisson:1937zz, 
Papapetrou:1951pa, Corinaldesi:1951pb,Taub1964, Dixon1970, Dixon1974, EhlersRudolph,
  2012CQGra..29e5012H}. The magnitude of the particle's
spin scales with the mass ratio as $s m/M$, where $0\leq s\leq 1$ is a dimensionless spin parameter for a compact
object and $m/M \ll 1$. Linear-in-spin corrections to geodesic motion thus enter at
the same order in the mass ratio as the gravitational self-force
\cite{2011LRR....14....7P}. While dissipative effects are dominated by
the gravitational radiation-reaction force \cite{2004PhRvD..69d4011B},
the influences of the spin and self-force on the conservative dynamics
could be equally important \cite{SP12,2004PhRvD..69d4011B}. We extend
the comparisons of Refs.~\cite{SP12,2004PhRvD..69d4011B,2011PhRvD..83b4027F} here to
include information beyond strictly circular orbits obtained from the
PA. In addition, we also compute explicitly the Lense-Thirring and spin
precession frequencies for small deviations from circular equatorial orbits.

The organization of this paper is as follows. In Sec.~\ref{sec:eobH},
we express the EOB Hamiltonian in spherical coordinates and in a generic 
fixed source frame. In Sec.~\ref{sec:eobcirc}, we specialize to aligned or 
antialigned spins and compute the angular momentum and frequency of 
circular orbits. Then, we apply the general method to compute libration frequencies to the EOB 
Hamiltonian in Sec.~\ref{sec:linearstability} and obtain the results for the epicyclic frequency. In 
Sec.~\ref{sec:EOB-NR} we compare them to NR data and discuss the efficiency of the EOB spin resummations in Sec.~\ref{sec:EOBspin}. In Sec.~\ref{sec:particle} we consider the case of a 
spinning particle on a nearly circular equatorial orbit in Kerr spacetime and derive explicit expressions for the PA 
and the precession frequencies for the orbital plane and spin vector. 
In Sec.~\ref{sec:SF} we specialize the spinning particle results to Schwarzschild and compare the spin-dipole and gravitational self-force contributions to the energy 
and PA. Finally, Sec.~\ref{sec:conclusion} contains 
our main conclusions. 

\vspace{0.2truecm}

Henceforth, greek letters denote spacetime indices and run over
$0,1,2,3$; latin letters from the middle of the alphabet $i,j,k$ are
spatial indices, while Latin letters from the beginning of the
alphabet $a,b,c$ are Minkowski spacetime indices. Summations over any
repeated indices are implied and square brackets around pairs of
indices indicate antisymmetrization, e.g., \mbox{$x^{[a}p^{b]}=(x^a p^b-x^b
p^a)/2$}. We use units with $G=c=1$ throughout. An asterisk when used
in a superscript denotes the dual of a tensor, e.g., 
\mbox{$R^*_{abcd}=\epsilon_{cd}^{\; \; \; fg}R_{abfg}/2$}, and
\mbox{$^*R_{abcd}=\epsilon_{ab}^{\; \; \; fg}R_{fgcd}/2$}, where
\mbox{$\epsilon_{0123}=1$} is the permutation symbol. The basis vectors of a
timelike tetrad are denoted by \mbox{$e_a^\mu$} and the Ricci rotation
coefficients by \mbox{$\omega_{ab}^{\; \; \; c}=e_a^\mu e_b^\nu e^c_{\nu; \mu}$}. A semicolon indicates the
covariant derivative and a comma denotes the partial
derivative. Boldface symbols stand for spatial vectors, vectorial
arrows denote four-vectors.  The notation for the quantities in the
EOB model follows that of Refs.~\cite{Barausse:2009aa,Barausse:2009xi}.


\section{Effective-one-body model}
\subsection{Hamiltonian in spherical coordinates}
\label{sec:eobH}

In the spinning EOB model of Refs.~\cite{Barausse:2009aa,Barausse:2009xi,Barausse:2011ys} 
(see also Refs.~\cite{Damour01c,Damour:024009,2011PhRvD..84h4028N,2013PhRvD..87l4036B} for 
a different implementation of the EOB Hamiltonian with spins), the dynamics of two 
black holes with masses $m_1$ and $m_2$ and spins $\bm{S}_1$ and $\bm{S}_2$ is generated by the Hamiltonian
\be
H_{\rm real}=M\sqrt{1+2\nu (H_{\rm eff}-1)}\,,\label{eq:Hreal}
\ee
where \mbox{$M=m_1+m_2$} and \mbox{$\nu=m_1 m_2/M^2$}. The Hamiltonian \mbox{$H_{\rm eff}$} describes an effective particle of mass \mbox{$\mu=\nu M$} and spin \mbox{${\bm S}_*$}
moving in a deformed, fully symmetrized (under the interchange of the body labels) Kerr metric 
with mass \mbox{$M$} and spin \mbox{$\bm{S}_{\rm Kerr}={\bm S}_1+{\bm S}_2$}. 

The constrained Hamiltonian of a spinning particle in Kerr spacetime was
calculated in Boyer-Lindquist coordinates in
Ref.~\cite{Barausse:2009aa}, using a spherical reference tetrad to
describe the dynamics of the particle's spin. In the flat space limit, such 
a description of the spins includes additional spin-orbit couplings besides the usual \mbox{$\bm{S}_* \cdot \bm{L}$} coupling. As discussed in Ref.~\cite{Barausse:2009xi}, the spherical gauge terms can be avoided when using the Cartesian components of the spins and the resulting Hamiltonian, derived in Cartesian quasi-isotropic coordinates, is more convenient for the mapping to an EOB model with two precessing spins. 
Below, we express this Hamiltonian explicitly in spherical coordinates for the orbital variables defined with
respect to the frame in which \mbox{${\bm S}_{\rm Kerr}$} is along the
$z$-axis. The structure of the re-expressed Hamiltonian will be
analogous to that in Ref.~\cite{Barausse:2009aa}, where it was
directly related to geometric and physical quantities. Here, however,
this structure is just a convenient way to arrange the terms.

 We use dimensionless spatial coordinates \mbox{${\bm x}$} and momenta \mbox{${\bm P}$} for the effective particle, with \mbox{${\bm x}$} in units of $M$ and \mbox{${\bm P}$} in units of $\mu$.
In the EOB spherical coordinates \mbox{$x^i=( r, \theta, \phi)$} with canonically conjugate specific
momenta \mbox{$P_i=(P_r, P_\theta,P_\phi)$} the effective Hamiltonian of Ref.~\cite{Barausse:2009xi} can be written as
\begin{subequations}
\label{eq:Heffform}
 \bea  H_{\rm eff}&=&H^{\rm NS}+H^{\rm S}+H^{\rm SS}, \label{eq:Hgeneric}\\
 H^{\rm NS}&=&\beta P_\phi+\alpha \sqrt{Q+2 \nu (4-3\nu) P_r^4/r^2}, \\
 H^{\rm S}&=&\left[\bm{F}_t+\left(\beta+\frac{\alpha \gamma P_\phi}{\sqrt{Q}}\right)\bm{F}_\phi\right]\cdot \bm{S}_*\nonumber\\
&&+\frac{\alpha
}{\sqrt{Q}}\left(\gamma^{rr}P_r \bm{F}_r+
\gamma^{\theta\theta}P_\theta \bm{F}_\theta\right)\cdot \bm{S}_* ,\ \ \ \ \ \ \ \ \ \ \ \ \ \ \ \label{eq:HwithF} 
\\
 H^{\rm SS }&=&\frac{1}{2 r^3} \left[3 (\bm{S}_* \cdot \hat{\bm{n}})^2-\bm{S}_*\cdot \bm{S}_*\right],
 \eea 
\end{subequations}
where \mbox{$\hat{ \bm n }$} is a radial unit vector. Here, the
following combinations of the specific momenta and spins appear 
\bes\bea Q&=&1+\gamma
P_\phi^2+\gamma^{rr}P_r^2+\gamma^{\theta \theta}P_\theta^2, \label{eq:Qdef}\\
\bm{ S}_*&=&{\bm{ \sigma}}_* \left[1+\nu f_{*}(r, \mathbf{P}) \right]+ \nu g_{*}(r, \mathbf{P}) {\bm{ \sigma}}, \label{eq:mapsstar}\\
{\bm{ \sigma}} &\equiv& \bm{S}_{\rm Kerr}= {\bm{S}}_1+{\bm{S}}_2 \ , \\
{\bm \sigma}_* &=&  \frac{m_2}{m_1}{\bm{S}}_1+\frac{m_1}{m_2}{\bm{S}}_2.
\eea \ees 
The functions \mbox{$f_*$} and \mbox{$g_*$} depend on the choice of identification between the PN spin terms and the EOB functions~\cite{Barausse:2011ys}.  
Except in the test particle limit, the effective particle's spin \mbox{$\bm{S}_*$} is not a canonical quantity but merely a function of canonical variables \mbox{$(\bm{x}, \bm{P}, \bm{S}_1, \bm{S}_2)$}. The metric functions are
\bes
\label{eq:potadm}
\bea \alpha &=& \frac{\sqrt{\Delta_t \Sigma}}{\sqrt{\Lambda_t}},\ \ \ \beta=\frac{2\sigma r}{\Lambda_t},\\
 \gamma &=&\frac{\Sigma}{\Lambda_t\sin^2\theta}, \ \ \
\gamma^{rr}=\frac{\Delta_r}{\Sigma}, \ \ \ \gamma^{\theta\theta}=\frac{1}{\Sigma} \ \ \
, \eea
\ees
where \bes 
\label{eq:various}
\bea
\sigma\equiv|\bm{\sigma}|&=&\sqrt{|\bm{S}_1|^2+|\bm{S}_2|^2+2\bm{S}_1\cdot \bm{S}_2}, \label{eq:magns}\\
\Delta_t&=&r^2 A(r)+\sigma^2, \\
\Sigma&=&r^2+\sigma^2 \cos^2\theta, \\
\Delta_r &=&\Delta_t D^{-1}(r), \\
\Lambda_t&=&(r^2+\sigma^2)^2-\sigma^2 \Delta_t \sin^2\theta. \eea
\ees 
The specific form of the potentials
\mbox{$A(r)$} and \mbox{$D^{-1}(r)$} depends on the choice of the EOB model. The vectors \mbox{$\bm{F}_t$, $\bm{F}_r$, $\bm{F}_\theta$ and $\bm{F}_\phi$} in Eq.~(\ref{eq:HwithF}) are given by
\begin{widetext}
\bes
\bea
\bm{F}_\phi &=&\cos\theta  \ \bm{\hat{ n}}+ \bm{\hat{ v}},\\
\bm{F}_t&=& \bm{\hat{n}} \ \frac{\sqrt{\gamma}\sqrt{ \gamma^{\theta\theta}}}{\sqrt{Q}}\left[\frac{P_\phi \alpha_{,\theta}(1+2\sqrt{Q})}{(1+\sqrt{Q})}-\alpha P_\phi \cot \theta-\frac{(1-2\sqrt{Q})\beta_{,\theta}}{2\gamma}\right]\nonumber\\
&+&\bm{\hat{ v}} \ \frac{\csc\theta\sqrt{\gamma^{rr}}}{\sqrt{\gamma}}\left[\frac{\gamma P_\phi \alpha_{,r}}{(1+\sqrt{Q})}+\frac{(2\sqrt{Q}-1)\beta_{,r}+\alpha P_\phi \gamma_{,r}}{2\sqrt{Q}}\right], \; \; \; \\
\bm{F}_r &=& -\bm{\hat{ n}} \ \frac{\sqrt{\gamma^{\theta\theta}}(\beta_{,\theta}P_r+\beta_{,r}P_\theta)}{2\alpha \sqrt{\gamma}(1+\sqrt{Q})}- \bm{\hat{ v}} \ \frac{\csc\theta\left(\beta_{,\theta} \gamma^{\theta \theta}P_\theta+2 P_r \gamma^{rr} \beta_{,r}\right)}{2\alpha \sqrt{\gamma} \sqrt{\gamma^{rr}}(1+\sqrt{Q})}- \bm{\hat{ \xi}} \ \frac{\csc\theta \sqrt {\gamma^{\theta \theta}}}{2\alpha \sqrt{\gamma^{rr}}}\left[\frac{2\sqrt{Q}\alpha_{,\theta}+P_\phi \beta_{,\theta}}{ (1+\sqrt{Q})}+\frac{\alpha \gamma^{\theta\theta}_{,\theta}}{\gamma^{\theta\theta}}\right],\ \ \\
\bm{F}_\theta &=& - \bm{\hat{ n}}  \ \frac{\sqrt{\gamma^{\theta \theta}}\beta_{,\theta}P_\theta}{\alpha \sqrt{\gamma}(1+\sqrt{Q})}- \bm{\hat{ v}} \ \frac{\csc\theta\sqrt{\gamma^{rr}} P_\theta \beta_{,r}}{2\alpha \sqrt{\gamma}(1+\sqrt{Q})}+\bm{\hat{ \xi}}  \csc\theta \ \bigg[1+\frac{\sqrt{\gamma^{rr}}}{2\alpha\sqrt{\gamma^{\theta \theta}}}\bigg(\frac{2\sqrt{Q} \alpha_{,r}+P_\phi \beta_{,r}}{ (1+\sqrt{Q})}+\frac{\alpha \gamma^{\theta \theta}_{ \; ,r}}{\gamma^{\theta \theta}} \bigg)\bigg]. \; \; \; \; 
\eea
\label{eq:Fsgeneric}
\ees
\end{widetext}
Here, the Cartesian unit vectors \mbox{$(\bm{\hat{ n}},\bm{\hat{ \xi}},\bm{\hat{ v}})$} are defined by
\be
\hat{\bm n}=\frac{{\bm x}}{r}, \ \ \ \ \ \ \ \hat{\bm \xi}=\hat e_{\rm Z}^{\rm \sigma}\times \hat {\bm n}, \ \ \ \ \ \ \  \hat {\bm v}=\hat{\bm n}\times \hat {\bm \xi},\label{eq:vectors}
\ee
where \mbox{$\hat{e}_{\rm Z}^{\rm \sigma}={\bm \sigma}/\sigma$} denotes the direction of the (deformed) 
Kerr spin. 

When the spins are precessing, the spherical coordinates tied to the
spin are no longer adequate for describing the motion in a fixed
frame. For this reason, Cartesian coordinates are used in current 
implementations of the EOB model for gravitational-wave--template construction~
\cite{Pan:2009wj,Taracchini2012,Pan:2013rra,Hinder:2013oqa,Barausse:2009xi,Barausse:2011ys}. 
The disadvantage of Cartesian coordinates is that the direct connection to
the binary geometry is obscured. An alternative geometric coordinate
choice that is analogous to the Keplerian orbital elements in celestial
mechanics was adapted to spinning binaries in
Refs. \cite{2005PhRvD..71b4039K,2009PhRvD..80l4034T,2013PhRvD..87f4035T} 
(see also Refs.~\cite{2010PhRvD..82j4031G,1998PhRvD..58l4001G,2009PhRvD..79d3016L}
for related parametrizations).  Below, we provide a brief description
of the modifications necessary to write the EOB Hamiltonian using the modified Keplerian coordinates 
that can be employed in future work related to template construction~\cite{Pan:2009wj,Taracchini2012,Pan:2013rra,Hinder:2013oqa}.  
References \cite{2005PhRvD..71b4039K,2009PhRvD..80l4034T,2013PhRvD..87f4035T}
provide the details and derivations relevant to this choice of variables. 

\begin{figure}
 \includegraphics[width=0.9\columnwidth]{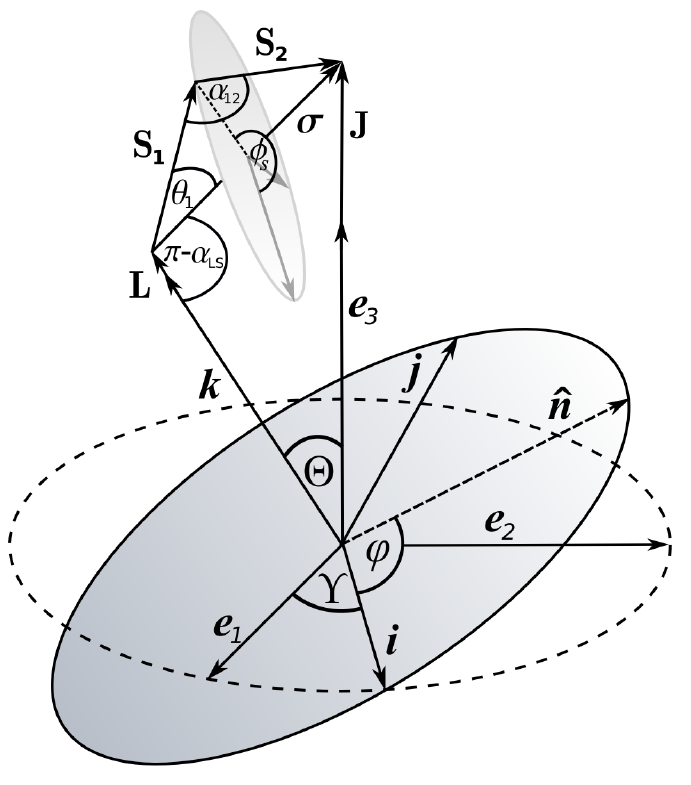}
\caption{{\textit{Parameter choice for the binary geometry.}} The \mbox{$\bm{e}_3$}-axis of the fixed frame is along the direction of the total angular momentum $\bm{J}$. The orbital plane is perpendicular to \mbox{$\bm{k}=\bm{L}/L$} and the vector $\bm{i}$ points to 
the intersection of the orbital plane and the plane normal to \mbox{$\bm{e}_3$}.
The total spin \mbox{$\bm{\sigma}$} and \mbox{$\bm{L}$} subtend the angle \mbox{$\alpha_{LS}$}. 
The projection of \mbox{$\bm{S}_1$} onto the \mbox{$(\bm{i}, \hat{e}_Z^\sigma \times \bm{i})$}-plane defines the angle \mbox{$\phi_S$}. }
\label{fig:angles}
\end{figure}

The dynamical variables in this
approach are the set of radial variables \mbox{$(r, P_r)$} and the magnitudes of the angular momenta together
with various angles parametrizing rotations from different
frames to a fixed reference frame \mbox{$({\bm e}_1,{\bm e}_2, {\bm
  e}_3)$}. These angles are illustrated in Fig.~\ref{fig:angles}  
and defined as follows. Without loss of generality, we can choose the orientation of
the fixed frame to be \mbox{$\bm{e}_3=\bm{J}/J$} because the total
angular momentum \mbox{${\bm J}={\bm L}+{\bm \sigma}$}, where
\mbox{$\bm{L}=\bm{x}\times \bm{P}$}, is conserved for the dynamics generated
by the EOB Hamiltonian. We introduce an orbital frame
\mbox{$(\bm{i},\bm{j},\bm{k})$} such that \mbox{$\bm{k}=\bm{L}/L$}. The
rotation between the two frames defines an inclination angle \mbox{$\Theta$} (the angle of the precession cone of $\bm{L}$ around $\bm{J}$) 
and an angle \mbox{$\Upsilon$} measuring the longitude of the line of nodes,
the intersection of the orbital plane with the fixed equatorial
plane. Specifically, \mbox{$\cos\Theta={\bm k}\cdot {\bm e}_3$} and
\mbox{$\cos\Upsilon={\bm i}\cdot {\bm e}_1$}. The relative separation
vector is given by \mbox{$\bm{x}=r \cos \varphi \, {\bm i}+r \sin\varphi \,
{\bm j}$}, where $\varphi$ is the azimuthal angle in the instantaneous
orbital plane. With these conventions the radial unit vector in the fixed frame is
\bea
\bm{\hat n}&=&\left(\cos \Upsilon  \cos \varphi -\cos \Theta  \sin
   \Upsilon  \sin \varphi , \right.\nonumber\\
&&\left.\cos \Theta  \cos \Upsilon
    \sin \varphi +\sin \Upsilon  \cos \varphi, \sin \Theta  \sin \varphi \right). \; \; \;\; \; \;  \label{eq:neuler}
\eea
The instantaneous direction of the total spin in the fixed frame can be expressed as
\be
\hat e_{\rm Z}^{\rm \sigma}=\left(\tilde w \sin \Upsilon,-\tilde w \cos
   \Upsilon ,\sin \Theta  \sin
   \alpha _{\text{LS}}+\cos \Theta  \cos
 \alpha_{\text{LS}}\right),  \; \;\; \; \label{eq:zeuler}
\ee
where \mbox{$\alpha_{\text{LS}}$} is the angle between ${\bm L}$ and ${\bm \sigma}$ (measured in the opposite sense to $\Theta$) with \mbox{${\bm L}\cdot {\bm \sigma}=L \sigma \cos \alpha_{\text{LS}}$}, and \mbox{$
\tilde w=\sin \Theta  \cos \alpha _{\text{LS}}-\cos \Theta  \sin \alpha _{\text{LS}}.$ }
The vectors \mbox{${\bm{\hat v}}$} and \mbox{$\hat {\bm \xi}$} can then be constructed by using Eqs.~(\ref{eq:neuler}) and (\ref{eq:zeuler}) in Eq.~(\ref{eq:vectors}). 
All occurrences of \mbox{$\cos\theta=\bm{\sigma}\cdot \bm{\hat n}$} in the metric functions and Hamiltonian should be replaced by 
\be
\cos\theta=\sin\alpha_{\rm LS} \sin\varphi . \label{eq:costheta}
\ee
Note that this differs from Ref.~\cite{2013PhRvD..87f4035T}, where the replacement for \mbox{$\cos\theta$} involves \mbox{$\Theta$} instead of \mbox{$\alpha_{\rm LS}$}, because the \mbox{$\theta$}-coordinate in the Hamiltonian is defined with respect to the direction of the deformed Kerr spin rather than the \mbox{$\bm{e}_3$}-axis in the fixed frame. 
The angular momenta appearing in Eqs.~(\ref{eq:Fsgeneric}) are given by
\be
P_\phi= L\cos\alpha_{\rm LS} , \; \; \; \; P_\theta \sin \theta = - L\cos\alpha_{\rm LS} \cos\varphi \label{eq:Peuler}
\ee
and the function $Q$ in this parametrization becomes
\be
Q=1+\frac{\Delta_rP_r^2}{\Sigma}+\frac{L^2}{\Lambda_t\Sigma} \frac{\Sigma^2\cos^2\alpha_{\rm LS}+\Lambda_t \cos^2\varphi \sin^2\alpha_{\rm LS}}{1-\sin^2\alpha_{\rm LS} \sin^2\varphi}.\label{eq:Qeuler}
\ee
The individual spins expressed in the fixed frame depend on several additional angles: $\theta_1$, defined by \mbox{$S_1 \sigma\cos\theta_1={\bm S}_1\cdot {\bm \sigma}\;$}; \mbox{$\theta_2=\theta_1+\alpha_{12}-\pi$}, where \mbox{$\alpha_{12}$} is the angle between \mbox{${\bm S}_1$} and \mbox{${\bm S}_2$}; and an azimuthal angle $\phi_S$ measured between $\bm{i}$ and the projection of \mbox{${\bm S}_1$} on the plane perpendicular to \mbox{$\hat e_{\rm Z}^{\rm \sigma}$}. The scalar products involving the spins \mbox{$\bm{S}_{\rm A}$} where $A=1,2$ that are needed for computing the terms containing \mbox{${\bm S}_*$} in the Hamiltonian are
\bes\label{eq:Sdots}
\bea
{\bm S}_{\rm A} \cdot \hat {\bm{n}}&=&S_{\rm A}\left(\cos\theta _{\rm A} \sin
   \varphi  \sin \alpha
   _{\text{LS}}-w\sin \theta _{\rm A} \right),\;\;\;\;\;\;\;\;\;\\
{\bm S}_{\rm A} \cdot \hat {\bm v}&=&S_{\rm A} \left[\cos \theta _{\rm A}
   \left(1-\sin ^2\varphi  \sin ^2\alpha
   _{\text{LS}}\right)\right.\nonumber\\
&&+\left. w\sin \theta
   _{\rm A} \sin \varphi  \sin \alpha
   _{\text{LS}} \right.],\\
{\bm S}_{\rm A} \cdot \hat {\bm \xi}&=&S_{\rm A} \sin \theta _{\rm A} \left(\sin
  \varphi  \cos \alpha
   _{\text{LS}} \cos \phi
   _S \right.\nonumber\\
&& \ \ \ \ \ \ \; \left.-\cos \varphi  \sin \phi
   _S\right),
\eea
\ees
where \mbox{$
w=\sin \varphi  \cos\alpha _{\text{LS}} \sin
   \phi _S+\cos\varphi  \cos\phi _S.
$} The angles \mbox{$\Theta, \theta_{\rm A}, \theta_S$} and \mbox{$\alpha_{\rm LS}, \alpha_{12}$} are functions of the magnitudes \mbox{$\sigma, L, J, S_1, S_2$} fixed by the instantaneous geometry of the binary.
From Eqs.~(\ref{eq:costheta})-(\ref{eq:Sdots}) it follows that the Hamiltonian in this parametrization has the form \mbox{$H_{\rm eff}(r,\varphi, P_r, L, J, \phi_S, \sigma, S_1, S_2)$}. 
The Poisson brackets are \mbox{$1=\{r,P_r\}=\{\Upsilon, J\}=\{\varphi,L\}=\{\phi_S,\sigma\}$}, the magnitudes \mbox{$S_1$} and \mbox{$S_2$} are conserved (their conjugate angles are cyclic coordinates) and the other angles are determined by geometric considerations. 

To compute the PA, we can without loss of generality specialize to equatorial orbits since the radial and precessional motions are independent. Motion in the equatorial plane requires that the spins be collinear to the orbital angular momentum, implying that \mbox{$\Theta=0$} and the other angles \mbox{$\theta_A, \, \alpha_{\rm LS}$} and \mbox{$\alpha_{12}$} either $0$ or $\pi$. In this case all the frames coincide and the \mbox{$\bm{\hat v}$}-components of Eqs.~(\ref{eq:Fsgeneric}) specialized to \mbox{$\theta=\pi/2$, $P_\theta=0$} can be used directly. The transformations discussed above would however be needed to compute the dragging of the nodes and the spin precessions in the EOB model using the method of Sec.~\ref{sec:linearstability}. 

\subsection{Specialization to equatorial orbits, angular momentum and frequency of circular orbits}
\label{sec:eobcirc}
The Hamiltonian for equatorial orbits obtained from the \mbox{$\bm{\hat v}$}-components of Eqs.~(\ref{eq:Fsgeneric}) with \mbox{$\theta=\pi/2, \, P_\theta=0$} simplifies to be
\bea &&H^{\rm eff}_{\rm eq}=\beta
P_\phi+\alpha \sqrt{Q}-\frac{S_*^2}{2 M r^3}\; \; \; \; \; \; \; \; \; \; \; \; \label{eq:Hequat}\\
&& \; \; \; \;  +S_*
\bigg[F_t^{\rm eq}+\left(\beta+\frac{\alpha\gamma
P_\phi}{\sqrt{Q}}\right) -\frac{
(\gamma^{rr})^{3/2}P_r^2 \beta_{,r}}{\sqrt{\gamma}
\sqrt{Q}(1+\sqrt{Q})}\bigg],\nonumber
\eea
where \mbox{$S_*=|\bm{S}_*|$} and
\be F_t^{\rm eq}=
\frac{\sqrt{\gamma^{rr}}}{2\sqrt{\gamma}}\left[\frac{2\gamma P_\phi
\alpha_{,r}}{(1+\sqrt{Q})}+\frac{\alpha P_\phi
\gamma_{,r}-\beta_{,r}}{\sqrt{Q}}+2\beta_{,r}\right]. \label{eq:Ftequat}\ee
For circular equatorial orbits, the third term on the second line of Eq.~(\ref{eq:Hequat}) vanishes and \mbox{$Q=1+r^2P_\phi^2/\Lambda_t$}. The resulting expression for the Hamiltonian agrees with Eq.~(C4) in Ref.~\cite{2011PhRvD..83d4044Y} when we substitute for the metric functions and for $Q$.
The quantity \mbox{$P_\phi$} for circular orbits is determined by solving
\mbox{$\partial H_{\rm eff}/\partial r=0$} which is explicitly
\bea
0&=&\beta_{,r} P_\phi+\alpha_{,r} \sqrt{Q}+\frac{\alpha\gamma_{,r}P_\phi^2}{2\sqrt{Q}}+S_{*}\frac{\partial F_t^{\rm eq}}{\partial r}+\frac{\partial S_{*}}{\partial r}
\frac{\partial H_{\rm eq}^{\rm eff}}{\partial S_*}\nonumber\\
&+&S_{*} \left[\beta_{,r}+\frac{(\alpha \gamma)_{,r} P_\phi}{\sqrt{Q}}-\frac{\alpha \gamma \gamma_{,r}P_\phi^3}{2 Q^{3/2}}\right]+\frac{3S_{*}^2}{2 r^4}.  \label{eq:Pphicirc}
\eea
The orbital frequency is
\bea 
&&H^{\rm circ}_{\rm real}\, \Omega_\phi =\frac{\partial H_{\rm eff}}{\partial
P_\phi}=\beta+\frac{\alpha \gamma P_\phi}{\sqrt{Q}}\nonumber\\
&& \; \; \; \; \; \;\;  \; \; \; \;  \; \; \; \; +S_*
\left[\frac{\alpha
\gamma}{\sqrt{Q}}-\frac{\alpha \gamma^2 P_\phi^2}{Q^{3/2}}+\frac{\partial F_t^{\rm eq}}{\partial P_\phi}\right]\nonumber\\
&& \; \; \; \; \; \; \;  \; \; \; \;  \; \; \; \; +\frac{\partial
S_*}{\partial P_\phi}  \left[\left(\beta+\frac{\alpha \gamma
P_\phi}{\sqrt{Q}}\right)+F_t^{\rm eq}-\frac{S_*}{ r^3}\right]\, , \; \; \; 
\; \; \; \; \; \; \; \; \; \label{eq:omegaphi}
\eea
where \mbox{$H_{\rm real}^{\rm circ}$} is the Hamiltonian (\ref{eq:Hreal}) with \mbox{$H_{\rm eff}= H^{\rm eff}_{\rm eq}|_{ P_r=0}$}. To obtain \mbox{$P_\phi(\Omega_\phi)$} we substitute the solution to Eq.~(\ref{eq:Pphicirc}) for \mbox{$P_\phi(r)$} into Eq.~(\ref{eq:omegaphi}), solve for \mbox{$r(\Omega_\phi)$} and use the result to compute \mbox{$P_\phi(r(\Omega_\phi))$}. We implemented these manipulations numerically using {\sc{mathematica}}. 

For nonspinning binaries, the metric potentials reduce to \mbox{$\alpha=\sqrt{A(r)}$}, \mbox{$\beta=0$}, \mbox{$\gamma=r^{-2}$}, \mbox{$\gamma^{rr}=A(r) D^{-1}(r)$} and \mbox{$\Lambda_t=r^4$}. The solution to Eq.~(\ref{eq:Pphicirc}) is then given explicitly by \mbox{$P_\phi^2=r^3 A'(r)/[2 A(r)-r A'(r)]$}. Using this in Eq.~(\ref{eq:omegaphi}) leads to \mbox{$H_{\rm real}^{\rm circ}\Omega_\phi=\sqrt{A'(r)}/\sqrt{2r}$}. These results agree with Eqs.~(4.5) and (4.10) in Ref.~\cite{Damour:2009sm} after converting the dependences on \mbox{$u=r^{-1}$}. 

\subsection{Libration frequencies for small deviations from equilibrium}
\label{sec:linearstability}
Having determined the quantities for equatorial orbits, we
now consider the epicyclic frequency of small perturbations to such
orbits. We employ a linear stability analysis to compute the radial
frequency using a general method that is also applicable to the
computations of the other frequencies in
Sec.~\ref{sec:particle}. Introducing the vector of canonical variables
\mbox{$y^L=(x^i, P_i, S_{1}^i,S_2^i)$}, the equations of motion derived from
the Hamiltonian (\ref{eq:Hreal}) take the form
\be \label{eq:ode} {\dot y}^L = f^L(y^K),  \ee
where the dot denotes \mbox{$d/dt$}. The generalization to systems with a higher dimensional phase space and different time evolution parameter is straightforward; one just replaces $t$ with the evolution parameter and all the vectors and matrices by their higher dimensional counterparts. 
We are interested in the behavior of solutions to Eq.~(\ref{eq:ode}) near an equilibrium configuration \mbox{$y^L_0$} corresponding to a circular equatorial orbit, where \mbox{$f^L(y_0^K)=0$} except for \mbox{$L=3$} which is Eq.~(\ref{eq:omegaphi}). Linearizing \mbox{$y^L= y_{0}^L+\zeta^L$}, where \mbox{$\zeta^L/y_0^L \ll 1$} represents a small deviation vector, leads to a set of linear differential equations with constant coefficients
\be \label{eq:lin} \dot{\zeta}^L =\bigg( \frac{\partial f^L}{\partial y^K} \Big|_{y=y_{0}}\bigg) \zeta^K + \mathcal{O}(\zeta^2). \ee
 We decompose the solutions to this system into the
eigenvalues and eigenvectors of the stability (Jacobian) matrix \mbox{$(\partial f^L/\partial y^K)$}. The eigenvalues $\lambda$ characterize the rate at which 
trajectories with small differences in initial conditions separate, since the eigensolutions to Eq.~(\ref{eq:lin}) are \mbox{$\sim e^{\pm \lambda t}$}. 
Complex values of the exponents $\lambda$ correspond to the
frequencies of libration about a stable equilibrium, while real, positive $\lambda$ either reflect the sensitivity to initial conditions of chaotic orbits or characterize the unstable direction of a hyperbolic point (e.g., a marginally stable orbit).

We compute Eq.~(\ref{eq:lin}) using the EOB Hamiltonian with the
circular equatorial orbit values for \mbox{$y_0$} and find the eigenvalues. The result for the radial
frequency is
\be
(H^{\rm circ}_{\rm real})^2\Omega_r^2= \frac{\partial^2 H^{\rm eff}_{\rm eq}}{\partial r^2}\frac{\partial^2 H^{\rm eff}_{\rm eq}}{\partial P_r^2} \bigg\rvert_{P_r=0}. \  \label{eq:omegar}
\ee
Note that Eqs.~(\ref{eq:Hequat}) and (\ref{eq:Qdef}) show that the term \mbox{$H_{,P_r P_r}$} in Eq.~(\ref{eq:omegar}) involves \mbox{$\gamma^{rr}$}, the only metric potential that depends on the EOB function \mbox{$D(r)$}. As can be verified by direct computation, this potential does not appear in the solutions for the circular orbit quantities from Eqs.~(\ref{eq:Pphicirc}) and (\ref{eq:omegaphi}). The expression  (\ref{eq:omegar}) is entirely equivalent to that obtained with the method based on perturbing the effective potential \cite{Damour:2009sm} which is determined by solving $H=E$ for \mbox{$P_r^2$.}   

We express the radial frequency (\ref{eq:omegar}) in terms of the gauge invariant frequency by using Eqs.~(\ref{eq:Pphicirc}) and (\ref{eq:omegaphi}) to eliminate $P_\phi$ and $r$. The angle of PA is related to Eqs.~(\ref{eq:omegar}) and (\ref{eq:omegaphi}) by 
\mbox{$\Delta \Phi_{\rm PA}=2\pi \vert { K} -1\vert $},
where 
\be
{ K}=\frac{\Omega_\phi}{\Omega_r}.\label{eq:kappadef}
\ee
Note that $K$ being the ratio of two frequencies is independent of the choice of time parametrization. The numerical values we obtain for $K$ 
will be used to compare with NR data in Sec.~\ref{sec:EOB-NR} below. For nonspinning binaries, Eq.~(\ref{eq:kappadef}) reduces to \mbox{$K^{-2}=D^{-1}(r)[rA(r) A''(r)/A'(r)-2r A'(r)+3 A(r)]$}, which agrees with Eq.~(5.19) of \cite{Damour:2009sm}.

The method described above can also be used to compute the 
precession frequencies for meridional oscillations of the orbital plane and for the spins in the case of small deviations from exact collinearity of the angular momenta. We will calculate these quantities explicitly for the case of a spinning particle in Sec.~\ref{sec:particle}. However, we do not provide these precession frequencies for the EOB model because they involve the same EOB potentials already present in the PA and data for comparisons with other approaches is currently lacking. In principle the precessions can be obtained by using in Eq.~(\ref{eq:lin}) the equations of motion for the variables discussed at the end of Sec. I A and finding the characteristic exponents. 

\begin{table*}
\begin{tabular}{ |rrr|rrr|rrr|rrr| c| }
\hline
& & & &  K & & & K+$\Delta$K& & & K-$\Delta$K& &\\
 q& $\chi_1$ & $\chi_2$&$a_0$& $a_1$  & $a_2$   &  $a_0$  &  $a_1$  & $ a_2 $ &  $a_0$  &  $a_1$  &   $a_2$ &$ M\Omega_\phi$ \\\hline
1 & $0.97$& $0.97$& 1.00764 &-3.9949 & -70.807 & 1.0065  & -3.9406 & -67.121 & 0.99418 & -2.7579 &-101.543 & [0.0169, 0.0344]\\
1& $0.95$& $0.95$ & 0.98830 &-2.2364 &-107.11  & 0.99952 & -3.2598 & -79.725 & 0.98340 & -1.7802 &-122.724 & [0.0184, 0.0318]\\
1 & 0.9 & 0.9     & 0.96487 &-0.3254 &-138.67  & 0.96828 & -0.5883 &-130.568 & 0.99319 & -2.6815 & -94.834 & [0.0200, 0.0310] \\
1 & 0.8 & 0.8     & 0.98881 &-1.8428 &-104.636 & 1.00304 & -3.1416 & -73.026 & 0.97868 & -0.9219 &-127.882 & [0.0177, 0.0317]\\
1 & 0.6 & 0.6     & 0.99923 &-2.0355 & -86.061 & 1.01226 & -3.1797 & -56.734 & 0.97886 & -0.2337 &-128.612 & [0.0190, 0.0310]\\
1 & -0.95 & -0.95 & 1.09949 &-7.4342 & 346.477 & 1.32874 &-33.6076 &1099.42  & 0.78660 & 26.3466 &-570.337 & [0.0177, 0.0260]\\
1 & -0.9 & -0.9   & 0.96722 & 6.4391 & -34.411 & 1.3842  &-38.2326 &1175.23  & 0.68089 & 37.9372 &-908.291 & [0.0177, 0.0240]\\
3 & 0.5 & 0.5     & 0.97678 & 0.2600 &-117.537 & 0.98929 & -0.8407 & -93.074 & 0.95453 &  2.1900 &-161.296 & [0.0180, 0.0320]\\
1 & 0 & 0         & 0.99555 & 0.5048 & -76.340 & 0.99679 &  0.2800 & -62.419 & 0.99430 &  0.7297 & -90.261 & [0.0120, 0.0320]\\
1 & 0.5 & 0       & 0.98950 & 0.2893 &-106.77  & 1.01884 & -3.0266 &  -8.075 & 0.95792 &  3.8258 &-210.184 & [0.0155, 0.0250]\\
1 & -0.5 & 0      & 0.93781 & 6.5575 &-171.793 & 1.2331  &-23.1674 & 588.235 & 0.84533 & 17.1947 &-486.223 & [0.0195, 0.0259]\\
1.5 & 0.5 & 0     & 0.97522 & 1.4335 &-139.448 & 1.03313 & -4.6662 &  30.687 & 0.92707 &  6.5007 &-281.776 & [0.0158, 0.0259]\\
1.5 & -0.5 & 0    & 0.99988 & 1.0478 & -30.022 & 1.00286 &  0.6444 & -15.295 & 0.99588 &  1.5908 & -49.196 & [0.0123, 0.0215]\\
3 & 0.5 & 0       & 1.00301 &-1.7336 & -65.616 & 1.02202 & -3.7818 &  -7.466 & 0.99159 & -0.4448 &-105.151 & [0.0164, 0.0287]\\
3 & -0.5 & 0      & 1.00559 & 0.7585 &  17.065 & 1.01162 &  0.0921 &  38.352 & 0.99854 &  1.5502 &  -8.129 & [0.0130, 0.0270]\\
5 & 0.5 & 0       & 0.99812 &-1.2905 & -76.358 & 0.99779 & -1.1426 & -79.709 & 0.99846 & -1.4383 & -73.008 & [0.0169, 0.0280]\\
5 & -0.5&0        & 1.02734 &-1.3157 & 101.025 & 1.03345 & -1.9245 & 117.851 & 1.02648 & -1.2087 &  95.786 & [0.0179, 0.0360]\\
8 & 0.5&0         & 0.97198 & 0.7119 &-114.923 & 0.98183 &  0.0285 &-102.411 & 0.96138 &  1.4528 &-128.537 & [0.0210, 0.0420]\\ 
8 & -0.5&0        & 1.02556 &-1.2578 & 130.85  & 1.05938 & -4.3455 & 203.072 & 0.99952 &  1.2217 &  69.698 & [0.0200, 0.0300]\\\hline
\end{tabular}
\caption{\textit{Periastron advance extracted from numerical simulations.}  The first three columns give mass-ratio and spin-projection onto the orbital angular momentum for the aligned-spin binary black hole simulations which are considered here.   The next three columns give the fitting parameters of Eq.~(\ref{eq:KNR-fit}) for the periastron advance, followed by fits of the lower and upper error-bounds.  The rightmost column indicates the frequency range within which each fit is valid.
 \label{tab:NR-fits}}
\end{table*}

\subsection{Reduction to the case of a spinning particle in Kerr spacetime}
\label{sec:tpl}
In this subsection we briefly outline the specialization of the EOB
results to the extreme mass-ratio case, where they describe a spinning
dipole in Kerr spacetime. The explicit expressions will be given in
Sec. \ref{sec:particle}, where we present a complementary approach
using the multipolar equations of motion. The EOB potentials for a
Kerr spacetime are \mbox{$A(r)=1-2/r$} and \mbox{$\, D^{-1}(r)=1$}, where distances
are in units of the black-hole mass parameter $M$. The Kerr spin 
reduces to \mbox{$\sigma=\pm aM^2$}, where $0\leq a \leq 1$ is the dimensionless Kerr spin
parameter and the upper (lower) signs correspond to prograde (retrograde) orbits. The spin \mbox{$S_*$} becomes \mbox{$S_*=\pm s m/M$}, where \mbox{$0\leq s\leq 1$} is
the particle's spin parameter, $m$ is its mass and the signs denote the relative orientation of the particle's spin and orbital angular momentum. Keeping only terms
up to linear order in \mbox{$S_*$}, we perturbatively solve for the
circular orbit quantities from Eqs.~(\ref{eq:Pphicirc}) and
(\ref{eq:omegaphi}). After
eliminating $P_\phi$ in favor of the conserved quantity \mbox{$J_z=P_\phi\pm S_*$}, substituting the Kerr metric
functions into the equatorial Hamiltonian of Eq.~(\ref{eq:Hequat}) and
using the perturbative circular orbit quantities we arrive at the
\mbox{$O(s)$} terms in Eq.~(\ref{eq:ratioofomega}) below.

\section{Comparison to numerical-relativity periastron advance}
\label{sec:EOB-NR}

\subsection{Numerical data}

Throughout this section, we use the notation \mbox{$q=m_1/m_2\ge 1$} for the
mass ratio and \mbox{$\chi_A=(\bm{S}_A \cdot \hat k)/m_A^2$} denotes the
spin-component along the orbital angular momentum (i.e. \mbox{$\chi_A<0$}
denotes a spin anti-parallel to \mbox{$\mathbf{L}$}).

We consider 19 numerical relativity simulations, performed with the
Spectral Einstein Code~\cite{SpECwebsite} (SpEC).  The simulations with
equal masses and equal spins (\mbox{$q=1, \chi_A=\chi_B$}) were presented
in Refs.~\cite{Hemberger2013a,Lovelace:2011nu,Lovelace:2010ne}; the
remaining simulations were presented in Ref.~\cite{2012arXiv1210.2958M}.
All simulations are also part of the SpEC binary black-hole simulation
catalog~\cite{2013arXiv1304.6077M}.  For all runs, only the inspiral
phase is used, with computational methods described
in~\cite{Chu2009,Scheel2009,Rinne2008b,Boyle2007,Scheel2006,Pfeiffer-Brown-etal:2007,Buchman:2012dw}.

Calculation of the periastron advance from the numerical simulations
is performed with the techniques described
in Refs.~\cite{LeTiecinprep,LeTiec:2011bk}.  In short, we compute the orbital frequency
\begin{equation}
\Omega(t) = \frac{|\mathbf{r}(t)\times\dot{\mathbf{r}}(t)|}{|\mathbf{r}(t)|^2},
\end{equation}
where \mbox{$\mathbf{r}(t)$} is the coordinate distance between the centers
of the apparent horizons of the two black holes.  Orbital eccentricity induces
oscillations into \mbox{$\Omega(t)$}, which are extracted by a suitable fit, from which
\mbox{$K_{\rm NR}$} is extracted as a function of orbital frequency \mbox{$M\Omega_\phi$}.  
To make the numerical data more easily usable, polynomial fits are performed of the form
\begin{equation}\label{eq:KNR-fit}
K_{\rm NR}(M\Omega_\phi) = 
\left[a_0+a_1(M\Omega_\phi)+a_2(M \Omega_\phi)^2\right]K_{\rm Schw},
\end{equation}
with \mbox{$K_{\rm Schw}=[1-6(M\Omega_\phi)^{2/3}]^{-1/2}$}.  The resulting
fits are listed in Table~\ref{tab:NR-fits}.  The accuracy with which
\mbox{$K_{\rm NR}$} can be computed depends sensitively on the orbital
eccentricity of the individual simulations, therefore we give error
bounds separately for each simulation.

\subsection{Comparison with the baseline effective-one-body model}

In this section we illustrate the results for the PA obtained from the
EOB Hamiltonian using Eqs.~(\ref{eq:kappadef}), (\ref{eq:omegar}),
(\ref{eq:Hequat})--(\ref{eq:omegaphi}), (\ref{eq:potadm}) and
(\ref{eq:various}). For the potentials \mbox{$A(r)$} and \mbox{$D^{-1}(r)$} that
appear in the metric functions in Eqs.~(\ref{eq:various}) we use the
3PN accurate Taylor series \cite{Barausse:2009xi}:
\mbox{$A=1-2/r+2\nu/r^3+\nu(94/3-41 \pi^2 /32)/r^4$} and
\mbox{$D^{-1}=1+6\nu/r^2+2(26-3\nu)\nu/r^3$}. We will show below that the PA
depends only weakly on the choice of the functions \mbox{$f_*$} and \mbox{$g_*$} in
the expression (\ref{eq:mapsstar}) for \mbox{$S_*$}. The most accurate
mapping that we use as the baseline model for the comparisons includes the 
3.5PN spin-orbit terms given in Eqs.~(51) and (52) of Ref.~\cite{Barausse:2011ys}, with all
the gauge parameters therein set to zero, \mbox{$a_j=b_j=0$} for
\mbox{$j=0,1,2,3$}. As such, the EOB model contains only information
available from PN theory and the test particle limit, without any
additional calibrations from NR. We verified that using the calibrated spinning EOB model~\cite{Taracchini2012} with the logarithmically resummed potentials $A$ and $D$ from Sec.~VE of Ref.~\cite{Barausse:2009xi} gives similar results in the low-frequency regime relevant here. We also include the predictions from PN theory in
the comparison, using the highest PN orders currently available: 3.5PN
order in the spin-orbit effects and 3PN order in the spin-spin effects 
(see the companion paper ~\cite{LeTiecinprep} for more details on the PA in PN theory). 
We find that including the 3PN spin-spin effects significantly improves the agreement with the NR data in all
cases.

\begin{figure}
  \includegraphics[width=0.9\columnwidth]{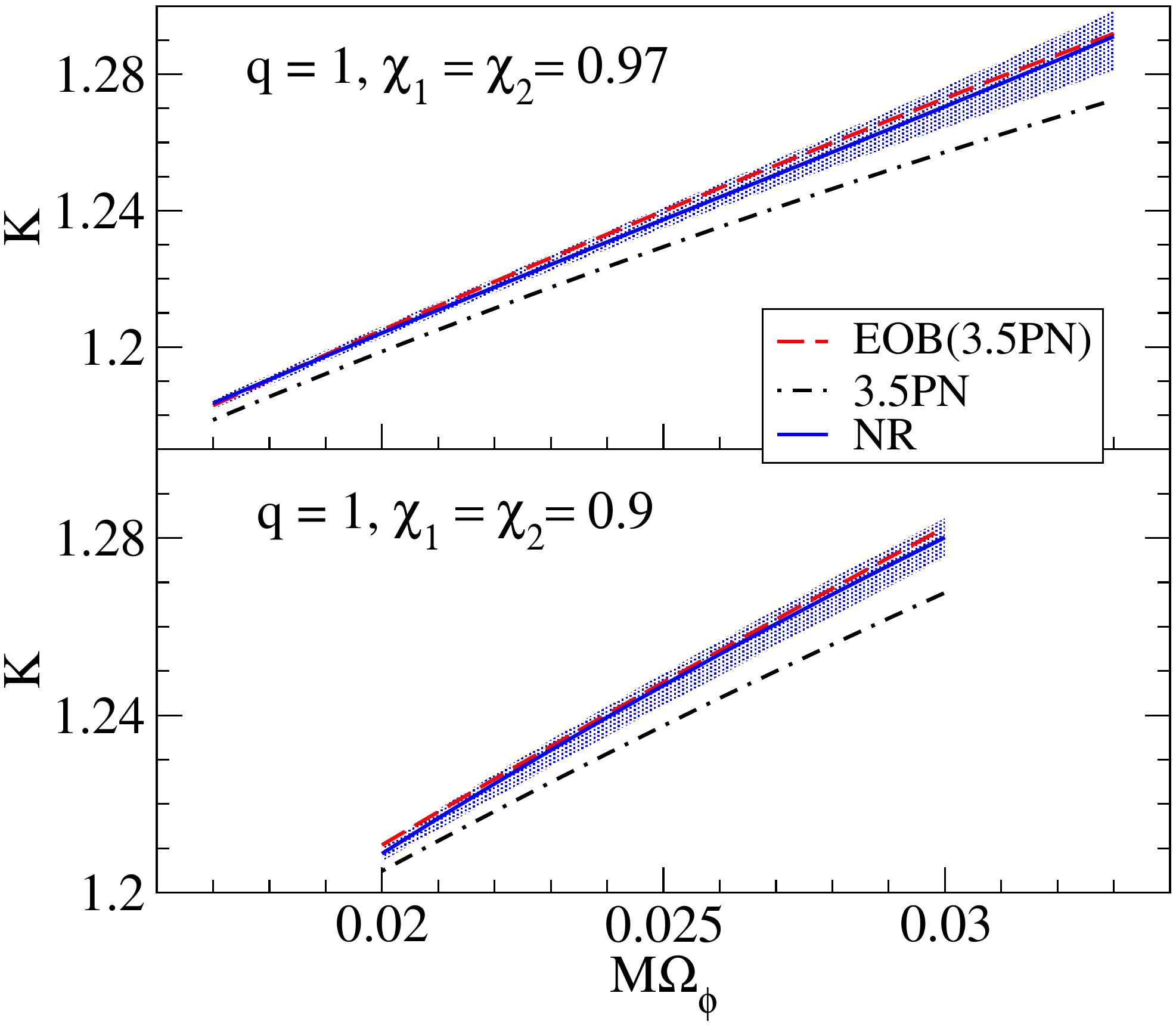}
 \includegraphics[width=0.9\columnwidth]{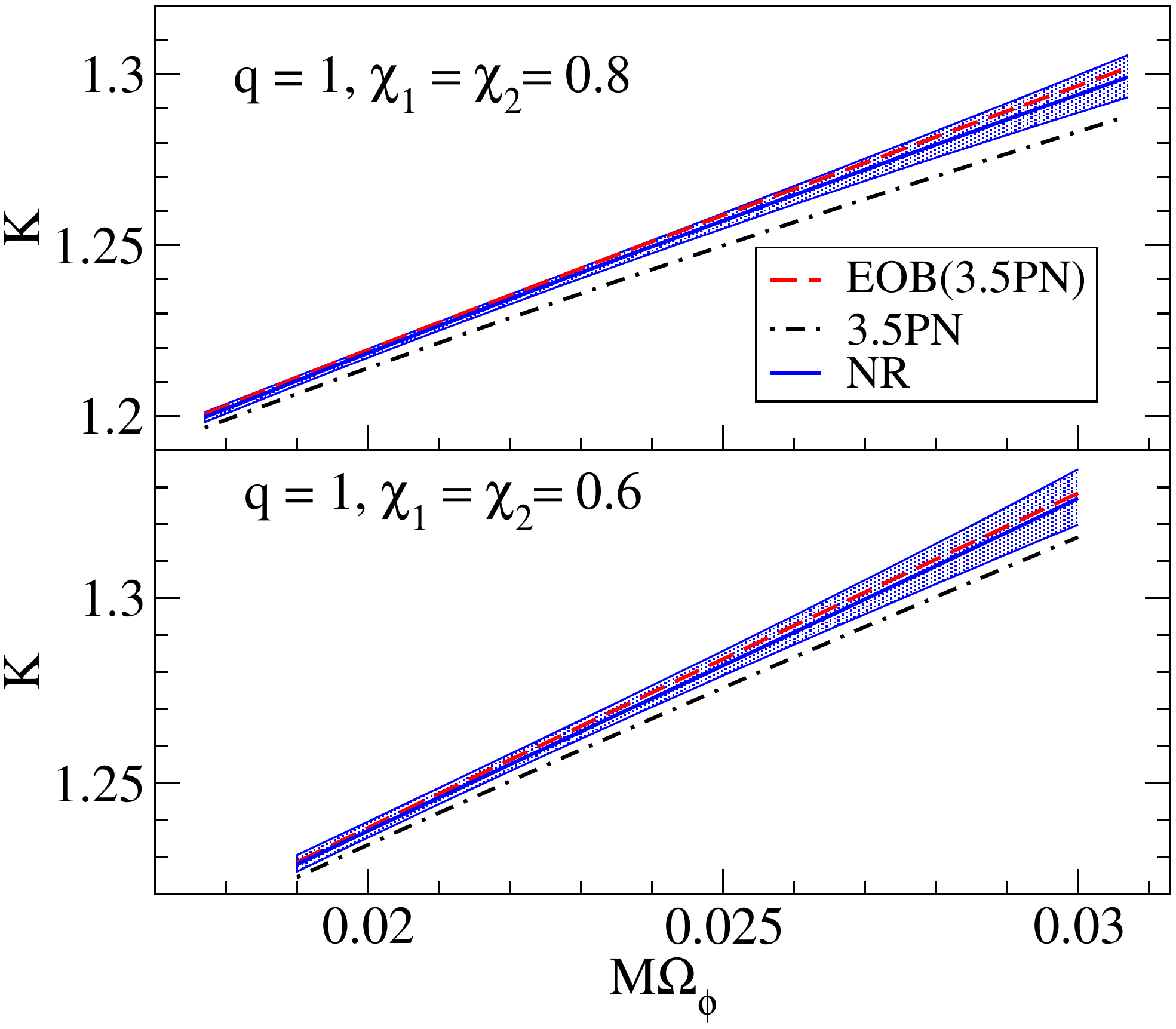}\\
\caption{\textit{Aligned spins, equal masses}. 
The EOB results are shown as red dashed lines. The solid blue curves are the fits to the NR data and the shaded region indicates the error estimate. The dash-dotted black curves are the PN predictions. 
}
\label{fig:aligned}
\end{figure}

\begin{figure}
\includegraphics[width=0.9\columnwidth]{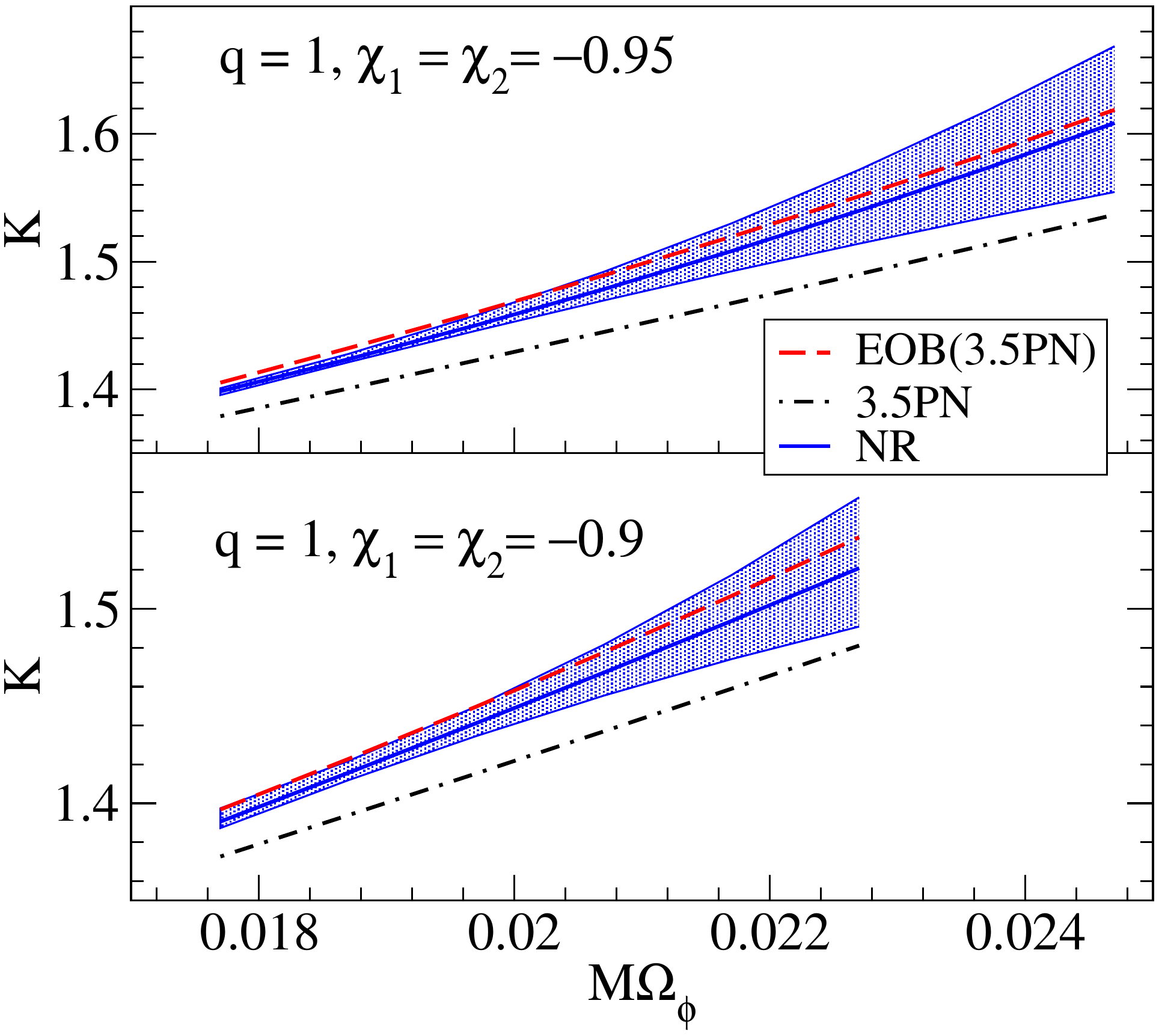}
\caption{\textit{Antialigned spins, equal masses}.}
\label{fig:antialigned}
\end{figure}

We show in Fig.~\ref{fig:aligned} the frequency ratio $K$ as a
function of the azimuthal frequency for equal-mass binaries with
aligned spins. Quite interestingly, the EOB prediction closely tracks the NR data over the
entire frequency range considered, even for the case of spins close to maximal 
\mbox{$\chi_1=\chi_2 = 0.97$}. The difference between EOB and NR
is within the estimated numerical error. The PN results, instead, are outside
the NR error bounds, with the discrepancy decreasing for lower
frequencies and lower spins. At low frequencies the curves should all
converge to $K=1$.
Two cases with antialigned spins, \mbox{$\chi_1=\chi_2=-0.95$} and
\mbox{$\chi_1=\chi_2=-0.9$} are displayed in Fig.~\ref{fig:antialigned},
again for equal masses. The agreement between EOB and NR in this case
is still very good, but we notice that the EOB prediction for the case
\mbox{$\chi_1=\chi_2=-0.95$} (\mbox{$\chi_1=\chi_2=-0.9$}) is slightly outside (coincides with) 
the numerical error for \mbox{$M\Omega_\phi \,\laq\, 0.02$}. We find that the small discrepancy, 
\mbox{$0.3\%$}, in the case \mbox{$\chi_1=\chi_2=-0.95$} does not change significantly when computing the EOB model at different 
PN orders. We plan to investigate this oddity in the future using black-hole simulations with antialigned spins  
and different spin magnitudes, and larger eccentricities. In fact, the extraction of the periastron advance 
for anti-aligned simulations is more delicate than for simulations with aligned spins. 
For the \mbox{$\chi_1=\chi_2=-0.95$} simulation, the eccentricity is $10^{-3}$, and to obtain a better 
estimate for the periastron advance a larger eccentricity is required.

\begin{figure}
\includegraphics[width=0.9\columnwidth]{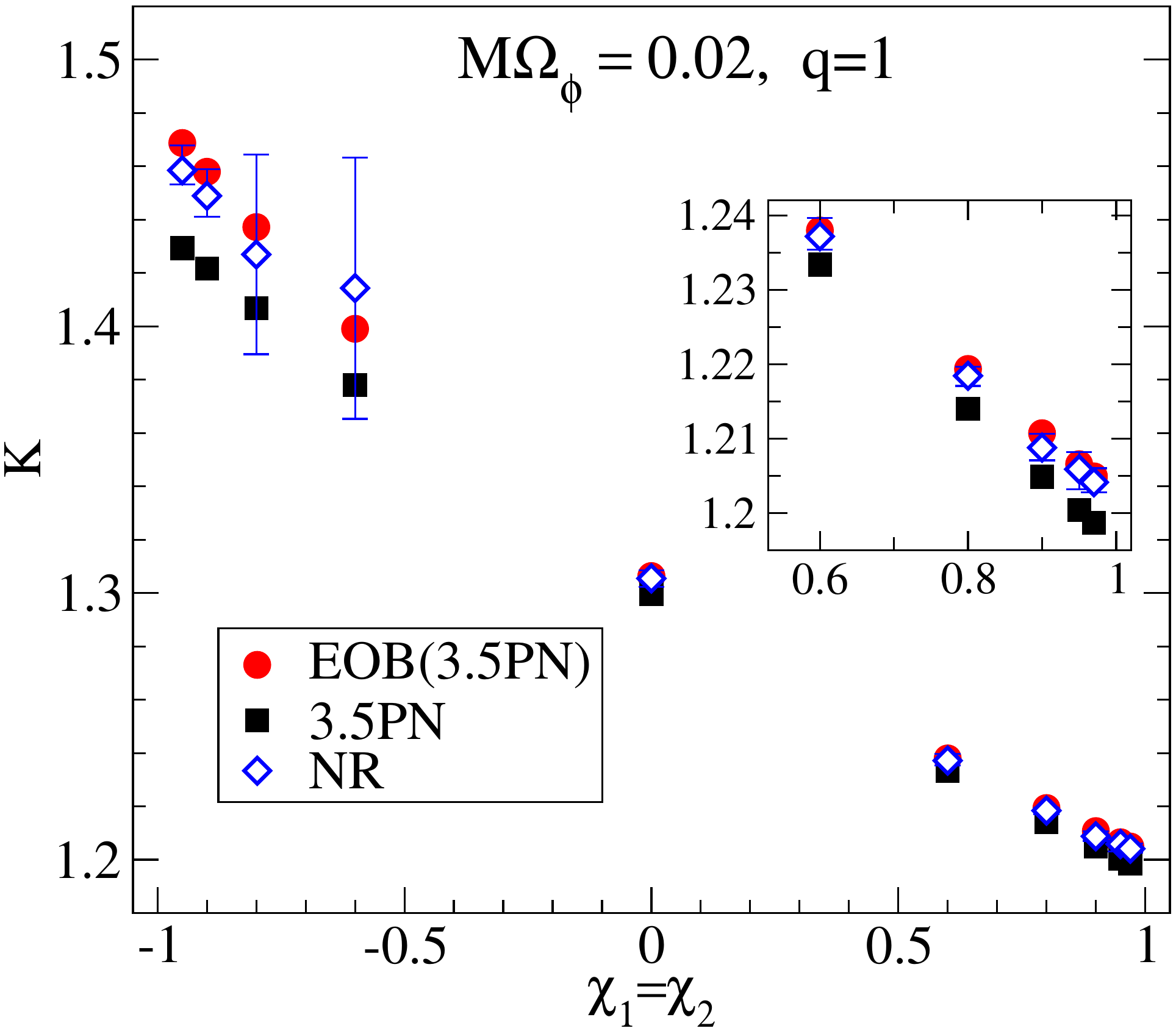}
\caption{\textit{Varying the spin parameter}. Binaries with equal spins and mass ratio $q=1$ at a fixed frequency $M\Omega_\phi=0.02$. 
The red circles are the EOB result, the blue diamonds are the NR data points, together with their error bounds. The black squares are the PN predictions. The inset shows an enlargement for spins $\ge 0.6$.}
\label{fig:fnsofchi}
\end{figure}

Since the comparisons above showed that the EOB and NR data for equal-mass binaries with equal spins agree over a range in frequency, we now pick a fiducial low frequency \mbox{$M\Omega_\phi=0.02$} and consider variations in the spin parameter, shown in Fig.~\ref{fig:fnsofchi}. The blue diamonds correspond to the NR data and its error bounds, the red circles are the EOB prediction and the black squares indicate the PN results. 
The increase in $K$ with decreasing spin parameter at fixed \mbox{$M\Omega_\phi$} is largely due to the fact that for smaller spins a binary at the fiducial frequency is closer to its innermost stable circular orbit (ISCO), where the PA diverges because the radial frequency goes to zero. This also accounts for the trend in the performance of the PN results in Fig.~\ref{fig:fnsofchi} which become increasingly accurate farther from the ISCO.
We confirmed this reasoning for the behavior of $K$ in the case of a test particle in Kerr spacetime (using the expressions given in Sec.~\ref{sec:particle} below). We find that when evaluated at a fixed circular orbit radius, \mbox{$K(\chi_{\rm Kerr})$} decreases with increasing spin, whereas when instead evaluated at a fixed distance from the ISCO, \mbox{$K(\chi_{\rm Kerr})$} monotonically increases as this location shifts towards the strong-field region.

\begin{figure}
\includegraphics[width=0.9\columnwidth]{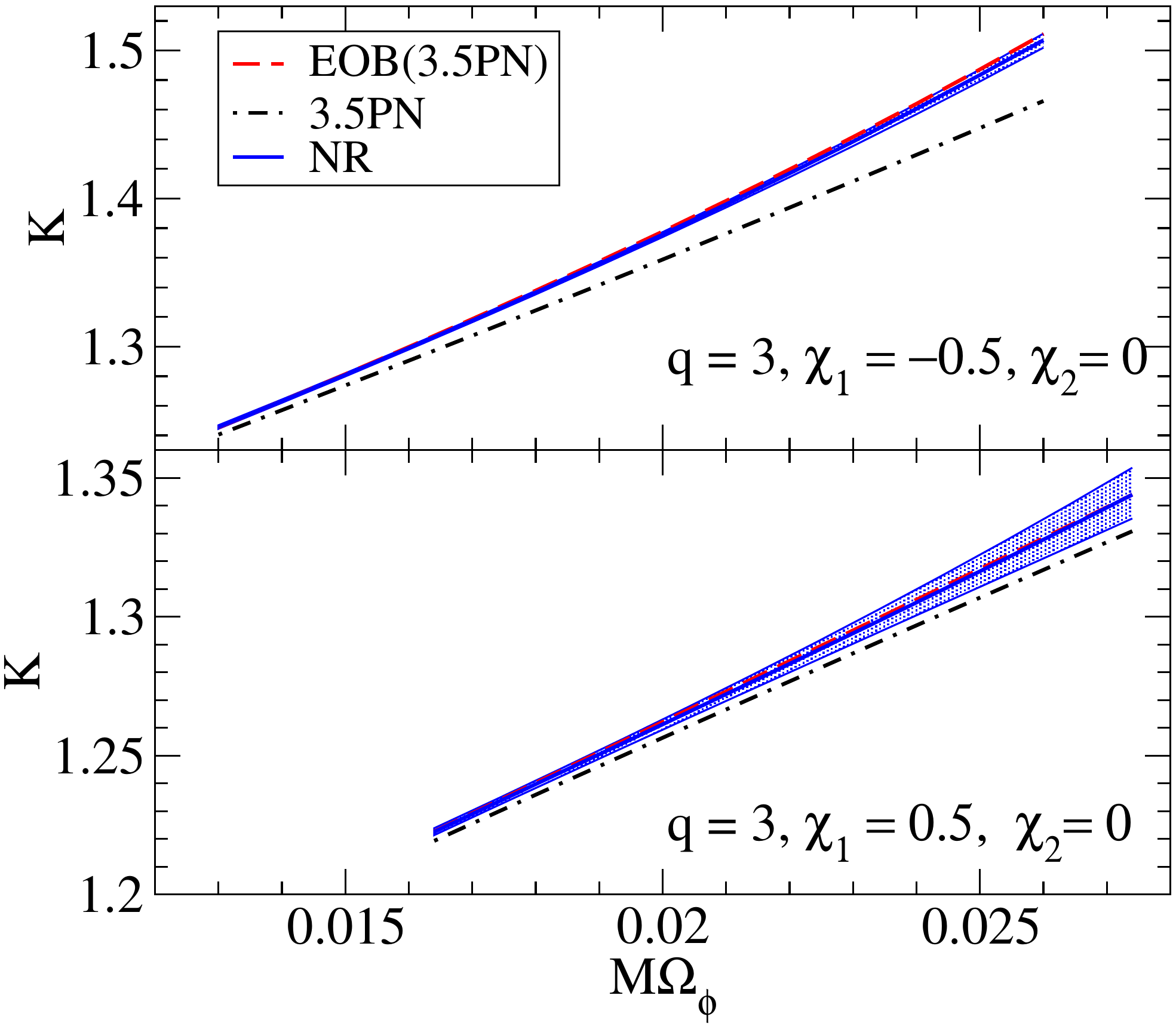}
\caption{\textit{Unequal masses}. Results for a binary with mass ratio $q=3$ and a single spin $\chi_1=\mp 0.5$. 
 }
\label{fig:unequalm}
\end{figure}

\begin{figure}
\includegraphics[width=0.9\columnwidth]{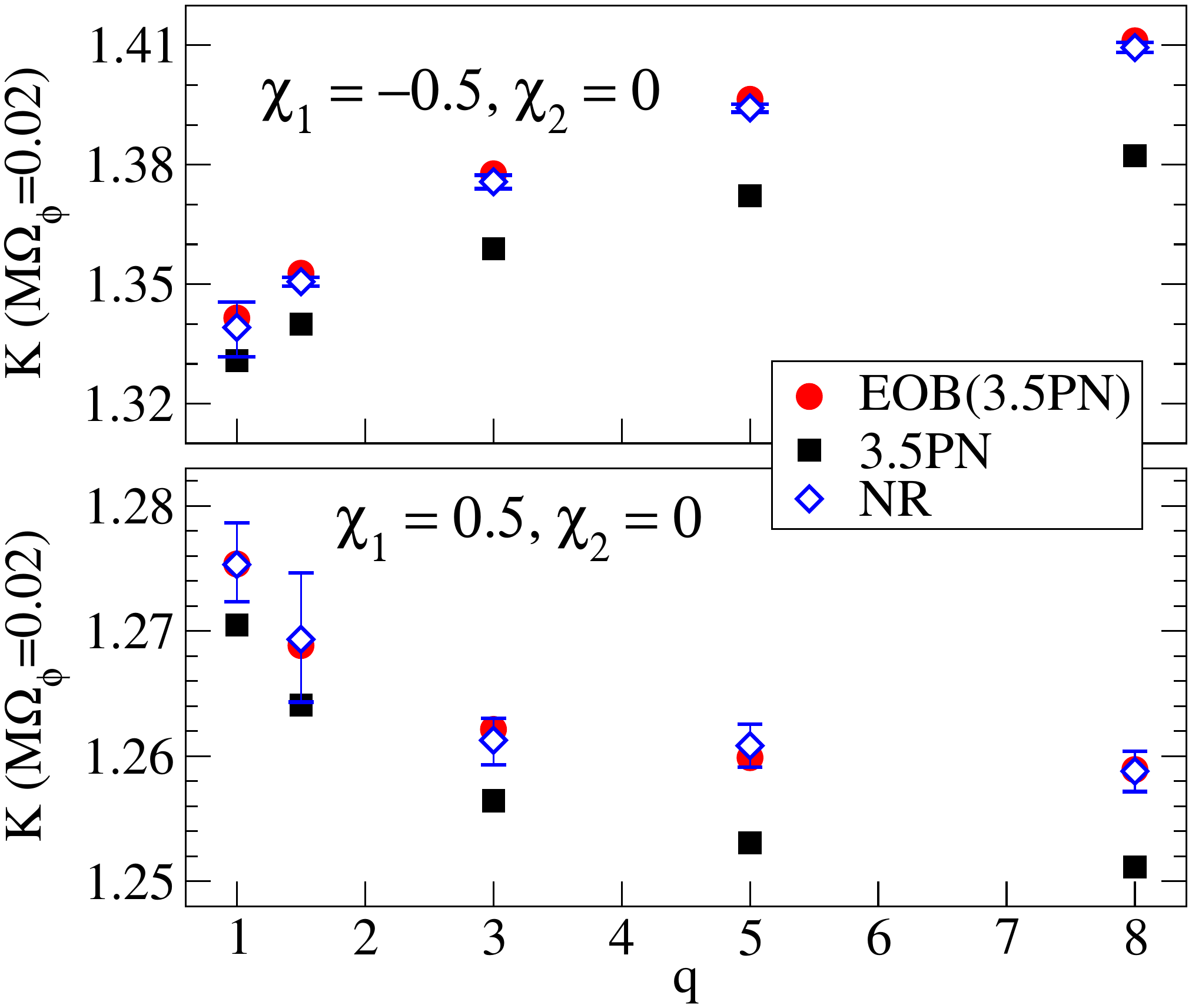}
\caption{\textit{Varying the mass ratio}. Results for binaries with a single spin $\chi_1=\mp 0.5$ at $M\Omega_\phi=0.02$.}
\label{fig:onespinq}
\end{figure}

Next, we consider the effect of varying the mass ratio. 
The values of $K$ for a binary with mass ratio $q=3$ and a single spin \mbox{$\chi_1=\mp 0.5$} are shown in Fig.~\ref{fig:unequalm} over a range of frequencies. Again, the EOB prediction tracks the NR data very closely. The results for $q=3$ with two spins \mbox{$\chi_1=\chi_2=0.5$} look very similar to the single spin cases in Fig.~\ref{fig:unequalm} and we do not show them here. We also verified that for
 binaries which have only one spinning component with \mbox{$\chi_1=\pm 0.5$}, the plots of $K$ versus \mbox{$\Omega_\phi$} are qualitatively very similar to those discussed above, showing excellent agreement for \mbox{$\chi_1=0.5$} and are marginally outside the error over a small low-frequency range in some of the cases with $\chi_1=-0.5$, and we do not include these figures here. Instead, in Fig.~\ref{fig:onespinq} we summarize the information for single spin systems with different mass ratios by evaluating them at the fiducial frequency. The EOB predictions are in good agreement with the NR data for all mass ratios. By contrast, the discrepancies between the PN predictions and the NR results increase for larger mass ratios, largely due to the closer proximity to the ISCO with increasing mass ratio. 


In summary, the comparisons with NR data show that throughout the
region of parameter space in frequencies, mass ratios and spins
considered here, the EOB model provides an excellent prediction for the PA,
while the PN results have larger differences to the NR data. This
re-affirms the utility of the EOB resummation for approximating the
conservative dynamics including nonprecessional spin effects.

\subsection{Varying the effective-one-body model}
\label{sec:EOBspin}
\begin{figure}
\includegraphics[width=0.9\columnwidth]{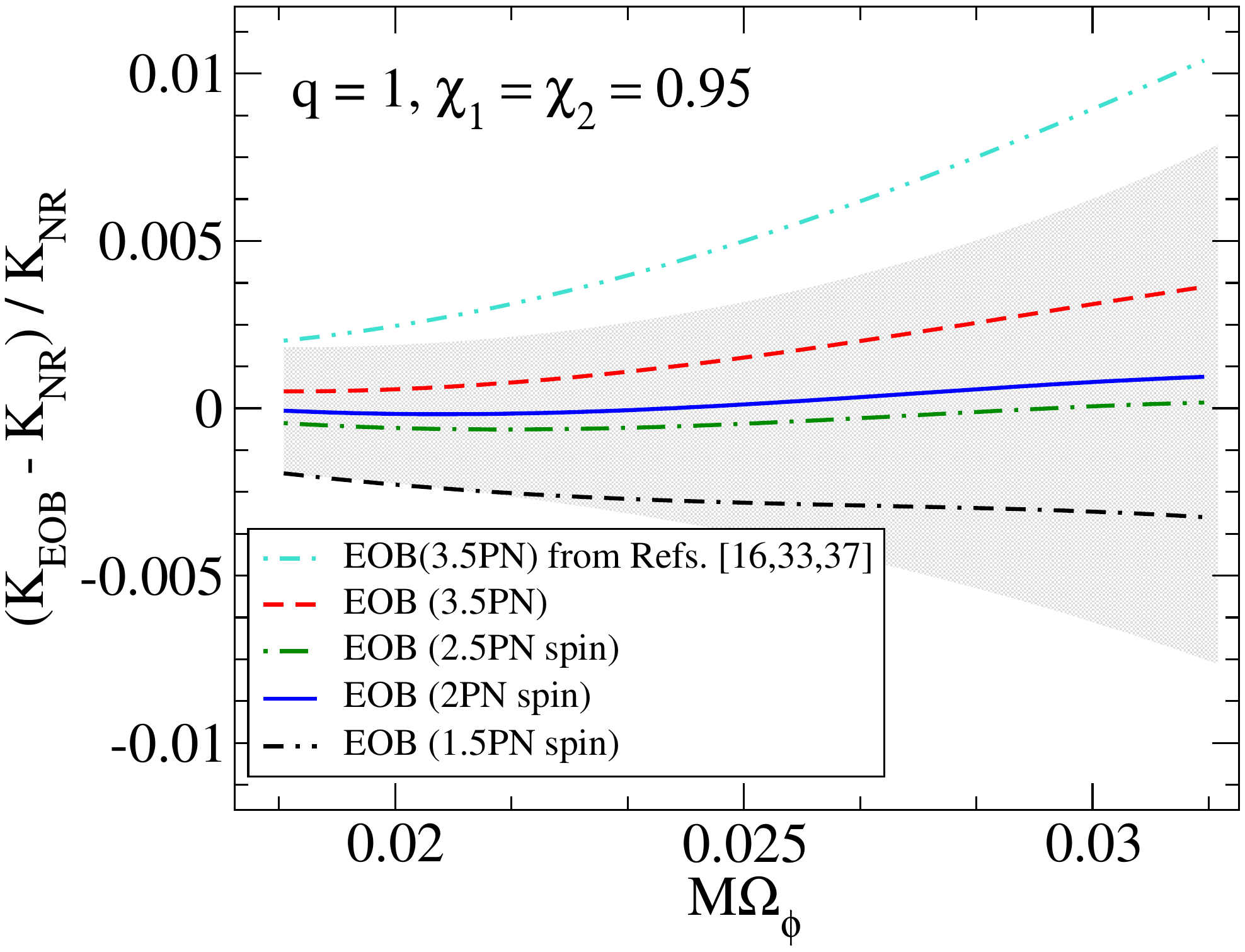}
\caption{\textit{Effect of varying the PN spin information included in the EOB model}. 
This figure shows, for the case $\chi_1=\chi_2=0.95$, $q=1$, the fractional differences in
  the PA prediction from the EOB model and the NR data when either we vary the PN order of the spin terms 
in the baseline EOB model or when we consider the EOB model of Refs.~\cite{Damour01c,Damour:024009,2011PhRvD..84h4028N} in which the terms linear in the effective particle's spin are not deformations of
the results for a spinning particle in Kerr.  All the EOB models employ the nonspinning 3PN terms. The gray shaded area indicates the uncertainty of the numerical data $K_{\rm NR}$.}
\label{fig:EOBPNorders}
\end{figure}

As mentioned before and discussed in detail in the companion paper ~\cite{LeTiecinprep}, 
the PN results show substantial improvement
when including higher-order spin terms in the model. On the other
hand, the EOB model is a resummation of all the PN terms to $O(\nu)$
in which the test-particle limit imposes the overall structure in two
ways: (i) the effective particle moves in a deformed Kerr spacetime
symmetrized with respect to the two-body masses and spins, and (ii)
the terms linear in the effective particle's spin are deformations of
the results for a spinning particle in Kerr.  
The PN information on the $\nu$ dependence enters via the deformations of the metric potentials
and the identification between the effective spin and the spins in the
binary. To assess the robustness of the EOB model under
  variations of the PN spin information we generated plots such as
  Fig.~\ref{fig:EOBPNorders} for all the data.  This figure is
  representative of the qualitative features in all cases considered. 
  It shows for a binary with  $\chi_1=\chi_2=0.95$, $q=1$, the fractional differences in the PA prediction from
  the EOB model and the NR data when including the spin terms at different PN order, while keeping the 
nonspinning terms through 3PN order.

We see that the model is fairly robust under variations in the PN spin information, the fractional corrections 
being small in all cases. Quite remarkably, for equal-mass binaries with aligned spins the EOB
  predictions using only the leading order 1.5PN spin-orbit effects (i.e. \mbox{$S_*=\sigma_*$})
  and 2PN spin-spin effects are well inside the numerical errors in all the $q=1$ cases considered. 
For larger mass ratios, the predictions from the leading-order spin couplings are still
close to the NR values but in several cases they are outside the error bounds by up to \mbox{$\sim 1\%$}. 
The additional variations when adding the 2.5PN spin-orbit effects are small but the 3.5PN spin-orbit terms lead to an appreciable improvement in these cases. As a representative example, for $q=3$, \mbox{$\chi_1=-0.5$, $\chi_2=0$} at \mbox{$M\Omega_\phi=0.02$}, the EOB model with only the 1.5PN (2.5PN) spin information is outside the error by \mbox{$\sim 0.17\%$}
  (\mbox{$\sim 0.16\%$}), while including also the 3.5PN spin-orbit information reduces the discrepancy to \mbox{$\sim 0.02\%$}.

 Given the improvements of the EOB resummations compared to the PN
  results and that idea (i) above is an immediate extension of the
  successful nonspinning EOB model, we now assess, in a preliminary
  fashion, the utility of the spin-resummation (ii) by comparing the EOB model 
used here and the NR data with the spinning EOB model of
  Refs.~\cite{Damour01c,Damour:024009,2011PhRvD..84h4028N}. The latter uses (i) with a different
  identification between the Kerr spin and the spins of the two bodies and it employs a 
substantially simpler effective spin coupling in the Hamiltonian because does not 
  enforce (ii). We compute $K$ predicted by the EOB model of Refs.~\cite{Damour01c,Damour:024009,2011PhRvD..84h4028N}  
  using Eqs.~(4.3)--(4.19) of Ref.~\cite{Damour:024009} with the Taylor
  series potentials $A$ and $D$ of Eqs.~(4.7) therein and the
  gyromagnetic coefficients from Eqs.~(29) and (30)
  in Ref.~\cite{2011PhRvD..84h4028N} with all the gauge parameters set to
  zero. This choice is similar to the one made in the baseline EOB
  model. We show the results in  Fig.~\ref{fig:EOBPNorders} and find that they are slightly 
  outside the NR error; the results for other systems with two aligned spins are
  qualitatively similar. Changes to the EOB model in Refs.~\cite{Damour01c,Damour:024009,2011PhRvD..84h4028N} 
  could likely eliminate this discrepancy but the purpose here is to employ
  similar choices for the basic inputs to gain insight into the
  efficiency of the models' underlying structure. The results of
  Fig.~\ref{fig:EOBPNorders} indicate that the more complicated
  spin-resummation imposed by (ii) is in fact a valuable feature of
  the EOB model in predicting the PA.

\section{Spinning particle in Kerr spacetime}
\label{sec:particle}
In this section we discuss an alternative computation of the PA for a spinning particle based on the multipolar equations of motion. This serves as a consistency check of the Hamiltonian results and enables the extension to higher order in \mbox{$S_*$}.  

For a body occupying a region 
in which the gravitational field varies slowly, the influence of its internal structure on its motion is encoded entirely in a collection of reduced multipole moments~\cite{Dixonmultipoles, Dixon1974}. These moments are defined as integrals over a hypersurface whose normal is specified with respect to a reference worldline \mbox{$z^\mu$} \cite{Beiglbock67CM}. The choice of reference worldline that defines the multipole moments is fixed by a spin supplementary condition (SSC) of the form \mbox{$ n_\mu S^{\mu\nu}=0$}, where \mbox{$S^{\mu\nu}$} is the spin tensor and \mbox{$n^\mu$} is a smooth timelike vector. This ensures that only the three physical spin degrees of freedom influence the motion (see Ref.~\cite{Kyrian:2007zz} for a discussion of the features of different choices for \mbox{$n^\mu$}).
The multipole moment expansion about the worldline reduces the four partial differential equations of stress energy conservation to ten ordinary differential equations for the momentum and spin components. Using a small parameter \mbox{$\epsilon \ll 1$} to indicate the scalings of the multipole moments, with each $\ell$-multipole being  \mbox{$O(\epsilon^\ell)$}, the equations of motion 
with the quadrupolar force and torque are~\cite{Dixon1974,EhlersRudolph,SP12}:
\bes\bea
&&\frac{DS^{\mu\nu}}{d\tau}=2 p^{[\mu}u^{\nu]}-\frac{4}{3}R_{\alpha\beta \gamma}^{\; \; \; \;  \; \; [\mu}J^{\nu]\gamma \alpha \beta}+O(\epsilon^3)\,, \label{eq:dsdtau}\\
&&\frac{Dp_\mu}{d\tau}= -\frac{1}{2}R_{\mu\nu\alpha\beta}u^\nu
S^{\alpha\beta}-\frac{1}{6}R_{\rho\alpha\beta\gamma; \mu}J^{ \rho\alpha\beta\gamma}+O(\epsilon^3).\; \; \; \; \;\; \; \;
\label{eq:dpdtau}\eea
\label{eq:odes}
\ees
Here, \mbox{$R_{\mu\alpha\beta\gamma}$} is the Riemann tensor, \mbox{$u^\mu=dx^\mu/d\tau$} is the tangent to the particle's worldline and \mbox{$\tau$} is a time parameter. 
 The quadrupole tensor \mbox{$J^{\mu\nu \alpha\beta}$} satisfies various symmetry and orthogonality relations~\cite{Dixon1974}, 
but its time dependence is set entirely by the body's internal dynamics. Equations~(\ref{eq:odes}) can only be solved once an equation of state for \mbox{$J^{\mu\nu \alpha\beta}$} has been specified. 

For the calculations in this subsection we perturbatively expand all the quantities to \mbox{$O(\epsilon^2)$}. We choose the covariant SSC \mbox{$S^{\mu\nu}p_\nu=0$} (the center of mass frame) and proper time as the evolution parameter. The momenta $p_\mu$ used in this subsection are related to the canonical momenta $P_\mu$ and canonical spin tensor \mbox{$\tilde S^{\alpha\beta}$} used in the Hamiltonian of Sec.~\ref{sec:eobH} by \mbox{$P_\mu=p_\mu+\tilde S^{\alpha\beta} \omega_{\mu\beta \alpha}/2$}, 
where \mbox{$\omega_{\mu \beta \alpha}$} are the spacetime components of the Ricci rotation coefficients. In terms of a tetrad frame \mbox{$e_a^\mu$} they are given by
\be
\omega_{a b}^{\; \; \, c}=e_a^\mu \, e_b^\nu \, e^c_{\nu; \mu}.
\ee
The notation adopted in Ref.~\cite{Barausse:2009aa} is \mbox{$E_{\mu \alpha\beta}=\omega_{\mu \beta \alpha}/2$}. The transformation law for the components of the spin tensor to canonical gauge are discussed in Refs.~\cite{Kyrian:2007zz,Steinhoff:2010zz} but will not be needed here because the final results will be expressed in terms of gauge invariant quantities. 

We use a model for the quadrupole tensor specialized to describe a spinning black hole and given by 
\be J^{\alpha \beta \gamma \delta}=-\frac{3}{(-p_\nu p^\nu)}p^{[\alpha}Q^{\beta][\gamma}p^{\delta]},
\ee
where the quadrupole tensor is
\be Q_{\alpha \beta}=
S_{\alpha \gamma}S_{\beta}^{\; \; \gamma}. 
\label{eq:Qab}
 \ee 
 The modifications necessary to model a non-vacuum compact object are explained in Ref.~\cite{SP12} and would require including the body's tidally induced mass and current quadrupole moments in \mbox{$J_{\alpha\beta\gamma\delta}$} as well as scaling Eq.~(\ref{eq:Qab}) by the rotational Love number. 
The spin tensor can be expressed in terms of a spin four-vector \mbox{$S_\mu$} as 
\be S^{\mu\nu}=-\frac{\epsilon^{\mu \nu \alpha\beta}S_\alpha p_\beta}{\sqrt{-p_\gamma p^\gamma}},
\ee
where \mbox{$\epsilon_{\mu\nu\alpha\beta}$} is the Levi-Civita tensor.

The multipolar equations of motion~(\ref{eq:odes}) admit several conserved quantities. For each Killing vector \mbox{$\xi^\mu=(\partial_t)^\mu,  (\partial_\phi)^\mu$} the quantities
\be
C_\xi=p_\mu \xi^\mu -\frac{1}{2} S^{\mu\nu}\nabla_\nu\xi_{\mu}, \label{eq:calc}
\ee
giving $E$ and $J_z$ 
are conserved to all multipole orders~\cite{EhlersRudolph}. 
To linear order in $\epsilon$ there also exists an extension of the Carter constant~\cite{1982RSPSA.385..229R,1993NuPhB.404...42G}, which makes the \mbox{$O(\epsilon)$} dynamics completely integrable. However, integrability is lost at \mbox{$O(\epsilon^2)$}.

To reduce all manipulations to operations in Minkowski space will work directly with the quantities projected onto a tetrad, choosing the frame 
\bea
e^0_\mu&=&\frac{\sqrt{\Delta}}{\sqrt{\Sigma}}\left(1,0,0,-a\sin^2\theta \right), \nonumber\\ 
e^1_\mu&=&\frac{\sqrt{\Sigma}}{\sqrt{\Delta}}\left(0,1,0,0\right), \; \; \; \; e^2_\mu= \left(0,0,\sqrt{\Sigma},0\right), \ \ \ \nonumber\\
e^3_\mu&=&\frac{\sin\theta}{\sqrt{\Sigma}}\left(-a, 0, 0, r^2+a^2\right), \; \; \; \label{eq:tetr}
\eea
where \mbox{$\Delta=r^2+a^2-2 r$} and we use dimensionless units where the Kerr mass parameter is $M=1$. 

The conserved energy and angular momentum (\ref{eq:calc}) are expressed in terms of the projections \mbox{$p^a=e^{a \mu}p_\mu$} as
\bes
\label{eq:conserved}
\bea
E&=&\frac{\sqrt{\Delta}}{\sqrt{\Sigma}}p^0+\frac{a\sin\theta}{\sqrt{\Sigma}} p^3+\frac{2ar\cos\theta }{\Sigma^2}S^{23}\nonumber\\
&&-\frac{r^2-a^2 \cos^2\theta}{\Sigma^2}S^{01},\\
J_z&=&\frac{a\sin^2\theta\sqrt{\Delta}}{\sqrt{\Sigma}}p^0+\frac{(r^2+a^2)\sin\theta}{\sqrt{\Sigma}}
p^3\nonumber\\
&&-\frac{a\sin^2\theta\left(r\Sigma+ r^2-a^2\cos^2\theta\right)}{\Sigma^2}S^{01}\nonumber\\
&&+\frac{r\sin\theta\sqrt{\Delta}}{\Sigma}S^{13} +\frac{a\sqrt{\Delta}\sin\theta\cos\theta }{\Sigma}S^{20}\nonumber\\
&&-\frac{\cos\theta}{\Sigma^2}\left[\Delta a^2\sin^2\theta  -(r^2+a^2)^2\right]S^{23}.
\eea
\ees

The tetrad projection of Eq.~(\ref{eq:dpdtau}) determines the evolution of the normalization \mbox{$(-p_a p^a)$}, which is no longer a constant at quadratic order in the spin~\cite{SP12}. 
Subtracting the nonconstant terms leads to a perturbatively conserved mass parameter given by 
\be m^2=-p_a p^a-\frac{1}{3}R_{bcdf}J^{bcdf}+O(\epsilon^3)\,.
\label{eq:pnorm}\ee

The quantity \mbox{$u^a$} is normalized to \mbox{$u^a u_a=-1$} since $\tau$ is the proper time. The evolution of \mbox{$\vec{u}$} has to be determined from its relationship with \mbox{$\vec{p}$} obtained by contracting Eq.~(\ref{eq:dsdtau}) with $p_a$ and rewriting the left-hand side using the preservation of the SSC,  \mbox{$u^\mu\nabla_\mu (S^{ab}p_b)=0$}. Substituting Eq.~(\ref{eq:dpdtau}) and expressing \mbox{$(p_ap^a)$} in terms of $m$ using Eq.~(\ref{eq:pnorm}) leads to~\cite{SP12}
\bea u^a&=&\frac{p^a}{m}\left(1+\frac{1}{2
m^2}Q^{bc}E_{bc}\right)
+\frac{1}{2m^3}S^{ab}R_{bcdf}p^c S^{df}\nonumber\\
&&-\frac{1}{m^3}R_{cdf}^{\;
\; \; \; \; a}Q^{fd}p^c, \; \; \; \; \; \; \; \; \label{eq:uofp} 
\eea
where \mbox{$ E_{bd}=R_{abcd}p^a p^c$}.

\subsection{Specialization to equatorial orbits}
In the following, we will rescale the momenta by the particle's mass $m$ and the angular momenta by $m M$ to work with dimensionless quantities as we did in Sec.~\ref{sec:eobH}. The particle's spin angular momentum is $s m^2$ and in the rescaled units it becomes
\be
S_*=s\frac{m}{M}, \ \ \ \ 0\leq s\leq 1.
\ee 
The values of $p^a$ for equatorial orbits are determined in terms of the conserved quantities $E,\, J_z$ by specializing Eqs.~(\ref{eq:conserved}) to \mbox{$\theta=\pi/2$}, \mbox{$S^{23}=S^{20}=0$}, 
\mbox{$S^{01}=S^2 p^3$}, \mbox{$S^{13}=-S^2 p^0$} since \mbox{$S^2=-{\rm sgn}{(S_*L_z)}\, S_*\, $} is the only nonvanishing spin component in this case. Here, \mbox{${\rm sgn}(S_*L_z)=\pm 1$} indicates if the particle's spin is aligned or antialigned with its orbital angular momentum. This leads to 
\bea\label{eq:pofej}
p^0 &=&\frac{(r^2+a^2)E-a J_z}{r\sqrt{\Delta}}\bigg(1+\frac{S_*^2}{r^3}\bigg)\nonumber\\
&&-\, {\rm sgn}(S_*L_z)\, S_* \frac{J_z-aE(r+1)}{r^2\sqrt{\Delta}},\\
p^3 &=&\frac{J_z -aE}{r}\bigg(1+\frac{S_*^2}{r^3}\bigg)- {\rm sgn}(S_*L_z)\, S_*
\frac{E}{r}. \nonumber \eea
We use these results in the normalization condition (\ref{eq:pnorm}), which for equatorial orbits reduces to (see Eq.~(73) of Ref.~\cite{SP12})\be -(p^0)^2+(p^1)^2+(p^3)^2
=-1+\frac{S_*^2}{r^5}\left[r^2+3(J_z-aE)^2\right]. \ \ \ \ \
\label{eq:pnormequat} \ee 
Solving Eq.~(\ref{eq:pnormequat}) for $p^1$ and setting the resulting expression
and its radial derivative to zero determines $E$ and $J_z$ as functions of the circular-orbit radius.
Next, the coordinate radius is eliminated in favor of the orbital frequency as follows. The projection of the four-velocity onto the equatorial tetrad is \mbox{$u^a=u^\mu e_\mu^a$} which gives for equatorial orbits
\be
\label{eq:utuphi}
u^t=\frac{(r^2+a^2)}{r\sqrt{\Delta}}u^0+\frac{a}{r}u^3, \ \ \ \ \\
 u^\phi =\frac{a}{r\sqrt{\Delta}}u^0+\frac{u^3}{r}. 
\ee
The orbital frequency is given by the ratio \mbox{$\Omega_\phi=u^\phi/u^t$}. To find the relationship of the tetrad components $u^0$ and $u^3$ with $E$ and $J_z$ we use Eq.~(\ref{eq:uofp}) written in terms of the spin vector:
\be
u^a=p^a(1+\frac{1}{2}Q^{cd}E_{cd})+{}^* R^{*a}_{\; \; \; bcd}S^b S^c p^d-p^c
R_{cdf}^{\; \; \; \; a}Q^{fd}\,. \label{eq:uofpspin}\ee
Here, \mbox{${}^*R^{*a}_{\, \;  \; bcd}$} is the left and right dual of the Riemann tensor computed from the contractions \mbox{$^*R^*_{abcd}=\epsilon_{ab}^{\, \; \; fg}\epsilon_{cd}^{\, \; \; lm}R_{fglm}/4$}. For circular equatorial orbits we evaluate Eq.~(\ref{eq:uofpspin}) using \mbox{$p^1=p^2=0$}, \mbox{$S^0=S^1=S^3=0$} to obtain \mbox{$u^a=p^a[1-S_*^2(1+3 (p^3)^2)/(2 r^3)]$} for $u^0$ and $u^3$ in terms of $p^0$ and $p^3$ and hence $E, \, J_z$ from~(\ref{eq:pnormequat}). We use this in the expression for the frequency, invert  perturbatively to find \mbox{$r(\Omega_\phi)$} and compute the following expressions for the conserved quantities as functions of \mbox{$\Omega_\phi$}:
\bes\label{eq:eandj}
\bea E(\Omega_\phi)&=&\frac{\pm a+(r_c-2) \sqrt{r_c}}{r_c^{3/4} \sqrt{\pm 2
   a+(r_c-3) \sqrt{r_c}}}\nonumber\\
&&-\frac{{\rm sgn}(S_*L_z)\, S_*
   \left(\pm \sqrt{r_c}-a\right)}{r_c^{9/4} \sqrt{\pm 2 a+(r_c-3)
   \sqrt{r_c}}}\nonumber\\
&& +\frac{S_*^2 \left(\pm 3 a+(r_c-4) \sqrt{r_c}\right)}{2
   r_c^{15/4} \sqrt{\pm 2 a+(r_c-3) \sqrt{r_c}}}
, \; \; \; \; \; \; \; \; \;   \label{eq:eofomega}\eea
\bea &&J_z(\Omega_\phi)=\frac{\pm \left(a^2\mp 2 a \sqrt{r_c}+r_c^2\right)}{\sqrt{\pm 2a+(r_c-3)\sqrt{r_c}} \ r_c^{3/4}}\nonumber\\
&& \; \; +\frac{{\rm sgn}(S_*L_z) \, S_*\left[a^2\mp a (1-3 r_c) \sqrt{r_c}+(r_c-4)
   r_c^2\right]}{r_c^{9/4} \sqrt{\pm 2 a+(r_c-3)
   \sqrt{r_c}}}\nonumber\\
&&\; \; \pm \frac{S_*^2 \left[3 a^2\pm 2 a \sqrt{r_c} (3 r_c-2)+r_c^2
   (2 r_c-7)\right]}{2 r_c^{15/4} \sqrt{\pm 2 a+(r_c-3)
   \sqrt{r_c}}}. \; \; \; \; \; 
 \label{eq:Jofomega}
\eea
\ees
 The upper/lower signs here refer to the value of \mbox{${\rm sgn}(a L_z)$}, signifying prograde/retrograde orbits relative to the Kerr spin. We use the notation
\be
r_c=\frac{ (1\mp a \Omega_\phi )^{2/3}}{\Omega_\phi ^{2/3}}\,. \; \; \; \; \; \; \label{eq:rcdef}
\ee
In the PN limit, the expansion of \mbox{$E(\Omega_\phi)$} for \mbox{$\Omega_\phi\to 0$} obtained from Eqs.~(\ref{eq:rcdef}) and (\ref{eq:eofomega})
reduces at \mbox{$O(S_*^2\Omega_\phi^2)$} to the result implied by the 
relations in Ref.~\cite{Racine2008},
and the terms at \mbox{$O(S_* \Omega_\phi^{5/3})$} and \mbox{$O(S_*\Omega_\phi^{7/3})$} agree with the corresponding pieces in Ref.~\cite{spinorbitNNLO}.

\subsection{Periastron advance and precession frequencies}
\label{subsec:resultstpl}
The equations of motion (\ref{eq:odes}) for a generic orbit expressed on the tetrad are  
\bes
\bea
\frac{dx^\mu}{d\tau}&=& u^\mu,\\
\frac{dp^a}{d\tau}&=& \omega_{bc}^{ \; \; \; \; a}u^b p^c+R^{* a}_{\; \; bcd}p^b S^cp^d+f^a,\\
\frac{dS^a}{d\tau}&=& \omega_{bc}^{ \; \; \; \; a}p^b S^c+p^a p^c p^f S^b S^d
R^*_{bcdf},\eea  
\ees
 where \mbox{$ f_a= e^{\mu}_aR_{bcdf; \mu}J^{bcdf}/6.
$} When performing the variation around circular orbits, we use that \mbox{$\delta x^\mu=e^\mu_a \delta x^a$} with 
\be
\frac{d\delta x^a}{d\tau}=-e^a_\mu e^\mu_{b,\nu}e^\nu_c u^c \delta x^b+\frac{\partial u^a}{\partial x^\mu}e^\mu_b \delta x^b+\frac{\partial u^a}{\partial p^b}\delta p^b+\frac{\partial u^a}{\partial S^b}\delta S^b.
\ee
We apply the method of Eq.~(\ref{eq:lin}) with \mbox{$\zeta=(\delta x^a, \delta p^a, \delta S^a)$} and use {\sc mathematica} to compute the characteristic polynomial for the Jacobian matrix. 

The characteristic polynomial factors into a radial and a mixed meridional and spin piece. Its solutions lead to three pairs of nontrivial eigenvalues \footnote{This number follows from the number of degrees of freedom after imposing all the constraints: the normalization and orthogonality conditions leave 8 degrees of freedom, but the conservation of $J_z$ and $E$ eliminates two of these} that we interpret as the periastron, nodal and gyroscope precession frequencies.  From the radial eigenvalue and after perturbatively substituting Eqs.~(\ref{eq:pofej}), (\ref{eq:eandj}) and $r(\Omega_\phi)$ we obtain

\bea
&&\frac{\Omega_r^2}{\Omega_\phi^2}=K^{-2}=\frac{-3 a^2\pm 8 a \sqrt{r_c}+(r_c-6) r_c}{r_c^2}\nonumber\\
&&\; \; \; +\frac{6 \, {\rm sgn}{(S_*L_z)} \, S_* \, \left(\pm\sqrt{r_c}- a\right) \left[(r_c-3)
   \sqrt{r_c}\pm 2 a\right]}{r_c^{7/2}}\nonumber\\
&& \; \; \; -  \frac{3S_*^2 \left[(r_c-7) \sqrt{r_c}\pm 6 a\right] \left[(r_c-3)
   \sqrt{r_c}\pm 2 a\right]}{r_c^5}. \; \; \; \;  \; \; \; \; \; \;  \label{eq:ratioofomega}
\eea
We will use the $O(S_*)$-piece of this result in Sec. \ref{sec:EOB-NR} to compare the effects of the spin dipole and the gravitational self-force.

One of the other pairs of eigenvalues characterizes the meridional frequency and is given as a function of $\Omega_\phi$ by

\bea
\frac{\Omega_\phi}{\Omega_\theta}&=&\frac{r_c}{\sqrt{3
   a^2\mp 4 a \sqrt{r_c}+r_c^2}}\nonumber\\
&-& \frac{3 a \, {\rm sgn}{(S_*L_z)}\, S_* \,  \left(\pm 2 a+ (r_c-3)
   \sqrt{r_c}\right)}{\sqrt{r_c} \left(3 a^2\mp 4 a
   \sqrt{r_c}+r_c^2\right)^{3/2}}\nonumber\\
&\pm & \frac{15 a^5 \, S_*^2\, \left(3 a^2+ r_c (7 r_c+4)\right)}{2
   r_c^{5/2} \left(\sqrt{r_c}\mp a\right)^2
   \left(3 a^2\mp 4 a \sqrt{r_c}+r_c^2\right)^{5/2}}\nonumber\\
&\pm & \frac{3 a \, S_*^2\, \left[a^2 (80r_c-7 r_c^2
   +4)+3 r_c^4-16r_c^3+26r_c^2\right]}{2
   r_c^{1/2} \left(\sqrt{r_c}\mp a\right)^2
   \left(3 a^2\mp 4 a \sqrt{r_c}+r_c^2\right)^{5/2}}\nonumber\\
&+ & \frac{6 a^4 \, S_*^2\,  \left(-22 a^2+ 3 r_c^2 (r_c-16)+13r_c\right)}{2
   r_c^{2} \left(\sqrt{r_c}\mp a\right)^2
   \left(3 a^2\mp 4 a \sqrt{r_c}+r_c^2\right)^{5/2}}\nonumber\\
&- &\frac{3 a^2 \, S_*^2\,  \left( r_c^3-11 r_c^2+39r_c+23\right)}{2
 \left(\sqrt{r_c}\mp a\right)^2
   \left(3 a^2\mp 4 a \sqrt{r_c}+r_c^2\right)^{5/2}}. \; \; \; 
\label{eq:omegatheta}
\eea
This Lense-Thirring precession is the analog of the PA for $\Omega_\theta$, giving a secular change in angle of $
\Delta \Phi_{\rm LT}=2\pi\vert \Omega_\phi/\Omega_\theta-1 \vert$. 
It is sometimes referred to as the angle of advance 
of the nodes of a circular orbit, where a node is an orbit's intersection point with the equatorial plane.
The weak-field limit of Eq.~(\ref{eq:omegatheta}) provides a useful check of the physical interpretation of the angular eigenvalues. Expanding the square root of the inverse of Eq.~(\ref{eq:omegatheta}) for small $\Omega_\phi$ gives the angular advance of the ascending node in the PN limit
$(2\pi)^{-1}\Delta \phi_{\rm LT}=\pm 2  a\Omega_\phi+O(\Omega_\phi^{4/3})$.
This agrees with the weak-field expressions for the Lense-Thirring effect around a rotating body~\cite{BarkerOConnell, LenseThirring} when we substitute the leading-order PN relation between 
$\Omega_\phi$ and $r$. As indicated by the $\pm$ signs for prograde/retrograde motion, the leading order effect is that the nodes are dragged in the sense of the Kerr spin angular momentum~\cite{PhysRevD.5.814}. 

The spin precession frequency is
\bea
\label{eq:spinprec}
\frac{\Omega_s}{\Omega_\phi}&=&\frac{\sqrt{\pm 2 a+r_c^{3/2}-3
   \sqrt{r_c}}}{r_c^{3/4}}\nonumber\\
&\mp & {\rm sgn} (S_*L_z) \, \frac{3 \,  S_* \, \left(3 a^2 \mp 4 a \sqrt{r_c}
   +r_c^2\right)}{2 r_c^{11/4} \sqrt{\pm 2 a+\sqrt{r_c} (r_c-3)}}\nonumber\\
&+ &\frac{S_*^2 a^2 \left(446 \sqrt{r_c}\mp 970 a+\left(3 r_c^2+9
   r_c-65\right) r_c^{3/2}\right)}{4
   r_c^{13/4} \left(\sqrt{r_c}\mp a \right)^2
   \left(\pm 2 a +\sqrt{r_c}
   (r_c-3)\right)^{3/2}}\nonumber\\
&+&\frac{S_*^2[9 (r_c-6) r_c^2+5 (23
   r_c-18)]}{4 r_c^{7/4} \left(\sqrt{r_c}\mp a
\right)^2 \left(\pm 2 a+\sqrt{r_c} (r_c-3)\right)^{3/2}}\nonumber\\
&+&\frac{a^4 S_*^2\left(30 a^2
   r_c-162 a^2+9 r_c^3-44 r_c^2+764 r_c\right)}{4
   r_c^{19/4} \left(\sqrt{r_c}\mp a\right)^2
   \left(\pm 2 a+\sqrt{r_c}
   (r_c-3)\right)^{3/2}}\nonumber\\
&+&\frac{S_*^2[(7-6 r_c) r_c^2+58
   r_c-108]}{8 r_c^{11/4} \left(\pm 2 a+\sqrt{r_c}
   (r_c-3)\right)^{3/2}}\nonumber\\
&-&\frac{S_*^2[15 a^4 (4 r_c-37)+2
   a^2 r_c^2 (9 r_c+91)]}{8 r_c^{19/4} \left(\pm 2 a+\sqrt{r_c}
   (r_c-3)\right)^{3/2}}
   \eea
In the weak-field limit, the expansion of Eq.~(\ref{eq:spinprec}) is
$
|\Omega_s/\Omega_\phi-1|=3\Omega_\phi^{2/3}/2+O(\Omega_\phi),
$
which is equivalent to the drift of a gyroscope computed in Refs.~\cite{Schiff, BarkerOConnell}.

As mentioned below Eq.~(\ref{eq:calc}), when consistently working to
linear order in the spin the particle motion is completely integrable~\cite{1982RSPSA.385..229R,1993NuPhB.404...42G}. However, studies of
the nonperturbative integrations of the equations of motion for
a spinning dipole found that the dynamics are formally chaotic for very
large spins and substantial orbital eccentricity~\cite{1997PhRvD..55.4848S,1998PhRvD..58b3005S,2003PhRvD..67b4005H,2003PhRvD..67j4023H}. In the
limit of circular equatorial orbits, traces of the onset of chaos were
found to persist as an instability in the meridional direction~\cite{1997PhRvD..55.4848S, 1998PhRvD..58b3005S} \footnote{Note that
  there are several typos in the matrix elements used to compute the
  Lyapunov exponents in Ref.~\cite{1998PhRvD..58b3005S}.}. The existence of this
instability was however limited to spin values of order $s\sim
O(M/m)\gg 1$, corresponding to $S_*=O(1)$, which are outside the regime
of validity of the spin dipole model used to determine the equations
of motion. Not surprisingly, we find that in the perturbative case
considered here, the frequency~(\ref{eq:omegatheta}) remains real and
the motion remains stable in the meridional direction until the last
stable orbit where $\Omega_r=0$.

\subsection{Comparison between spin dipole  and gravitational self-force in Schwarzschild}
\label{sec:SF}
For extreme mass-ratio binaries the leading order corrections to geodesic motion are linear in the mass ratio and are due to two effects: the influence of the particle's spin dipole and gravitational self-force (SF) corrections. Although both effects enter at the same order in $m/M$, the magnitude of their imprint on observables could be very different, e.g., dissipative effects are dominated by gravitational radiation reaction. For the conservative dynamics, previous comparisons focused on circular orbits and include the ISCO shift~\cite{2011PhRvD..83b4027F} and the relationship $E(P_\phi)$ or $E(J_z)$ \cite{SP12}. For a Schwarzschild background, we complement the studies of Ref.~\cite{SP12} by using $E(\Omega_\phi)$, thus specifying the identification between nonspinning and spinning configurations in terms of the observable frequency $\Omega_\phi$. This bypasses the subtlety that such a comparison at a fixed value of the conserved quantity $J_z$ corresponds to comparing different orbital configurations. 
We 
also extend this comparison to the post-geodesic effects in $K$ obtained from the results of Sec. III and Ref.~\cite{LeTiec:2011bk}.

\begin{figure}
\includegraphics[width=0.9\columnwidth,angle=0]{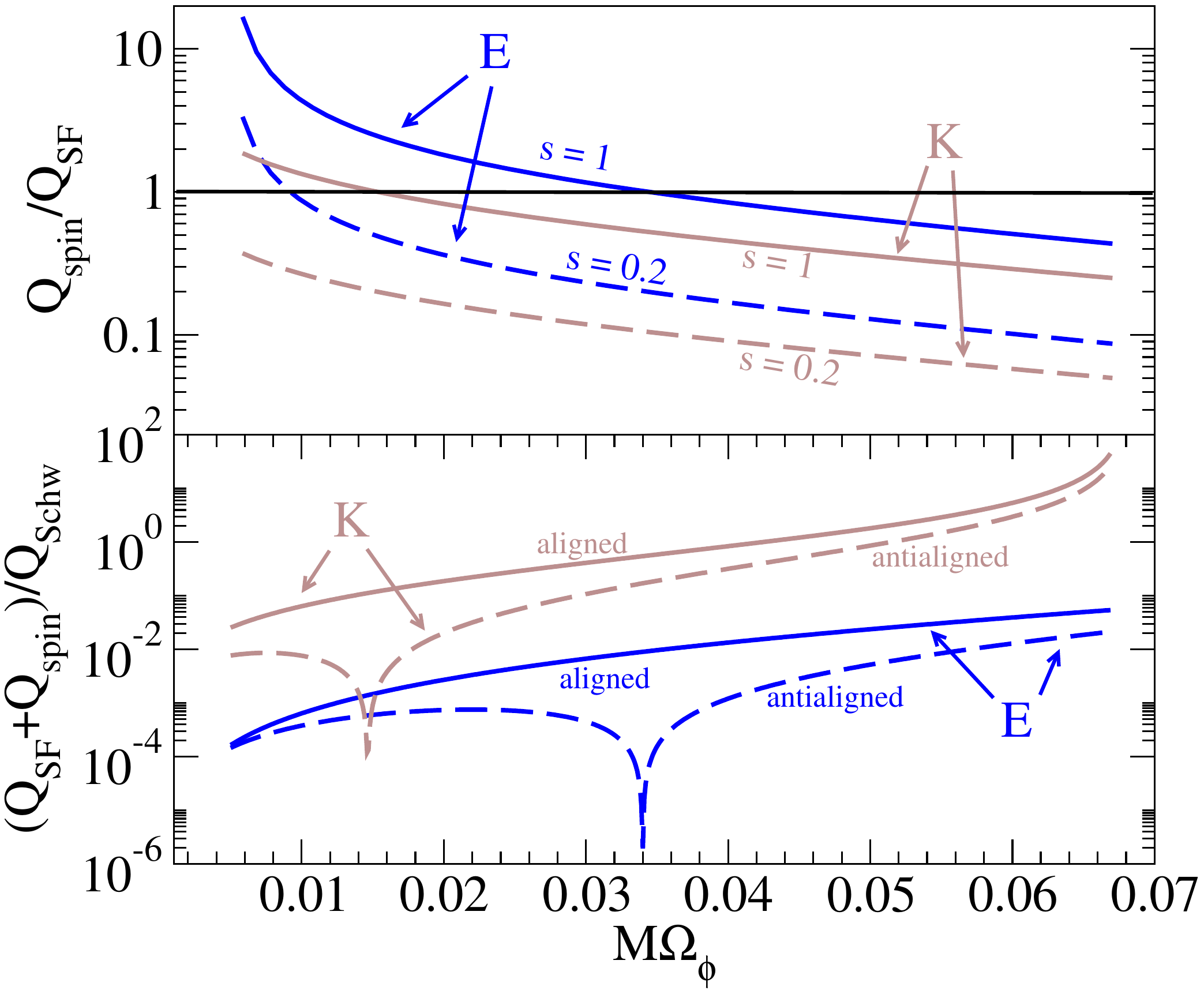}
\caption{\textit{Post-geodesic effects in Schwarzschild}. Gravitational self-force and spin-dipole effects at $O(\nu)$ in the quantities $Q=(E(\Omega_\phi),K(\Omega_\phi))$ for a Schwarzschild background. 
\textit{Upper panel}: Relative importance of conservative spin dipole and SF effects for strictly ($Q=E$, solid lines) and nearly ($Q=K$, dashed lines) circular orbits. 
\textit{Lower panel}: Combined $O(\nu)$ effect for maximally spinning particles with $s=1$. Solid lines are for ${\rm sgn}(S_*L_z)=+1$, dashed lines for ${\rm sgn}(S_*L_z)=-1$. 
} 
\label{fig:tpl}
\end{figure}

Figure~\ref{fig:tpl} illustrates the spin dipole and SF effects on
$E(\Omega_\phi)$ and $K(\Omega_\phi)$ in the following way. We write
Eq.~(\ref{eq:eofomega}) specialized to the Schwarzschild case ($a=0$) 
as $E(\Omega_\phi)=E_{\rm Schw}+S_*\,E_{\rm spin}+O(S_*^2)$, where $E_{\rm Schw}$ 
is the Schwarzschild geodesic term from the first line of Eq.~(\ref{eq:eofomega}) 
and $E_{\rm spin}$ is the term from the second line of Eq.~(\ref{eq:eofomega}) and express
$K$ from Eq.~(\ref{eq:ratioofomega}) in a similar fashion. For the 
SF contributions, we use the $O(\nu)$ term in $E(\Omega_\phi)$ from Eq.~(3b) in
Ref.~\cite{LeTiec:2011dp} and the $O(q)$-term in $K$ from Eq.~(6) of
Ref.~\cite{LeTiec:2011bk}. We recall that the convention of Ref.~\cite{LeTiec:2011bk} is $q=m/M=\nu+O(\nu^2)$ and use the notation $Q$ for either of the quantities $E$ or $K$. 

In the upper panel of Fig.~\ref{fig:tpl} 
we quantify the relative importance of SF and spin-dipole contributions. We 
show the ratios $E_{\rm spin}/E_{\rm SF}$ and $K_{\rm spin}/K_{\rm SF}$ when 
$s=1$ (solid lines) and $s =0.2$ (dashed lines). We see that the maximum spin-dipole contribution  
to the energy (PA) can be larger than the SF contribution when $M \Omega_\phi\,\laq\, 0.035$ 
($M \Omega_\phi\,\laq\, 0.015$). Moreover, the SF contributions increase more rapidly with the orbital 
frequency than the spin-dipole effects and become dominant in the strong-field regime. 
In the lower panel of Fig.~\ref{fig:tpl} we evaluate the total $O(\nu)$ fractional corrections. Shown are the ratios $(E_{\rm SF} + E_{\rm spin})/E_{\rm Schw}$ and 
$(K_{\rm SF} + K_{\rm spin})/K_{\rm Schw}$ when $s=1$, for the 
aligned (solid lines) and antialigned (dashed lines) configurations. The scale on the $y$-axis thus gives only the dimensionless coefficient at $O(\nu)$ and must be multiplied by $\nu$ to evaluate the net physical fractional corrections. 
We notice that whereas the SF and spin-dipole contributions to the 
energy are quite small, they are much more important in the PA. 
Notably, they become comparable to $\nu K_{\rm Schw}$ when the orbital frequency is 
larger than $M \Omega_\phi \simeq 0.03$. Also, in the antialigned case the contributions 
from SF and spin dipole can cancel each other and the net effects are smaller than in the aligned case at the same frequency. 

\section{Conclusions}
\label{sec:conclusion}

We calculated the periastron advance for binaries with aligned spins
in the limit of circular orbits from the EOB Hamiltonian and for a
spinning particle in a Kerr background. We focused on the gauge
invariant ratio $K$ of the azimuthal and radial frequencies of the
motion.

For the EOB model, first we wrote explicitly the spinning EOB
Hamiltonian for generic binaries in spherical coordinates and detailed the modifications necessary to express the quantities in a fixed source frame. The spherical coordinates adapted to the binary's geometry could be a useful tool in future studies of precessing
binaries in the EOB framework~\cite{Pan:2013rra}. This form of the Hamiltonian could also
be used in the future for computing the gauge-invariant expressions for the
orbit-averaged precession frequencies of the orbital plane and the
spins for small deviations from equatorial orbits (aligned
spins). Then, after reducing the dynamics to equatorial orbits and
nonprecessing spins, we derived an implicit relation for $K$ in terms
of partial derivatives of the Hamiltonian and of solutions to
algebraic equations that determine circular-orbit quantities.  We
evaluated these results numerically for a 3.5PN EOB model and used
them to compare with PN predictions and NR data from
Ref.~\cite{Mroueinprep}. The EOB model that we employed is not
calibrated to any numerical-relativity simulation.  Quite remarkably,
throughout the region of parameter space in frequencies, mass ratios
and spins covered by the NR data, the discrepancies between the EOB
and NR results are within the NR errors in all cases except for 
$\chi_1=\chi_2=-0.95$ and $\chi_1=\chi_2=-0.9$. In the former case the EOB 
prediction is slightly outside (by $0.3\%$) the numerical error for $M\Omega_\phi \,\laq\, 0.02$; 
in the latter case it coincides with the numerical error for $M\Omega_\phi \,\laq\, 0.02$. This 
quirk will be investigated in the future using black-hole simulations with antialigned spins  
and different spin magnitudes, and larger eccentricities. The differences to the PN 
predictions were larger and are discussed in detail in Ref.~\cite{LeTiecinprep}.  

We also found that the EOB results for equal-mass binaries are quite
stable when varying the PN spin information.  All these results
confirm the utility of the EOB approach for approximating not only the
gravitational waveforms~\cite{Pan:2009wj,Taracchini2012,Pan:2013rra,Hinder:2013oqa},
but also the conservative dynamics in the presence of nonprecessional
spin effects, even for nearly extremal spins.  Moreover, the excellent agreement also confirms the 
two main ideas underlying the EOB approach with spins, that is, the mapping of the
two-body dynamics of spinning particles onto the dynamics of a
spinning particle in a (deformed) Kerr spacetime and the resummation of 
all PN terms linear in the spin of the effective particle. In fact, for equal-mass binaries 
the EOB predictions are in very good agreement with the numerical results already when only the leading PN
spin-orbit and spin-spin effects are included in the EOB model. 

Although the frequency range considered here is well within the
adiabatic regime, the NR data include a small contribution from
gravitational radiation reaction. The PA in the EOB model with dissipation 
could be computed from the EOB orbital frequency by applying the same method employed in the 
NR simulations to extract the PA. We defer this study to the future. 

In future work, we will use the EOB results obtained in this paper to
obtain gauge invariant expressions for the $O(\nu)$ terms in the
metric potentials for binaries with spins. This could be useful for
improving the spinning EOB model using information from SF
calculations, analogous to the nonspinning case considered in 
Refs.~\cite{LeTiec:2011dp,Barausse:2011ys}.

For a spinning particle in a Kerr spacetime, we obtained explicit gauge invariant
expressions for $K$ and for the meridional and spin 
precession frequencies in the circular equatorial limit and including terms quadratic in the particle's spin. 
Specializing the results for the energy and PA to a Schwarzschild background we compared the spin-dipole and SF effects, which both scale linearly with the mass ratio. We found that the spin dipole dominates 
at low frequency and, depending on the particle's spin, could be nonnegligible compared to the conservative SF even at higher frequencies. We used these results to quantify the dimensionless coefficients associated with the post-geodesic effects at linear order in the mass ratio, which were substantially smaller for the energy than for the PA. 
The extension of our comparisons to a Kerr spacetime, which includes more intricate spin interactions, is immediate once the SF results for the PA in Kerr become available. Such studies could provide information on which physical effects to include when modeling the conservative dynamics of small mass ratio systems to a desired accuracy. In addition, our results for $K$ could inform the description of binaries with less extreme mass ratios. For example, they could be used to improve analytical and phenomenological models as done in  Ref.~\cite{LeTiecinprep} and as a benchmark for higher-order PN spin terms when they become available. 

\begin{acknowledgments}
  We thank Andrea Taracchini and Yi Pan for help with implementing the EOB
  Hamiltonian and Enrico Barausse, Alexandre Le Tiec, Yi Pan and
  Andrea Taracchini for useful interactions and
  comments. A.B. acknowledges partial support from NSF Grants
 No.~PHY-0903631 and No.~PHY-1208881, and NASA Grant
  No.~NNX09AI81G. T.H. acknowledges support from NSF Grants
 No.~PHY-0903631 and No.~PHY-1208881 and the Maryland Center for
  Fundamental Physics.  A.M. and H.P. acknowledge support from NSERC of
  Canada, from the Canada Research Chairs Program, and from the
  Canadian Institute for Advanced Research. We acknowledge support from the Sherman Fairchild Foundation, from NSF grants No.~PHY-0969111 and No.~PHYS-1005426 at Cornell and from NSF Grants No.~PHY-1068881 and No.~PHY-1005655 at Caltech. The numerical relativity simulations were performed at the GPC
  supercomputer at the SciNet HPC Consortium~\cite{scinet}; SciNet is
  funded by: the Canada Foundation for Innovation (CFI) under the
  auspices of Compute Canada; the Government of Ontario; Ontario
  Research Fund--Research Excellence; and the University of
  Toronto. Further computations were performed on the Caltech computer
  cluster Zwicky, which is funded by the Sherman Fairchild Foundation
 and the NSF MRI-R2 grant No.~PHY-0960291, on SHC at Caltech, which is supported by the Sherman Fairchild Foundation, and on the NSF XSEDE network under grant No.~TG-PHY990007N.
\end{acknowledgments}

\appendix

\section{Useful quantities for a  spinning particle in Kerr spacetime}

For circular equatorial orbits, the quantities entering the
relationship between $u^a$ and $p^a$ reduce to 
\bes \bea
Q^{ab}E_{ab}&=&-\frac{S_*^2}{ r^{3} }\left[(p^0)^4+(p^0p^3)^2-2 (p^3)^4 \right],\; \; \; \;  \\
{}^* R^{*a}_{\; \; \; bcd}S^b S^c p^d &=& \frac{S_*^2}{r^3}(- p^0, \, 0, \, 0, \, 2  p^3 ),\\
p^c
R_{cdf}^{\; \; \; \; a}Q^{fd} &=&\frac{S_*^2}{r^{3}}\bigg\{ p^0  [ (p^3)^2-(p^0)^2]\delta_a^0\nonumber\\
&& \; \; + 2 p^3 [ (p^0)^2-(p^3)^2]\delta_a^3\, \bigg\}.  \eea \ees

The rotation coefficients used in our calculation are given by
\bes\bea
\omega_{00}^{\; \; \; 2}&=&\omega_{03}^{\; \; \; 0}=-\omega_{11}^{\; \; \; 2}=\omega_{12}^{\; \; \; 1}=\frac{a^2 \cos\theta\sin\theta}{\Sigma^{3/2}}\,,\nonumber\\
-\omega_{02}^{\; \; 3}&=& -\omega_{30}^{\; \;  2}=-\omega_{32}^{\; \;  0}=\omega_{03}^{\; \;  2}=\omega_{20}^{\; \;  3}=\omega_{23}^{\; \;  0}=\frac{a\sqrt{\Delta}\cos\theta}{\Sigma^{3/2}}\,,\nonumber\\
\omega_{32}^{\; \; \; 3}&=&-\omega_{33}^{\; \; \; 2}=-\frac{(r^2+a^2)\cos\theta}{\Sigma^{3/2}\sin\theta}\,,\nonumber\\
\omega_{00}^{\; \; \; 1}&=&\omega_{01}^{\; \; \; 0}=\frac{r(a^2-r)-a^2 \cos^2\theta(r-1)}{\sqrt{\Delta}\Sigma^{3/2}}\,,\nonumber\\
\omega_{03}^{\; \; \; 1}&=&\omega_{10}^{\; \; \; 3}=\omega_{13}^{\; \; \; 0}=\omega_{30}^{\; \; \; 1}=\omega_{31}^{\; \; \; 0}=-\omega_{01}^{\; \; \; 3}=\frac{a r \sin\theta}{\Sigma^{3/2}}\,,\nonumber\\
\omega_{21}^{\; \; \; 2}&=&-\omega_{22}^{\; \; \; 1}=\omega_{31}^{\;
  \; \; 3}=-\omega_{33}^{\; \; \;
  1}=-\frac{r\sqrt{\Delta}}{\Sigma^{3/2}}.\nonumber 
\eea \ees


\begin{thebibliography}{10}%
\makeatletter
\providecommand \@ifxundefined [1]{%
 \ifx #1\undefined \expandafter \@firstoftwo
 \else \expandafter \@secondoftwo
\fi
}%
\providecommand \@ifnum [1]{%
 \ifnum #1\expandafter \@firstoftwo
 \else \expandafter \@secondoftwo
\fi
}%
\providecommand \enquote [1]{``#1''}%
\providecommand \bibnamefont  [1]{#1}%
\providecommand \bibfnamefont [1]{#1}%
\providecommand \citenamefont [1]{#1}%
\providecommand\href[0]{\@sanitize\@href}%
\providecommand\@href[1]{\endgroup\@@startlink{#1}\endgroup\@@href}%
\providecommand\@@href[1]{#1\@@endlink}%
\providecommand \@sanitize [0]{\begingroup\catcode`\&12\catcode`\#12\relax}%
\@ifxundefined \pdfoutput {\@firstoftwo}{%
 \@ifnum{\z@=\pdfoutput}{\@firstoftwo}{\@secondoftwo}%
}{%
 \providecommand\@@startlink[1]{\leavevmode}%
 \providecommand\@@endlink[0]{}%
}{%
 \providecommand\@@startlink[1]{%
  \leavevmode
  \pdfstartlink
   attr{/Border[0 0 1 ]/H/I/C[0 1 1]}%
   user{/Subtype/Link/A<</Type/Action/S/URI/URI(#1)>>}%
  \relax
 }%
 \providecommand\@@endlink[0]{\pdfendlink}%
}%
\providecommand \url  [0]{\begingroup\@sanitize \@url }%
\providecommand \@url [1]{\endgroup\@href {#1}{\urlprefix}}%
\providecommand \urlprefix [0]{URL }%
\providecommand \Eprint[0]{\href }%
\@ifxundefined \urlstyle {%
  \providecommand \doi [1]{doi:\discretionary{}{}{}#1}%
}{%
  \providecommand \doi [0]{doi:\discretionary{}{}{}\begingroup
  \urlstyle{rm}\Url }%
}%
\providecommand \doibase [0]{http://dx.doi.org/}%
\providecommand \Doi[1]{\href{\doibase#1}}%
\providecommand \bibAnnote [3]{%
  \BibitemShut{#1}%
  \begin{quotation}\noindent
    \textsc{Key:}\ #2\\\textsc{Annotation:}\ #3%
  \end{quotation}%
}%
\providecommand \bibAnnoteFile [2]{%
  \IfFileExists{#2}{\bibAnnote {#1} {#2} {\input{#2}}}{}%
}%
\providecommand \typeout [0]{\immediate \write \m@ne }%
\providecommand \selectlanguage [0]{\@gobble}%
\providecommand \bibinfo [0]{\@secondoftwo}%
\providecommand \bibfield [0]{\@secondoftwo}%
\providecommand \translation [1]{[#1]}%
\providecommand \BibitemOpen[0]{}%
\providecommand \bibitemStop [0]{}%
\providecommand \bibitemNoStop [0]{.\EOS\space}%
\providecommand \EOS [0]{\spacefactor3000\relax}%
\providecommand \BibitemShut [1]{\csname bibitem#1\endcsname}%
\bibitem{PAsolar}%
  \BibitemOpen
  \bibfield{author}{%
  \bibinfo {author} {\bibfnamefont{E.~V.}\ \bibnamefont{{Pitjeva}}},\ }%
  \bibfield{journal}{%
  \Doi{10.1134/1.1922533}{\bibinfo {journal} {Astron. Lett.}}\ }%
  \textbf{\bibinfo {volume} {31}},\ \bibinfo {pages} {340} (\bibinfo {year}
  {2005})%
  \bibAnnoteFile{NoStop}{PAsolar}%
\bibitem{standish1992orbital}%
  \BibitemOpen
  \bibfield{author}{%
  \bibinfo {author} {\bibfnamefont{E.~M.}\ \bibnamefont{{Standish}}}, \bibinfo
  {author} {\bibfnamefont{X.~X.}\ \bibnamefont{{Newhall}}}, \bibinfo {author}
  {\bibfnamefont{J.~G.}\ \bibnamefont{{Williams}}},\ and\ \bibinfo {author}
  {\bibfnamefont{D.~K.}\ \bibnamefont{{Yeomans}}},\ }%
  in\ \emph{\bibinfo {booktitle} {Explanatory Supplement to the Astronomical
  Almanac}},\ \bibinfo {editor} {edited by\ \bibinfo {editor}
  {\bibfnamefont{P.~K.}\ \bibnamefont{Seidelmann}}}\ (\bibinfo {publisher}
  {University Science Books Mill Valley, CA},\ \bibinfo {year} {1992})\ pp.\
  \bibinfo {pages} {279--323}%
  \bibAnnoteFile{NoStop}{standish1992orbital}%
\bibitem{1967PhRvL..18..313D}%
  \BibitemOpen
  \bibfield{author}{%
  \bibinfo {author} {\bibfnamefont{R.~H.}\ \bibnamefont{{Dicke}}}\ and\
  \bibinfo {author} {\bibfnamefont{H.~M.}\ \bibnamefont{{Goldenberg}}},\ }%
  \bibfield{journal}{%
  \Doi{10.1103/PhysRevLett.18.313}{\bibinfo {journal} {Phys.\ Rev.\ Lett.}}\ }%
  \textbf{\bibinfo {volume} {18}},\ \bibinfo {pages} {313} (\bibinfo {year}
  {1967})%
  \bibAnnoteFile{NoStop}{1967PhRvL..18..313D}%
\bibitem{2011EPJH...36..407R}%
  \BibitemOpen
  \bibfield{author}{%
  \bibinfo {author} {\bibfnamefont{J.~P.}\ \bibnamefont{{Rozelot}}}\ and\
  \bibinfo {author} {\bibfnamefont{C.}~\bibnamefont{{Damiani}}},\ }%
  \bibfield{journal}{%
  \Doi{10.1140/epjh/e2011-20017-4}{\bibinfo {journal} {Europ. Phys. J. H}}\ }%
  \textbf{\bibinfo {volume} {36}},\ \bibinfo {pages} {407} (\bibinfo {year}
  {2011})%
  \bibAnnoteFile{NoStop}{2011EPJH...36..407R}%
\bibitem{1990grg..conf..313S}%
  \BibitemOpen
  \bibfield{author}{%
  \bibinfo {author} {\bibfnamefont{I.~I.}\ \bibnamefont{{Shapiro}}},\ }%
  in\ \emph{\bibinfo {booktitle} {General Relativity and Gravitation}},\
  \bibinfo {editor} {edited by\ \bibinfo {editor}
  {\bibfnamefont{N.}~\bibnamefont{{Ashby}}}, \bibinfo {editor}
  {\bibfnamefont{D.~F.}\ \bibnamefont{{Bartlett}}},\ and\ \bibinfo {editor}
  {\bibfnamefont{W.}~\bibnamefont{{Wyss}}}}\  (\bibinfo {publisher}
  {Cambridge University Press, Cambridge, UK},\ \bibinfo {year} {1990}) \ p.\
  \bibinfo {pages} {313}%
  \bibAnnoteFile{NoStop}{1990grg..conf..313S}%
\bibitem{1999AJ....117..587P}%
  \BibitemOpen
  \bibfield{author}{%
  \bibinfo {author} {\bibfnamefont{A.~V.}\ \bibnamefont{{Petrova}}}\ and\
  \bibinfo {author} {\bibfnamefont{V.~V.}\ \bibnamefont{{Orlov}}},\ }%
  \bibfield{journal}{%
  \Doi{10.1086/300671}{\bibinfo {journal} {Astron. J.}}\ }%
  \textbf{\bibinfo {volume} {117}},\ \bibinfo {pages} {587} (\bibinfo {year}
  {1999})%
  \bibAnnoteFile{NoStop}{1999AJ....117..587P}%
\bibitem{DamourSchaferPA2PN}%
  \BibitemOpen
  \bibfield{author}{%
  \bibinfo {author} {\bibfnamefont{T.}~\bibnamefont{{Damour}}}\ and\ \bibinfo
  {author} {\bibfnamefont{G.}~\bibnamefont{{Schafer}}},\ }%
  \bibfield{journal}{%
  \Doi{10.1007/BF02828697}{\bibinfo {journal} {Nuovo Cimento B Serie}}\ }%
  \textbf{\bibinfo {volume} {101}},\ \bibinfo {pages} {127} (\bibinfo {year}
  {1988})%
  \bibAnnoteFile{NoStop}{DamourSchaferPA2PN}%
\bibitem{2005ApJ...629..979L}%
  \BibitemOpen
  \bibfield{author}{%
  \bibinfo {author} {\bibfnamefont{J.~M.}\ \bibnamefont{{Lattimer}}}\ and\
  \bibinfo {author} {\bibfnamefont{B.~F.}\ \bibnamefont{{Schutz}}},\ }%
  \bibfield{journal}{%
  \Doi{10.1086/431543}{\bibinfo {journal} {\apj}}\ }%
  \textbf{\bibinfo {volume} {629}},\ \bibinfo {pages} {979} (\bibinfo {year}
  {2005}),\ \Eprint{http://arxiv.org/abs/astro-ph/0411470}{arXiv:0411470}%
  \bibAnnoteFile{NoStop}{2005ApJ...629..979L}%
\bibitem{PhysRevD.66.044002}%
  \BibitemOpen
  \bibfield{author}{%
  \bibinfo {author} {\bibfnamefont{K.}~\bibnamefont{Glampedakis}}\ and\
  \bibinfo {author} {\bibfnamefont{D.}~\bibnamefont{Kennefick}},\ }%
  \bibfield{journal}{%
  \bibinfo {journal} {Phys. Rev. D}\ }%
  \textbf{\bibinfo {volume} {66}},\ \bibinfo {pages} {044002} (\bibinfo {year}
  {2002})%
  \bibAnnoteFile{NoStop}{PhysRevD.66.044002}%
\bibitem{2007CQGra..24S..83P}%
  \BibitemOpen
  \bibfield{author}{%
  \bibinfo {author} {\bibfnamefont{F.}~\bibnamefont{{Pretorius}}}\ and\
  \bibinfo {author} {\bibfnamefont{D.}~\bibnamefont{{Khurana}}},\ }%
  \bibfield{journal}{%
  \Doi{10.1088/0264-9381/24/12/S07}{\bibinfo {journal} {Classical Quantum
  Gravity}}\ }%
  \textbf{\bibinfo {volume} {24}},\ \bibinfo {pages} {83} (\bibinfo {year}
  {2007}),\ \Eprint{http://arxiv.org/abs/gr-qc/0702084}{arXiv:gr-qc/0702084}%
  \bibAnnoteFile{NoStop}{2007CQGra..24S..83P}%
\bibitem{Blanchet2006}%
  \BibitemOpen
  \bibfield{author}{%
  \bibinfo {author} {\bibfnamefont{L.}~\bibnamefont{Blanchet}},\ }%
  \bibfield{journal}{%
  \bibinfo {journal} {Living Rev.~Relativity}\ }%
  \textbf{\bibinfo {volume} {9}},\ \bibinfo {pages} {4} (\bibinfo {year}
  {2006})%
  \bibAnnoteFile{NoStop}{Blanchet2006}%
\bibitem{2011LRR....14....7P}%
  \BibitemOpen
  \bibfield{author}{%
  \bibinfo {author} {\bibfnamefont{E.}~\bibnamefont{{Poisson}}}, \bibinfo
  {author} {\bibfnamefont{A.}~\bibnamefont{{Pound}}},\ and\ \bibinfo {author}
  {\bibfnamefont{I.}~\bibnamefont{{Vega}}},\ }%
  \bibfield{journal}{%
  \Doi{10.12942/lrr-2011-7}{\bibinfo {journal} {Living Rev.~Relativity}}\ }%
  \textbf{\bibinfo {volume} {14}},\ \bibinfo {pages} {7} (\bibinfo {year}
  {2011}),\ \Eprint{http://arxiv.org/abs/1102.0529}{arXiv:1102.0529}%
  \bibAnnoteFile{NoStop}{2011LRR....14....7P}%
\bibitem{Buonanno99}%
  \BibitemOpen
  \bibfield{author}{%
  \bibinfo {author} {\bibfnamefont{A.}~\bibnamefont{Buonanno}}\ and\ \bibinfo
  {author} {\bibfnamefont{T.}~\bibnamefont{Damour}},\ }%
  \bibfield{journal}{%
  \Doi{10.1103/PhysRevD.59.084006}{\bibinfo {journal} {\prd}}\ }%
  \textbf{\bibinfo {volume} {59}},\ \bibinfo {pages} {084006} (\bibinfo {year}
  {1999}),\ \Eprint{http://arxiv.org/abs/gr-qc/9811091}{arXiv:gr-qc/9811091}%
  \bibAnnoteFile{NoStop}{Buonanno99}%
\bibitem{Buonanno00}%
  \BibitemOpen
  \bibfield{author}{%
  \bibinfo {author} {\bibfnamefont{A.}~\bibnamefont{Buonanno}}\ and\ \bibinfo
  {author} {\bibfnamefont{T.}~\bibnamefont{Damour}},\ }%
  \bibfield{journal}{%
  \Doi{10.1103/PhysRevD.62.064015}{\bibinfo {journal} {\prd}}\ }%
  \textbf{\bibinfo {volume} {62}},\ \bibinfo {pages} {064015} (\bibinfo {year}
  {2000}),\ \Eprint{http://arxiv.org/abs/gr-qc/0001013}{arXiv:gr-qc/0001013}%
  \bibAnnoteFile{NoStop}{Buonanno00}%
\bibitem{2000PhRvD..62h4011D}%
  \BibitemOpen
  \bibfield{author}{%
  \bibinfo {author} {\bibfnamefont{T.}~\bibnamefont{Damour}}, \bibinfo {author}
  {\bibfnamefont{P.}~\bibnamefont{Jaranowski}},\ and\ \bibinfo {author}
  {\bibfnamefont{G.}~\bibnamefont{Sch\"afer}},\ }%
  \bibfield{journal}{%
  \Doi{10.1103/PhysRevD.62.084011}{\bibinfo {journal} {Phys.\ Rev.\ D}}\ }%
  \textbf{\bibinfo {volume} {62}},\ \bibinfo {pages} {084011} (\bibinfo {year}
  {2000})%
  \bibAnnoteFile{NoStop}{2000PhRvD..62h4011D}%
\bibitem{Damour01c}%
  \BibitemOpen
  \bibfield{author}{%
  \bibinfo {author} {\bibfnamefont{T.}~\bibnamefont{Damour}},\ }%
  \bibfield{journal}{%
  \Doi{10.1103/PhysRevD.64.124013}{\bibinfo {journal} {\prd}}\ }%
  \textbf{\bibinfo {volume} {64}},\ \bibinfo {pages} {124013} (\bibinfo {year}
  {2001}),\ \Eprint{http://arxiv.org/abs/gr-qc/0103018}{arXiv:gr-qc/0103018}%
  \bibAnnoteFile{NoStop}{Damour01c}%
\bibitem{DJS2000}%
  \BibitemOpen
  \bibfield{author}{%
  \bibinfo {author} {\bibfnamefont{T.}~\bibnamefont{{Damour}}}, \bibinfo
  {author} {\bibfnamefont{P.}~\bibnamefont{{Jaranowski}}},\ and\ \bibinfo
  {author} {\bibfnamefont{G.}~\bibnamefont{{Sch{\"a}fer}}},\ }%
  \bibfield{journal}{%
  \Doi{10.1103/PhysRevD.62.044024}{\bibinfo {journal} {Phys.\ Rev.\ D}}\ }%
  \textbf{\bibinfo {volume} {62}},\ \bibinfo {pages} {044024} (\bibinfo {year}
  {2000})%
  \bibAnnoteFile{NoStop}{DJS2000}%
\bibitem{Damour:2007nc}%
  \BibitemOpen
  \bibfield{author}{%
  \bibinfo {author} {\bibfnamefont{T.}~\bibnamefont{Damour}}, \bibinfo {author}
  {\bibfnamefont{P.}~\bibnamefont{Jaranowski}},\ and\ \bibinfo {author}
  {\bibfnamefont{G.}~\bibnamefont{Sch{\"a}fer}},\ }%
  \bibfield{journal}{%
  \Doi{10.1103/PhysRevD.77.064032}{\bibinfo {journal} {\prd}}\ }%
  \textbf{\bibinfo {volume} {77}},\ \bibinfo {pages} {064032} (\bibinfo {year}
  {2008}),\ \Eprint{http://arxiv.org/abs/0711.1048}{arXiv:0711.1048}%
  \bibAnnoteFile{NoStop}{Damour:2007nc}%
\bibitem{KG3PNandSO}%
  \BibitemOpen
  \bibfield{author}{%
  \bibinfo {author} {\bibfnamefont{C.}~\bibnamefont{{K{\"o}nigsd{\"o}rffer}}}\
  and\ \bibinfo {author} {\bibfnamefont{A.}~\bibnamefont{{Gopakumar}}},\ }%
  \bibfield{journal}{%
  \Doi{10.1103/PhysRevD.71.024039}{\bibinfo {journal} {\prd}}\ }%
  \textbf{\bibinfo {volume} {71}},\ \bibinfo {eid} {024039} (\bibinfo {year}
  {2005}),\ \Eprint{http://arxiv.org/abs/gr-qc/0501011}{arXiv:gr-qc/0501011}%
  \bibAnnoteFile{NoStop}{KG3PNandSO}%
\bibitem{1979Ap&SS..62..185A}%
  \BibitemOpen
  \bibfield{author}{%
  \bibinfo {author} {\bibfnamefont{G.}~\bibnamefont{{Antonacopoulos}}}\ and\
  \bibinfo {author} {\bibfnamefont{E.}~\bibnamefont{{Tsoupakis}}},\ }%
  \bibfield{journal}{%
  \Doi{10.1007/BF00643911}{\bibinfo {journal} {Astrophys. Space. Sci.}}\ }%
  \textbf{\bibinfo {volume} {62}},\ \bibinfo {pages} {185} (\bibinfo {year}
  {1979})%
  \bibAnnoteFile{NoStop}{1979Ap&SS..62..185A}%
\bibitem{1989PThPh..81..679O}%
  \BibitemOpen
  \bibfield{author}{%
  \bibinfo {author} {\bibfnamefont{T.}~\bibnamefont{{Ohta}}}\ and\ \bibinfo
  {author} {\bibfnamefont{T.}~\bibnamefont{{Kimura}}},\ }%
  \bibfield{journal}{%
  \Doi{10.1143/PTP.81.679}{\bibinfo {journal} {Progr. Theor. Phys.}}\ }%
  \textbf{\bibinfo {volume} {81}},\ \bibinfo {pages} {679} (\bibinfo {year}
  {1989})%
  \bibAnnoteFile{NoStop}{1989PThPh..81..679O}%
\bibitem{1938MNRAS..98..734C}%
  \BibitemOpen
  \bibfield{author}{%
  \bibinfo {author} {\bibfnamefont{T.~G.}\ \bibnamefont{{Cowling}}},\ }%
  \bibfield{journal}{%
  \bibinfo {journal} {Mon. Not. Royal Astr. Soc.}\ }%
  \textbf{\bibinfo {volume} {98}},\ \bibinfo {pages} {734} (\bibinfo {year}
  {1938})%
  \bibAnnoteFile{NoStop}{1938MNRAS..98..734C}%
\bibitem{1980MNRAS.193..603P}%
  \BibitemOpen
  \bibfield{author}{%
  \bibinfo {author} {\bibfnamefont{J.}~\bibnamefont{{Papaloizou}}}\ and\
  \bibinfo {author} {\bibfnamefont{J.~E.}\ \bibnamefont{{Pringle}}},\ }%
  \bibfield{journal}{%
  \bibinfo {journal} {Mon. Not. Royal Astr. Soc.}\ }%
  \textbf{\bibinfo {volume} {193}},\ \bibinfo {pages} {603} (\bibinfo {year}
  {1980})%
  \bibAnnoteFile{NoStop}{1980MNRAS.193..603P}%
\bibitem{2013LRR....16....1A}%
  \BibitemOpen
  \bibfield{author}{%
  \bibinfo {author} {\bibfnamefont{M.~A.}\ \bibnamefont{{Abramowicz}}}\ and\
  \bibinfo {author} {\bibfnamefont{P.~C.}\ \bibnamefont{{Fragile}}},\ }%
  \bibfield{journal}{%
  \Doi{10.12942/lrr-2013-1}{\bibinfo {journal} {Living Rev.~Relativity}}\ }%
  \textbf{\bibinfo {volume} {16}},\ \bibinfo {pages} {1} (\bibinfo {year}
  {2013}),\
  \Eprint{http://arxiv.org/abs/1104.5499}{arXiv:1104.5499}%
  \bibAnnoteFile{NoStop}{2013LRR....16....1A}%
\bibitem{2002CQGra..19.2743S}%
  \BibitemOpen
  \bibfield{author}{%
  \bibinfo {author} {\bibfnamefont{W.}~\bibnamefont{{Schmidt}}},\ }%
  \bibfield{journal}{%
  \Doi{10.1088/0264-9381/19/10/314}{\bibinfo {journal} {Classical \ Quantum
  Gravity}}\ }%
  \textbf{\bibinfo {volume} {19}},\ \bibinfo {pages} {2743} (\bibinfo {year}
  {2002}),\ \Eprint{http://arxiv.org/abs/gr-qc/0202090}{arXiv:gr-qc/0202090}%
  \bibAnnoteFile{NoStop}{2002CQGra..19.2743S}%
\bibitem{2013PhRvL.111b1101H}%
  \BibitemOpen
  \bibfield{author}{%
  \bibinfo {author} {\bibfnamefont{S.}~\bibnamefont{{Hergt}}}, \bibinfo
  {author} {\bibfnamefont{A.}~\bibnamefont{{Shah}}},\ and\ \bibinfo {author}
  {\bibfnamefont{G.}~\bibnamefont{{Sch{\"a}fer}}},\ }%
  \bibfield{journal}{%
  \Doi{10.1103/PhysRevLett.111.021101}{\bibinfo {journal} {\prl}}\ }%
  \textbf{\bibinfo {volume} {111}},\ \bibinfo {eid} {021101} (\bibinfo {year}
  {2013}),\ \Eprint{http://arxiv.org/abs/1303.6829}{arXiv:1303.6829}%
  \bibAnnoteFile{NoStop}{2013PhRvL.111b1101H}%
\bibitem{Barack:2011ed}%
  \BibitemOpen
  \bibfield{author}{%
  \bibinfo {author} {\bibfnamefont{L.}~\bibnamefont{Barack}}\ and\ \bibinfo
  {author} {\bibfnamefont{N.}~\bibnamefont{Sago}},\ }%
  \bibfield{journal}{%
  \Doi{10.1103/PhysRevD.83.084023}{\bibinfo {journal} {\prd}}\ }%
  \textbf{\bibinfo {volume} {83}},\ \bibinfo {pages} {084023} (\bibinfo {year}
  {2011}),\ \Eprint{http://arxiv.org/abs/1101.3331}{arXiv:1101.3331}%
  \bibAnnoteFile{NoStop}{Barack:2011ed}%
\bibitem{Damour:2009sm}%
  \BibitemOpen
  \bibfield{author}{%
  \bibinfo {author} {\bibfnamefont{T.}~\bibnamefont{Damour}},\ }%
  \bibfield{journal}{%
  \Doi{10.1103/PhysRevD.81.024017}{\bibinfo {journal} {\prd}}\ }%
  \textbf{\bibinfo {volume} {81}},\ \bibinfo {pages} {024017} (\bibinfo {year}
  {2010}),\ \Eprint{http://arxiv.org/abs/0910.5533}{arXiv:0910.5533}%
  \bibAnnoteFile{NoStop}{Damour:2009sm}%
\bibitem{Barack:2010ny}%
  \BibitemOpen
  \bibfield{author}{%
  \bibinfo {author} {\bibfnamefont{L.}~\bibnamefont{Barack}}, \bibinfo {author}
  {\bibfnamefont{T.}~\bibnamefont{Damour}},\ and\ \bibinfo {author}
  {\bibfnamefont{N.}~\bibnamefont{Sago}},\ }%
  \bibfield{journal}{%
  \Doi{10.1103/PhysRevD.82.084036}{\bibinfo {journal} {\prd}}\ }%
  \textbf{\bibinfo {volume} {82}},\ \bibinfo {pages} {084036} (\bibinfo {year}
  {2010}),\ \Eprint{http://arxiv.org/abs/1008.0935}{arXiv:1008.0935}%
  \bibAnnoteFile{NoStop}{Barack:2010ny}%
\bibitem{Barausse:2011dq}%
  \BibitemOpen
  \bibfield{author}{%
  \bibinfo {author} {\bibfnamefont{E.}~\bibnamefont{{Barausse}}}, \bibinfo
  {author} {\bibfnamefont{A.}~\bibnamefont{{Buonanno}}},\ and\ \bibinfo
  {author} {\bibfnamefont{A.}~\bibnamefont{{Le Tiec}}},\ }%
  \bibfield{journal}{%
  \bibinfo {journal} {\prd}\ }%
  \textbf{\bibinfo {volume} {85}},\ \bibinfo {pages} {064010} (\bibinfo {year}
  {2012}),\ \Eprint{http://arxiv.org/abs/1111.5610}{arXiv:1111.5610}%
  \bibAnnoteFile{NoStop}{Barausse:2011dq}%
\bibitem{Mroue:2010re}%
  \BibitemOpen
  \bibfield{author}{%
  \bibinfo {author} {\bibfnamefont{A.~H.}\ \bibnamefont{Mroue}}, \bibinfo
  {author} {\bibfnamefont{H.~P.}\ \bibnamefont{Pfeiffer}}, \bibinfo {author}
  {\bibfnamefont{L.~E.}\ \bibnamefont{Kidder}},\ and\ \bibinfo {author}
  {\bibfnamefont{S.~A.}\ \bibnamefont{Teukolsky}},\ }%
  \bibfield{journal}{%
  \Doi{10.1103/PhysRevD.82.124016}{\bibinfo {journal} {Phys.\ Rev.\ D}}\ }%
  \textbf{\bibinfo {volume} {82}},\ \bibinfo {pages} {124016} (\bibinfo {year}
  {2010}),\ \Eprint{http://arxiv.org/abs/1004.4697}{arXiv:1004.4697}%
  \bibAnnoteFile{NoStop}{Mroue:2010re}%
\bibitem{LeTiec:2011bk}%
  \BibitemOpen
  \bibfield{author}{%
  \bibinfo {author} {\bibfnamefont{A.}~\bibnamefont{Le~Tiec}}, \bibinfo
  {author} {\bibfnamefont{A.~H.}\ \bibnamefont{Mroue}}, \bibinfo {author}
  {\bibfnamefont{L.}~\bibnamefont{Barack}}, \bibinfo {author}
  {\bibfnamefont{A.}~\bibnamefont{Buonanno}}, \bibinfo {author}
  {\bibfnamefont{H.~P.}\ \bibnamefont{Pfeiffer}}, \emph{et~al.},\ }%
  \bibfield{journal}{%
  \Doi{10.1103/PhysRevLett.107.141101}{\bibinfo {journal} {Phys.\ Rev.\
  Lett.}}\ }%
  \textbf{\bibinfo {volume} {107}},\ \bibinfo {pages} {141101} (\bibinfo {year}
  {2011}),\ \Eprint{http://arxiv.org/abs/1106.3278}{arXiv:1106.3278}%
  \bibAnnoteFile{NoStop}{LeTiec:2011bk}%
\bibitem{Damour:024009}%
  \BibitemOpen
  \bibfield{author}{%
  \bibinfo {author} {\bibfnamefont{T.}~\bibnamefont{Damour}}, \bibinfo {author}
  {\bibfnamefont{P.}~\bibnamefont{Jaranowski}},\ and\ \bibinfo {author}
  {\bibfnamefont{G.}~\bibnamefont{Schafer}},\ }%
  \bibfield{journal}{%
  \Doi{10.1103/PhysRevD.78.024009}{\bibinfo {journal} {\prd}}\ }%
  \textbf{\bibinfo {volume} {78}},\ \bibinfo {pages} {024009} (\bibinfo {year}
  {2008}),\ \Eprint{http://arxiv.org/abs/0803.0915}{arXiv:0803.0915}%
  \bibAnnoteFile{NoStop}{Damour:024009}%
\bibitem{Barausse:2009aa}%
  \BibitemOpen
  \bibfield{author}{%
  \bibinfo {author} {\bibfnamefont{E.}~\bibnamefont{Barausse}}, \bibinfo
  {author} {\bibfnamefont{E.}~\bibnamefont{Racine}},\ and\ \bibinfo {author}
  {\bibfnamefont{A.}~\bibnamefont{Buonanno}},\ }%
  \bibfield{journal}{%
  \Doi{10.1103/PhysRevD.80.104025}{\bibinfo {journal} {\prd}}\ }%
  \textbf{\bibinfo {volume} {80}},\ \bibinfo {pages} {104025} (\bibinfo {year}
  {2009}),\ \Eprint{http://arxiv.org/abs/0907.4745}{arXiv:0907.4745}%
  \bibAnnoteFile{NoStop}{Barausse:2009aa}%
\bibitem{Barausse:2009xi}%
  \BibitemOpen
  \bibfield{author}{%
  \bibinfo {author} {\bibfnamefont{E.}~\bibnamefont{Barausse}}\ and\ \bibinfo
  {author} {\bibfnamefont{A.}~\bibnamefont{Buonanno}},\ }%
  \bibfield{journal}{%
  \Doi{10.1103/PhysRevD.81.084024}{\bibinfo {journal} {\prd}}\ }%
  \textbf{\bibinfo {volume} {81}},\ \bibinfo {pages} {084024} (\bibinfo {year}
  {2010}),\ \Eprint{http://arxiv.org/abs/0912.3517}{arXiv:0912.3517}%
  \bibAnnoteFile{NoStop}{Barausse:2009xi}%
\bibitem{Barausse:2011ys}%
  \BibitemOpen
  \bibfield{author}{%
  \bibinfo {author} {\bibfnamefont{E.}~\bibnamefont{Barausse}}\ and\ \bibinfo
  {author} {\bibfnamefont{A.}~\bibnamefont{Buonanno}},\ }%
  \bibfield{journal}{%
  \Doi{10.1103/PhysRevD.84.104027}{\bibinfo {journal} {\prd}}\ }%
  \textbf{\bibinfo {volume} {84}},\ \bibinfo {pages} {104027} (\bibinfo {year}
  {2011}),\ \Eprint{http://arxiv.org/abs/1107.2904}{arXiv:1107.2904}%
  \bibAnnoteFile{NoStop}{Barausse:2011ys}%
\bibitem{2011PhRvD..84h4028N}%
  \BibitemOpen
  \bibfield{author}{%
  \bibinfo {author} {\bibfnamefont{A.}~\bibnamefont{{Nagar}}},\ }%
  \bibfield{journal}{%
  \Doi{10.1103/PhysRevD.84.084028}{\bibinfo {journal} {\prd}}\ }%
  \textbf{\bibinfo {volume} {84}},\ \bibinfo {eid} {084028} (\bibinfo {year}
  {2011}),\ \Eprint{http://arxiv.org/abs/1106.4349}{arXiv:1106.4349}%
  \bibAnnoteFile{NoStop}{2011PhRvD..84h4028N}%
\bibitem{2013PhRvD..87l4036B}%
  \BibitemOpen
  \bibfield{author}{%
  \bibinfo {author} {\bibfnamefont{S.}~\bibnamefont{{Balmelli}}}\ and\ \bibinfo
  {author} {\bibfnamefont{P.}~\bibnamefont{{Jetzer}}},\ }%
  \bibfield{journal}{%
  \Doi{10.1103/PhysRevD.87.124036}{\bibinfo {journal} {\prd}}\ }%
  \textbf{\bibinfo {volume} {87}},\ \bibinfo {eid} {124036} (\bibinfo {year}
  {2013}),\ \Eprint{http://arxiv.org/abs/1305.5674}{arXiv:1305.5674}%
  \bibAnnoteFile{NoStop}{2013PhRvD..87l4036B}%
\bibitem{Steinhoff:2008zr}%
  \BibitemOpen
  \bibfield{author}{%
  \bibinfo {author} {\bibfnamefont{J.}~\bibnamefont{Steinhoff}}, \bibinfo
  {author} {\bibfnamefont{G.}~\bibnamefont{Sch{\"a}fer}},\ and\ \bibinfo
  {author} {\bibfnamefont{S.}~\bibnamefont{Hergt}},\ }%
  \bibfield{journal}{%
  \Doi{10.1103/PhysRevD.77.104018}{\bibinfo {journal} {Phys.~Rev.~D}}\ }%
  \textbf{\bibinfo {volume} {77}},\ \bibinfo {pages} {104018} (\bibinfo {year}
  {2008}),\ \Eprint{http://arxiv.org/abs/0805.3136}{arXiv:0805.3136}%
  \bibAnnoteFile{NoStop}{Steinhoff:2008zr}%
\bibitem{Pan:2013rra}%
  \BibitemOpen
  \bibfield{author}{%
  \bibinfo {author} {\bibfnamefont{Y.}~\bibnamefont{{Pan}}}, \bibinfo {author}
  {\bibfnamefont{A.}~\bibnamefont{{Buonanno}}}, \bibinfo {author}
  {\bibfnamefont{A.}~\bibnamefont{{Taracchini}}}, \bibinfo {author}
  {\bibfnamefont{L.~E.}\ \bibnamefont{{Kidder}}}, \bibinfo {author}
  {\bibfnamefont{A.~H.}\ \bibnamefont{{Mroue}}}, \bibinfo {author}
  {\bibfnamefont{H.~P.}\ \bibnamefont{{Pfeiffer}}}, \bibinfo {author}
  {\bibfnamefont{M.~A.}\ \bibnamefont{{Scheel}}},\ and\ \bibinfo {author}
  {\bibfnamefont{B.}~\bibnamefont{{Szilagyi}}},\ }%
  \bibfield{journal}{%
  \bibinfo {journal} {ArXiv e-prints}}%
   (\bibinfo {year} {2013}),\
  \Eprint{http://arxiv.org/abs/1307.6232}{arXiv:1307.6232}%
  \bibAnnoteFile{NoStop}{Pan:2013rra}%
\bibitem{Mroueinprep}%
  \BibitemOpen
  \bibfield{author}{%
  \bibinfo {author} {\bibfnamefont{A.~H.}\ \bibnamefont{{Mrou{\'e}}}}, \bibinfo
  {author} {\bibfnamefont{G.}~\bibnamefont{Lovelace}},\ and\ \bibinfo {author}
  {\bibfnamefont{H.~P.}\ \bibnamefont{Pfeiffer}},\ }%
  \bibfield{journal}{%
  \bibinfo {journal} {in preparation}}%
   (\bibinfo {year} {2013})%
  \bibAnnoteFile{NoStop}{Mroueinprep}%
\bibitem{2012arXiv1210.2958M}%
  \BibitemOpen
  \bibfield{author}{%
  \bibinfo {author} {\bibfnamefont{A.~H.}\ \bibnamefont{{Mrou{\'e}}}}\ and\
  \bibinfo {author} {\bibfnamefont{H.~P.}\ \bibnamefont{{Pfeiffer}}},\ }%
  \bibfield{journal}{%
  \bibinfo {journal} {ArXiv e-prints}}%
   (\bibinfo {year} {2012}),\
  \Eprint{http://arxiv.org/abs/1210.2958}{arXiv:1210.2958}%
  \bibAnnoteFile{NoStop}{2012arXiv1210.2958M}%
\bibitem{Hemberger2013}%
  \BibitemOpen
  \bibfield{author}{%
  \bibinfo {author} {\bibfnamefont{D.~A.}\ \bibnamefont{{Hemberger}}}, \bibinfo
  {author} {\bibfnamefont{M.~A.}\ \bibnamefont{{Scheel}}}, \bibinfo {author}
  {\bibfnamefont{L.~E.}\ \bibnamefont{{Kidder}}}, \bibinfo {author}
  {\bibfnamefont{B.}~\bibnamefont{{Szil{\'a}gyi}}}, \bibinfo {author}
  {\bibfnamefont{G.}~\bibnamefont{{Lovelace}}}, \bibinfo {author}
  {\bibfnamefont{N.~W.}\ \bibnamefont{{Taylor}}},\ and\ \bibinfo {author}
  {\bibfnamefont{S.~A.}\ \bibnamefont{{Teukolsky}}},\ }%
  \bibfield{journal}{%
  \Doi{10.1088/0264-9381/30/11/115001}{\bibinfo {journal} {Classical \ Quantum
  Gravity}}\ }%
  \textbf{\bibinfo {volume} {30}},\ \bibinfo {eid} {115001} (\bibinfo {year}
  {2013}),\ \Eprint{http://arxiv.org/abs/1211.6079}{arXiv:1211.6079}%
  \bibAnnoteFile{NoStop}{Hemberger2013}%
\bibitem{2013arXiv1304.6077M}%
  \BibitemOpen
  \bibfield{author}{%
  \bibinfo {author} {\bibfnamefont{A.~H.}\ \bibnamefont{{Mroue}}}, \bibinfo
  {author} {\bibfnamefont{M.~A.}\ \bibnamefont{{Scheel}}}, \bibinfo {author}
  {\bibfnamefont{B.}~\bibnamefont{{Szilagyi}}}, \bibinfo {author}
  {\bibfnamefont{H.~P.}\ \bibnamefont{{Pfeiffer}}}, \bibinfo {author}
  {\bibfnamefont{M.}~\bibnamefont{{Boyle}}}, \bibinfo {author}
  {\bibfnamefont{D.~A.}\ \bibnamefont{{Hemberger}}}, \bibinfo {author}
  {\bibfnamefont{L.~E.}\ \bibnamefont{{Kidder}}}, \bibinfo {author}
  {\bibfnamefont{G.}~\bibnamefont{{Lovelace}}}, \bibinfo {author}
  {\bibfnamefont{S.}~\bibnamefont{{Ossokine}}}, \bibinfo {author}
  {\bibfnamefont{N.~W.}\ \bibnamefont{{Taylor}}}, \bibinfo {author}
  {\bibfnamefont{A.}~\bibnamefont{{Zenginoglu}}}, \bibinfo {author}
  {\bibfnamefont{L.~T.}\ \bibnamefont{{Buchman}}}, \bibinfo {author}
  {\bibfnamefont{T.}~\bibnamefont{{Chu}}}, \bibinfo {author}
  {\bibfnamefont{E.}~\bibnamefont{{Foley}}}, \bibinfo {author}
  {\bibfnamefont{M.}~\bibnamefont{{Giesler}}}, \bibinfo {author}
  {\bibfnamefont{R.}~\bibnamefont{{Owen}}},\ and\ \bibinfo {author}
  {\bibfnamefont{S.~A.}\ \bibnamefont{{Teukolsky}}},\ }%
  \bibfield{journal}{%
  \bibinfo {journal} {ArXiv e-prints}}%
   (\bibinfo {year} {2013}),\
  \Eprint{http://arxiv.org/abs/1304.6077}{arXiv:1304.6077}%
  \bibAnnoteFile{NoStop}{2013arXiv1304.6077M}%
\bibitem{Mathisson:1937zz}%
  \BibitemOpen
  \bibfield{author}{%
  \bibinfo {author} {\bibfnamefont{M.}~\bibnamefont{Mathisson}},\ }%
  \bibfield{journal}{%
  \bibinfo {journal} {Acta Phys.Polon.}\ }%
  \textbf{\bibinfo {volume} {6}},\ \bibinfo {pages} {163} (\bibinfo {year}
  {1937})%
  \bibAnnoteFile{NoStop}{Mathisson:1937zz}%
\bibitem{Papapetrou:1951pa}%
  \BibitemOpen
  \bibfield{author}{%
  \bibinfo {author} {\bibfnamefont{A.}~\bibnamefont{Papapetrou}},\ }%
  \bibfield{journal}{%
  \Doi{10.1098/rspa.1951.0200}{\bibinfo {journal} {Proc.\ Roy.\ Soc.\ Lond.\
  A}}\ }%
  \textbf{\bibinfo {volume} {209}},\ \bibinfo {pages} {248} (\bibinfo {year}
  {1951})%
  \bibAnnoteFile{NoStop}{Papapetrou:1951pa}%
\bibitem{Corinaldesi:1951pb}%
  \BibitemOpen
  \bibfield{author}{%
  \bibinfo {author} {\bibfnamefont{E.}~\bibnamefont{Corinaldesi}}\ and\
  \bibinfo {author} {\bibfnamefont{A.}~\bibnamefont{Papapetrou}},\ }%
  \bibfield{journal}{%
  \Doi{10.1098/rspa.1951.0201}{\bibinfo {journal} {Proc.\ Roy.\ Soc.\ Lond.\
  A}}\ }%
  \textbf{\bibinfo {volume} {209}},\ \bibinfo {pages} {259} (\bibinfo {year}
  {1951})%
  \bibAnnoteFile{NoStop}{Corinaldesi:1951pb}%
\bibitem{Taub1964}%
  \BibitemOpen
  \bibfield{author}{%
  \bibinfo {author} {\bibfnamefont{A.~H.}\ \bibnamefont{{Taub}}},\ }%
  \bibfield{journal}{%
  \Doi{10.1063/1.1704055}{\bibinfo {journal} {J.\ Math.\ Phys.}}\ }%
  \textbf{\bibinfo {volume} {5}},\ \bibinfo {pages} {112} (\bibinfo {year}
  {1964})%
  \bibAnnoteFile{NoStop}{Taub1964}%
\bibitem{Dixon1970}%
  \BibitemOpen
  \bibfield{author}{%
  \bibinfo {author} {\bibfnamefont{W.~G.}\ \bibnamefont{{Dixon}}},\ }%
  \bibfield{journal}{%
  \Doi{10.1098/rspa.1970.0020}{\bibinfo {journal} {Proc.\ Roy.\ Soc.\ Lond.\
  A}}\ }%
  \textbf{\bibinfo {volume} {314}},\ \bibinfo {pages} {499} (\bibinfo {year}
  {1970})%
  \bibAnnoteFile{NoStop}{Dixon1970}%
\bibitem{Dixon1974}%
  \BibitemOpen
  \bibfield{author}{%
  \bibinfo {author} {\bibfnamefont{W.~G.}\ \bibnamefont{Dixon}},\ }%
  \bibfield{journal}{%
  \bibinfo {journal} {Phil.\ Trans.\ Roy.\ Soc.\ Lond.\ A}\ }%
  \textbf{\bibinfo {volume} {277}},\ \bibinfo {pages} {59} (\bibinfo {year}
  {1974})%
  \bibAnnoteFile{NoStop}{Dixon1974}%
\bibitem{EhlersRudolph}%
  \BibitemOpen
  \bibfield{author}{%
  \bibinfo {author} {\bibfnamefont{J.}~\bibnamefont{{Ehlers}}}\ and\ \bibinfo
  {author} {\bibfnamefont{E.}~\bibnamefont{{Rudolph}}},\ }%
  \bibfield{journal}{%
  \Doi{10.1007/BF00763547}{\bibinfo {journal} {Gen.\ Rel.\ Grav.}}\ }%
  \textbf{\bibinfo {volume} {8}},\ \bibinfo {pages} {197} (\bibinfo {year}
  {1977})%
  \bibAnnoteFile{NoStop}{EhlersRudolph}%
\bibitem{2012CQGra..29e5012H}%
  \BibitemOpen
  \bibfield{author}{%
  \bibinfo {author} {\bibfnamefont{A.~I.}\ \bibnamefont{{Harte}}},\ }%
  \bibfield{journal}{%
  \Doi{10.1088/0264-9381/29/5/055012}{\bibinfo {journal} {Classical \ Quantum
  Grav.}}\ }%
  \textbf{\bibinfo {volume} {29}},\ \bibinfo {eid} {055012} (\bibinfo {year}
  {2012}),\ \Eprint{http://arxiv.org/abs/1103.0543}{arXiv:1103.0543}%
  \bibAnnoteFile{NoStop}{2012CQGra..29e5012H}%
\bibitem{2004PhRvD..69d4011B}%
  \BibitemOpen
  \bibfield{author}{%
  \bibinfo {author} {\bibfnamefont{L.~M.}\ \bibnamefont{{Burko}}},\ }%
  \bibfield{journal}{%
  \Doi{10.1103/PhysRevD.69.044011}{\bibinfo {journal} {\prd}}\ }%
  \textbf{\bibinfo {volume} {69}},\ \bibinfo {eid} {044011} (\bibinfo {year}
  {2004}),\ \Eprint{http://arxiv.org/abs/gr-qc/0308003}{arXiv:gr-qc/0308003}%
  \bibAnnoteFile{NoStop}{2004PhRvD..69d4011B}%
\bibitem{SP12}%
  \BibitemOpen
  \bibfield{author}{%
  \bibinfo {author} {\bibfnamefont{J.}~\bibnamefont{{Steinhoff}}}\ and\
  \bibinfo {author} {\bibfnamefont{D.}~\bibnamefont{{Puetzfeld}}},\ }%
  \bibfield{journal}{%
  \Doi{10.1103/PhysRevD.86.044033}{\bibinfo {journal} {\prd}}\ }%
  \textbf{\bibinfo {volume} {86}},\ \bibinfo {eid} {044033} (\bibinfo {year} {2012}),\
  \Eprint{http://arxiv.org/abs/1205.3926}{arXiv:1205.3926 [gr-qc]}%
  \bibAnnoteFile{NoStop}{SP12}%
\bibitem{2011PhRvD..83b4027F}%
  \BibitemOpen
  \bibfield{author}{%
  \bibinfo {author} {\bibfnamefont{M.}~\bibnamefont{{Favata}}},\ }%
  \bibfield{journal}{%
  \Doi{10.1103/PhysRevD.83.024027}{\bibinfo {journal} {\prd}}\ }%
  \textbf{\bibinfo {volume} {83}},\ \bibinfo {eid} {024027} (\bibinfo {year}
  {2011}),\ \Eprint{http://arxiv.org/abs/1008.4622}{arXiv:1008.4622}%
  \bibAnnoteFile{NoStop}{2011PhRvD..83b4027F}%
\bibitem{Pan:2009wj}%
  \BibitemOpen
  \bibfield{author}{%
  \bibinfo {author} {\bibfnamefont{Y.}~\bibnamefont{Pan}}, \bibinfo {author}
  {\bibfnamefont{A.}~\bibnamefont{Buonanno}}, \bibinfo {author}
  {\bibfnamefont{L.~T.}\ \bibnamefont{Buchman}}, \bibinfo {author}
  {\bibfnamefont{T.}~\bibnamefont{Chu}}, \bibinfo {author}
  {\bibfnamefont{L.~E.}\ \bibnamefont{Kidder}}, \emph{et~al.},\ }%
  \bibfield{journal}{%
  \Doi{10.1103/PhysRevD.81.084041}{\bibinfo {journal} {\prd}}\ }%
  \textbf{\bibinfo {volume} {81}},\ \bibinfo {pages} {084041} (\bibinfo {year}
  {2010}),\ \Eprint{http://arxiv.org/abs/0912.3466}{arXiv:0912.3466}%
  \bibAnnoteFile{NoStop}{Pan:2009wj}%
\bibitem{Taracchini2012}%
  \BibitemOpen
  \bibfield{author}{%
  \bibinfo {author} {\bibfnamefont{A.}~\bibnamefont{Taracchini}}, \bibinfo
  {author} {\bibfnamefont{Y.}~\bibnamefont{Pan}}, \bibinfo {author}
  {\bibfnamefont{A.}~\bibnamefont{Buonanno}}, \bibinfo {author}
  {\bibfnamefont{E.}~\bibnamefont{Barausse}}, \bibinfo {author}
  {\bibfnamefont{M.}~\bibnamefont{Boyle}}, \bibinfo {author}
  {\bibfnamefont{T.}~\bibnamefont{Chu}}, \bibinfo {author}
  {\bibfnamefont{G.}~\bibnamefont{Lovelace}}, \bibinfo {author}
  {\bibfnamefont{H.~P.}\ \bibnamefont{Pfeiffer}},\ and\ \bibinfo {author}
  {\bibfnamefont{M.~A.}\ \bibnamefont{Scheel}},\ }%
  \bibfield{journal}{%
  \Doi{10.1103/PhysRevD.86.024011}{\bibinfo {journal} {\prd}}\ }%
  \textbf{\bibinfo {volume} {86}},\ \bibinfo {pages} {024011} (\bibinfo {year}
  {2012})%
  \bibAnnoteFile{NoStop}{Taracchini2012}%
\bibitem{Hinder:2013oqa}%
  \BibitemOpen
  \bibfield{author}{%
  \bibinfo {author} {\bibfnamefont{I.}~\bibnamefont{Hinder}} \emph{et~al.}
  (\bibinfo {collaboration} {The NRAR Collaboration})}%
   (\bibinfo {year} {2013}),\
  \Eprint{http://arxiv.org/abs/1307.5307}{arXiv:1307.5307}%
  \bibAnnoteFile{NoStop}{Hinder:2013oqa}%
\bibitem{2005PhRvD..71b4039K}%
  \BibitemOpen
  \bibfield{author}{%
  \bibinfo {author} {\bibfnamefont{C.}~\bibnamefont{{K{\"o}nigsd{\"o}rffer}}}\
  and\ \bibinfo {author} {\bibfnamefont{A.}~\bibnamefont{{Gopakumar}}},\ }%
  \bibfield{journal}{%
  \Doi{10.1103/PhysRevD.71.024039}{\bibinfo {journal} {\prd}}\ }%
  \textbf{\bibinfo {volume} {71}},\ \bibinfo {eid} {024039} (\bibinfo {year}
  {2005}),\ \Eprint{http://arxiv.org/abs/gr-qc/0501011}{gr-qc/0501011}%
  \bibAnnoteFile{NoStop}{2005PhRvD..71b4039K}%
\bibitem{2009PhRvD..80l4034T}%
  \BibitemOpen
  \bibfield{author}{%
  \bibinfo {author} {\bibfnamefont{M.}~\bibnamefont{{Tessmer}}},\ }%
  \bibfield{journal}{%
  \Doi{10.1103/PhysRevD.80.124034}{\bibinfo {journal} {\prd}}\ }%
  \textbf{\bibinfo {volume} {80}},\ \bibinfo {eid} {124034} (\bibinfo {year}
  {2009}),\ \Eprint{http://arxiv.org/abs/0910.5931}{arXiv:0910.5931}%
  \bibAnnoteFile{NoStop}{2009PhRvD..80l4034T}%
\bibitem{2013PhRvD..87f4035T}%
  \BibitemOpen
  \bibfield{author}{%
  \bibinfo {author} {\bibfnamefont{M.}~\bibnamefont{{Tessmer}}}, \bibinfo
  {author} {\bibfnamefont{J.}~\bibnamefont{{Steinhoff}}},\ and\ \bibinfo
  {author} {\bibfnamefont{G.}~\bibnamefont{{Sch{\"a}fer}}},\ }%
  \bibfield{journal}{%
  \Doi{10.1103/PhysRevD.87.064035}{\bibinfo {journal} {\prd}}\ }%
  \textbf{\bibinfo {volume} {87}},\ \bibinfo {eid} {064035} (\bibinfo {year}
  {2013}),\ \Eprint{http://arxiv.org/abs/1301.3665}{arXiv:1301.3665}%
  \bibAnnoteFile{NoStop}{2013PhRvD..87f4035T}%
\bibitem{2010PhRvD..82j4031G}%
  \BibitemOpen
  \bibfield{author}{%
  \bibinfo {author} {\bibfnamefont{L.~{\'A}.}\ \bibnamefont{{Gergely}}},\ }%
  \bibfield{journal}{%
  \Doi{10.1103/PhysRevD.82.104031}{\bibinfo {journal} {\prd}}\ }%
  \textbf{\bibinfo {volume} {82}},\ \bibinfo {eid} {104031} (\bibinfo {year}
  {2010}),\ \Eprint{http://arxiv.org/abs/1005.5330}{arXiv:1005.5330}%
  \bibAnnoteFile{NoStop}{2010PhRvD..82j4031G}%
\bibitem{1998PhRvD..58l4001G}%
  \BibitemOpen
  \bibfield{author}{%
  \bibinfo {author} {\bibfnamefont{L.~{\'A}.}\ \bibnamefont{{Gergely}}},
  \bibinfo {author} {\bibfnamefont{Z.~I.}\ \bibnamefont{{Perj{\'e}s}}},\ and\
  \bibinfo {author} {\bibfnamefont{M.}~\bibnamefont{{Vas{\'u}th}}},\ }%
  \bibfield{journal}{%
  \Doi{10.1103/PhysRevD.58.124001}{\bibinfo {journal} {\prd}}\ }%
  \textbf{\bibinfo {volume} {58}},\ \bibinfo {eid} {124001} (\bibinfo {year}
  {1998}),\ \Eprint{http://arxiv.org/abs/gr-qc/9808063}{gr-qc/9808063}%
  \bibAnnoteFile{NoStop}{1998PhRvD..58l4001G}%
\bibitem{2009PhRvD..79d3016L}%
  \BibitemOpen
  \bibfield{author}{%
  \bibinfo {author} {\bibfnamefont{J.}~\bibnamefont{{Levin}}}\ and\ \bibinfo
  {author} {\bibfnamefont{R.}~\bibnamefont{{Grossman}}},\ }%
  \bibfield{journal}{%
  \Doi{10.1103/PhysRevD.79.043016}{\bibinfo {journal} {\prd}}\ }%
  \textbf{\bibinfo {volume} {79}},\ \bibinfo {eid} {043016} (\bibinfo {year}
  {2009}),\ \Eprint{http://arxiv.org/abs/0809.3838}{arXiv:0809.3838}%
  \bibAnnoteFile{NoStop}{2009PhRvD..79d3016L}%
\bibitem{2011PhRvD..83d4044Y}%
  \BibitemOpen
  \bibfield{author}{%
  \bibinfo {author} {\bibfnamefont{N.}~\bibnamefont{{Yunes}}}, \bibinfo
  {author} {\bibfnamefont{A.}~\bibnamefont{{Buonanno}}}, \bibinfo {author}
  {\bibfnamefont{S.~A.}\ \bibnamefont{{Hughes}}}, \bibinfo {author}
  {\bibfnamefont{Y.}~\bibnamefont{{Pan}}}, \bibinfo {author}
  {\bibfnamefont{E.}~\bibnamefont{{Barausse}}}, \bibinfo {author}
  {\bibfnamefont{M.~C.}\ \bibnamefont{{Miller}}},\ and\ \bibinfo {author}
  {\bibfnamefont{W.}~\bibnamefont{{Throwe}}},\ }%
  \bibfield{journal}{%
  \Doi{10.1103/PhysRevD.83.044044}{\bibinfo {journal} {\prd}}\ }%
  \textbf{\bibinfo {volume} {83}},\ \bibinfo {eid} {044044} (\bibinfo {year}
  {2011}),\ \Eprint{http://arxiv.org/abs/1009.6013}{arXiv:1009.6013}%
  \bibAnnoteFile{NoStop}{2011PhRvD..83d4044Y}%
\bibitem{SpECwebsite}%
  \BibitemOpen
  \bibinfo {howpublished} {\url{http://www.black-holes.org/SpEC.html}}%
  \bibAnnoteFile{NoStop}{SpECwebsite}%
\bibitem{Hemberger2013a}%
  \BibitemOpen
  \bibfield{author}{%
  \bibinfo {author} {\bibfnamefont{D.~A.}\ \bibnamefont{{Hemberger}}}, \bibinfo
  {author} {\bibfnamefont{G.}~\bibnamefont{{Lovelace}}}, \bibinfo {author}
  {\bibfnamefont{T.~J.}\ \bibnamefont{{Loredo}}}, \bibinfo {author}
  {\bibfnamefont{L.~E.}\ \bibnamefont{{Kidder}}}, \bibinfo {author}
  {\bibfnamefont{M.~A.}\ \bibnamefont{{Scheel}}}, \bibinfo {author}
  {\bibfnamefont{B.}~\bibnamefont{{Szil{\'a}gyi}}}, \bibinfo {author}
  {\bibfnamefont{N.~W.}\ \bibnamefont{{Taylor}}},\ and\ \bibinfo {author}
  {\bibfnamefont{S.~A.}\ \bibnamefont{{Teukolsky}}},\ }%
  \bibfield{journal}{%
  \bibinfo {journal} {ArXiv e-prints}}%
   (\bibinfo {year} {2013}),\
  \Eprint{http://arxiv.org/abs/1305.5991}{arXiv:1305.5991}%
  \bibAnnoteFile{NoStop}{Hemberger2013a}%
\bibitem{Lovelace:2011nu}%
  \BibitemOpen
  \bibfield{author}{%
  \bibinfo {author} {\bibfnamefont{G.}~\bibnamefont{Lovelace}}, \bibinfo
  {author} {\bibfnamefont{M.}~\bibnamefont{Boyle}}, \bibinfo {author}
  {\bibfnamefont{M.~A.}\ \bibnamefont{Scheel}},\ and\ \bibinfo {author}
  {\bibfnamefont{B.}~\bibnamefont{Szil\'{a}gyi}},\ }%
  \bibfield{journal}{%
  \Doi{10.1088/0264-9381/29/4/045003}{\bibinfo {journal} {Classical\ Quantum
  Grav.}}\ }%
  \textbf{\bibinfo {volume} {29}},\ \bibinfo {pages} {045003} (\bibinfo {year}
  {2012}),\ \Eprint{http://arxiv.org/abs/1110.2229}{arXiv:1110.2229}%
  \bibAnnoteFile{NoStop}{Lovelace:2011nu}%
\bibitem{Lovelace:2010ne}%
  \BibitemOpen
  \bibfield{author}{%
  \bibinfo {author} {\bibfnamefont{G.}~\bibnamefont{Lovelace}}, \bibinfo
  {author} {\bibfnamefont{M.~A.}\ \bibnamefont{Scheel}},\ and\ \bibinfo
  {author} {\bibfnamefont{B.}~\bibnamefont{Szilagyi}},\ }%
  \bibfield{journal}{%
  \Doi{10.1103/PhysRevD.83.024010}{\bibinfo {journal} {\prd}}\ }%
  \textbf{\bibinfo {volume} {83}},\ \bibinfo {pages} {024010} (\bibinfo {year}
  {2011}),\ \Eprint{http://arxiv.org/abs/1010.2777}{arXiv:1010.2777}%
  \bibAnnoteFile{NoStop}{Lovelace:2010ne}%
\bibitem{Chu2009}%
  \BibitemOpen
  \bibfield{author}{%
  \bibinfo {author} {\bibfnamefont{T.}~\bibnamefont{Chu}}, \bibinfo {author}
  {\bibfnamefont{H.~P.}\ \bibnamefont{Pfeiffer}},\ and\ \bibinfo {author}
  {\bibfnamefont{M.~A.}\ \bibnamefont{Scheel}},\ }%
  \bibfield{journal}{%
  \Doi{10.1103/PhysRevD.80.124051}{\bibinfo {journal} {\prd}}\ }%
  \textbf{\bibinfo {volume} {80}},\ \bibinfo {pages} {124051} (\bibinfo {year}
  {2009}),\ \Eprint{http://arxiv.org/abs/0909.1313}{arXiv:0909.1313}%
  \bibAnnoteFile{NoStop}{Chu2009}%
\bibitem{Scheel2009}%
  \BibitemOpen
  \bibfield{author}{%
  \bibinfo {author} {\bibnamefont{{M.~A. Scheel, M. Boyle, T. Chu, L.~E. Kidder, K.~D.
  Matthews and H.~P. Pfeiffer}}},\ }%
  \bibfield{journal}{%
  \bibinfo {journal} {Phys.\ Rev.\ D}\ }%
  \textbf{\bibinfo {volume} {79}},\ \bibinfo {pages} {024003} (\bibinfo {year}
  {2009}),\ \Eprint{http://arxiv.org/abs/0810.1767}{arXiv:0810.1767}%
  \bibAnnoteFile{NoStop}{Scheel2009}%
\bibitem{Rinne2008b}%
  \BibitemOpen
  \bibfield{author}{%
  \bibinfo {author} {\bibfnamefont{O.}~\bibnamefont{Rinne}}, \bibinfo {author}
  {\bibfnamefont{L.~T.}\ \bibnamefont{Buchman}}, \bibinfo {author}
  {\bibfnamefont{M.~A.}\ \bibnamefont{Scheel}},\ and\ \bibinfo {author}
  {\bibfnamefont{H.~P.}\ \bibnamefont{Pfeiffer}},\ }%
  \bibfield{journal}{%
  \bibinfo {journal} {Classical\ Quantum Grav.}\ }%
  \textbf{\bibinfo {volume} {26}},\ \bibinfo {pages} {075009} (\bibinfo {year}
  {2009})%
  \bibAnnoteFile{NoStop}{Rinne2008b}%
\bibitem{Boyle2007}%
  \BibitemOpen
  \bibfield{author}{%
  \bibinfo {author} {\bibfnamefont{M.}~\bibnamefont{Boyle}}, \bibinfo {author}
  {\bibfnamefont{D.~A.}\ \bibnamefont{Brown}}, \bibinfo {author}
  {\bibfnamefont{L.~E.}\ \bibnamefont{Kidder}}, \bibinfo {author}
  {\bibfnamefont{A.~H.}\ \bibnamefont{Mrou{\'e}}}, \bibinfo {author}
  {\bibfnamefont{H.~P.}\ \bibnamefont{Pfeiffer}}, \bibinfo {author}
  {\bibfnamefont{M.~A.}\ \bibnamefont{Scheel}}, \bibinfo {author}
  {\bibfnamefont{G.~B.}\ \bibnamefont{Cook}},\ and\ \bibinfo {author}
  {\bibfnamefont{S.~A.}\ \bibnamefont{Teukolsky}},\ }%
  \bibfield{journal}{%
  \Doi{10.1103/PhysRevD.76.124038}{\bibinfo {journal} {Phys.\ Rev.\ D}}\ }%
  \textbf{\bibinfo {volume} {76}},\ \bibinfo {eid} {124038} (\bibinfo {year}
  {2007})%
  \bibAnnoteFile{NoStop}{Boyle2007}%
\bibitem{Scheel2006}%
  \BibitemOpen
  \bibfield{author}{%
  \bibinfo {author} {\bibfnamefont{M.~A.}\ \bibnamefont{Scheel}}, \bibinfo
  {author} {\bibfnamefont{H.~P.}\ \bibnamefont{Pfeiffer}}, \bibinfo {author}
  {\bibfnamefont{L.}~\bibnamefont{Lindblom}}, \bibinfo {author}
  {\bibfnamefont{L.~E.}\ \bibnamefont{Kidder}}, \bibinfo {author}
  {\bibfnamefont{O.}~\bibnamefont{Rinne}},\ and\ \bibinfo {author}
  {\bibfnamefont{S.~A.}\ \bibnamefont{Teukolsky}},\ }%
  \bibfield{journal}{%
  \bibinfo {journal} {Phys.\ Rev.\ D}\ }%
  \textbf{\bibinfo {volume} {74}},\ \bibinfo {pages} {104006} (\bibinfo {year}
  {2006})%
  \bibAnnoteFile{NoStop}{Scheel2006}%
\bibitem{Pfeiffer-Brown-etal:2007}%
  \BibitemOpen
  \bibfield{author}{%
  \bibinfo {author} {\bibfnamefont{H.~P.}\ \bibnamefont{Pfeiffer}}, \bibinfo
  {author} {\bibfnamefont{D.~A.}\ \bibnamefont{Brown}}, \bibinfo {author}
  {\bibfnamefont{L.~E.}\ \bibnamefont{Kidder}}, \bibinfo {author}
  {\bibfnamefont{L.}~\bibnamefont{Lindblom}}, \bibinfo {author}
  {\bibfnamefont{G.}~\bibnamefont{Lovelace}},\ and\ \bibinfo {author}
  {\bibfnamefont{M.~A.}\ \bibnamefont{Scheel}},\ }%
  \bibfield{journal}{%
  \bibinfo {journal} {Classical\ Quantum Grav.}\ }%
  \textbf{\bibinfo {volume} {24}},\ \bibinfo {pages} {S59} (\bibinfo {year}
  {2007})%
  \bibAnnoteFile{NoStop}{Pfeiffer-Brown-etal:2007}%
\bibitem{Buchman:2012dw}%
  \BibitemOpen
  \bibfield{author}{%
  \bibinfo {author} {\bibfnamefont{L.~T.}\ \bibnamefont{Buchman}}, \bibinfo
  {author} {\bibfnamefont{H.~P.}\ \bibnamefont{Pfeiffer}}, \bibinfo {author}
  {\bibfnamefont{M.~A.}\ \bibnamefont{Scheel}},\ and\ \bibinfo {author}
  {\bibfnamefont{B.}~\bibnamefont{Szilagyi}},\ }%
  \bibfield{journal}{%
  \Doi{10.1103/PhysRevD.86.084033}{\bibinfo {journal} {Phys.\ Rev.\ D}}\ }%
  \textbf{\bibinfo {volume} {86}},\ \bibinfo {pages} {084033} (\bibinfo {year}
  {2012}),\ \Eprint{http://arxiv.org/abs/1206.3015}{arXiv:1206.3015}%
  \bibAnnoteFile{NoStop}{Buchman:2012dw}%
\bibitem{LeTiecinprep}%
  \BibitemOpen
  \bibfield{author}{%
  \bibinfo {author} {\bibfnamefont{A.}~\bibnamefont{Le~Tiec}}, \bibinfo
  {author} {\bibfnamefont{A.}~\bibnamefont{Buonanno}}, \bibinfo {author}
  {\bibfnamefont{A.~H.}\ \bibnamefont{{Mrou{\'e}}}}, \bibinfo {author}
  {\bibfnamefont{H.~P.}\ \bibnamefont{Pfeiffer}},  \bibinfo {author}
  {\bibfnamefont{D.~A.}~\bibnamefont{Hemberger}},\ and\  \bibinfo {author}
  {\bibfnamefont{G.}~\bibnamefont{Lovelace}},\ }%
  \bibfield{journal}{%
  \Eprint{http://arxiv.org/abs/1309.0541}{arXiv:1309.0541}}%
   (\bibinfo {year} {2013})%
  \bibAnnoteFile{NoStop}{LeTiecinprep}%
\bibitem{Dixonmultipoles}%
  \BibitemOpen
  \bibfield{author}{%
  \bibinfo {author} {\bibfnamefont{W.~G.}\ \bibnamefont{{Dixon}}},\ }%
  \bibfield{journal}{%
  \Doi{10.1007/BF02412488}{\bibinfo {journal} {Gen.\ Rel.\ Grav.}}\ }%
  \textbf{\bibinfo {volume} {4}},\ \bibinfo {pages} {199} (\bibinfo {year}
  {1973})%
  \bibAnnoteFile{NoStop}{Dixonmultipoles}%
\bibitem{Beiglbock67CM}%
  \BibitemOpen
  \bibfield{author}{%
  \bibinfo {author} {\bibfnamefont{W.}~\bibnamefont{{Beiglb{\"o}ck}}},\ }%
  \bibfield{journal}{%
  \Doi{10.1007/BF01646841}{\bibinfo {journal} {Comm.\ Mathem.\ Phys.}}\ }%
  \textbf{\bibinfo {volume} {5}},\ \bibinfo {pages} {106} (\bibinfo {year}
  {1967})%
  \bibAnnoteFile{NoStop}{Beiglbock67CM}%
\bibitem{Kyrian:2007zz}%
  \BibitemOpen
  \bibfield{author}{%
  \bibinfo {author} {\bibfnamefont{K.}~\bibnamefont{Kyrian}}\ and\ \bibinfo
  {author} {\bibfnamefont{O.}~\bibnamefont{Semerak}},\ }%
  \bibfield{journal}{%
  \bibinfo {journal} {Mon.\ Not.\ Roy.\ Astron.\ Soc.}\ }%
  \textbf{\bibinfo {volume} {382}},\ \bibinfo {pages} {1922} (\bibinfo {year}
  {2007})%
  \bibAnnoteFile{NoStop}{Kyrian:2007zz}%
\bibitem{Steinhoff:2010zz}%
  \BibitemOpen
  \bibfield{author}{%
  \bibinfo {author} {\bibfnamefont{J.}~\bibnamefont{Steinhoff}},\ }%
  \bibfield{journal}{%
  \bibinfo {journal} {Ann.\ Phys.}\ }%
  \textbf{\bibinfo {volume} {523}},\ \bibinfo {pages} {296} (\bibinfo {year}
  {2011}),\ \Eprint{http://arxiv.org/abs/1106.4203}{arXiv:1106.4203}%
  \bibAnnoteFile{NoStop}{Steinhoff:2010zz}%
\bibitem{1982RSPSA.385..229R}%
  \BibitemOpen
  \bibfield{author}{%
  \bibinfo {author} {\bibfnamefont{R.}~\bibnamefont{{R{\"u}diger}}},\ }%
  \bibfield{journal}{%
  \bibinfo {journal} {Proc.\ Roy.\ Soc.\ Lond.\ A}\ }%
  \textbf{\bibinfo {volume} {385}},\ \bibinfo {pages} {229} (\bibinfo {year}
  {1983})%
  \bibAnnoteFile{NoStop}{1982RSPSA.385..229R}%
\bibitem{1993NuPhB.404...42G}%
  \BibitemOpen
  \bibfield{author}{%
  \bibinfo {author} {\bibfnamefont{G.~W.}\ \bibnamefont{{Gibbons}}}, \bibinfo
  {author} {\bibfnamefont{R.~H.}\ \bibnamefont{{Rietdijk}}},\ and\ \bibinfo
  {author} {\bibfnamefont{J.~W.}\ \bibnamefont{{van Holten}}},\ }%
  \bibfield{journal}{%
  \Doi{10.1016/0550-3213(93)90472-2}{\bibinfo {journal} {Nucl. Phys. B}}\ }%
  \textbf{\bibinfo {volume} {404}},\ \bibinfo {pages} {42} (\bibinfo {year}
  {1993}),\ \Eprint{http://arxiv.org/abs/hep-th/9303112}{hep-th/9303112}%
  \bibAnnoteFile{NoStop}{1993NuPhB.404...42G}%
\bibitem{Racine2008}%
  \BibitemOpen
  \bibfield{author}{%
  \bibinfo {author} {\bibfnamefont{E.}~\bibnamefont{Racine}}, \bibinfo {author}
  {\bibfnamefont{A.}~\bibnamefont{Buonanno}},\ and\ \bibinfo {author}
  {\bibfnamefont{L.~E.}\ \bibnamefont{Kidder}},\ }%
  \bibfield{journal}{%
  \bibinfo {journal} {Phys. Rev. D}\ }%
  \textbf{\bibinfo {volume} {80}},\ \bibinfo {pages} {044010} (\bibinfo {year}
  {2009})%
  \bibAnnoteFile{NoStop}{Racine2008}%
\bibitem{spinorbitNNLO}%
  \BibitemOpen
  \bibfield{author}{%
  \bibinfo {author} {\bibfnamefont{A.}~\bibnamefont{{Boh{\'e}}}}, \bibinfo
  {author} {\bibfnamefont{S.}~\bibnamefont{{Marsat}}}, \bibinfo {author}
  {\bibfnamefont{G.}~\bibnamefont{{Faye}}},\ and\ \bibinfo {author}
  {\bibfnamefont{L.}~\bibnamefont{{Blanchet}}}\ }%
 \bibfield{journal}{%
  \bibinfo {journal} {Classical Quantum Gravity}\ }%
  \textbf{\bibinfo {volume} {30}},\ \bibinfo {pages} {075017} (\bibinfo {year}
  {2013}),\ \Eprint{http://arxiv.org/abs/1212.5520}{arXiv:1212.5520}%
  \bibAnnoteFile{NoStop}{spinorbitNNLO}%
\bibitem{BarkerOConnell}%
  \BibitemOpen
  \bibfield{author}{%
  \bibinfo {author} {\bibfnamefont{B.~M.}\ \bibnamefont{Barker}}\ and\ \bibinfo
  {author} {\bibfnamefont{R.~F.}\ \bibnamefont{O'Connell}},\ }%
  \bibfield{journal}{%
  \Doi{10.1103/PhysRevD.12.329}{\bibinfo {journal} {Phys. Rev. D}}\ }%
  \textbf{\bibinfo {volume} {12}},\ \bibinfo {pages} {329} (\bibinfo {year}
  {1975})%
  \bibAnnoteFile{NoStop}{BarkerOConnell}%
\bibitem{LenseThirring}%
  \BibitemOpen
  \bibfield{author}{%
  \bibinfo {author} {\bibfnamefont{B.}~\bibnamefont{{Mashoon}}}, \bibinfo
  {author} {\bibfnamefont{F.~W.}\ \bibnamefont{{Hehl}}},\ and\ \bibinfo
  {author} {\bibfnamefont{D.~S.}\ \bibnamefont{{Theiss}}},\ }%
  \bibfield{journal}{%
  \Doi{10.1007/BF00762913}{\bibinfo {journal} {Gen. Rel. Grav.}}\ }%
  \textbf{\bibinfo {volume} {16}},\ \bibinfo {pages} {711} (\bibinfo {year}
  {1984})%
  \bibAnnoteFile{NoStop}{LenseThirring}%
\bibitem{PhysRevD.5.814}%
  \BibitemOpen
  \bibfield{author}{%
  \bibinfo {author} {\bibfnamefont{D.~C.}\ \bibnamefont{Wilkins}},\ }%
  \bibfield{journal}{%
  \bibinfo {journal} {Phys. Rev. D}\ }%
  \textbf{\bibinfo {volume} {5}},\ \bibinfo {pages} {814} (\bibinfo {year}
  {1972})%
  \bibAnnoteFile{NoStop}{PhysRevD.5.814}%
\bibitem{Schiff}%
  \BibitemOpen
  \bibfield{author}{%
  \bibinfo {author} {\bibfnamefont{L.~I.}\ \bibnamefont{Schiff}},\ }%
  \bibfield{journal}{%
  \Doi{10.1103/PhysRevLett.4.215}{\bibinfo {journal} {Phys. Rev. Lett.}}\ }%
  \textbf{\bibinfo {volume} {4}},\ \bibinfo {pages} {215} (\bibinfo {year}
  {1960})%
  \bibAnnoteFile{NoStop}{Schiff}%
\bibitem{1997PhRvD..55.4848S}%
  \BibitemOpen
  \bibfield{author}{%
  \bibinfo {author} {\bibfnamefont{S.}~\bibnamefont{{Suzuki}}}\ and\ \bibinfo
  {author} {\bibfnamefont{K.-I.}\ \bibnamefont{{Maeda}}},\ }%
  \bibfield{journal}{%
  \Doi{10.1103/PhysRevD.55.4848}{\bibinfo {journal} {\prd}}\ }%
  \textbf{\bibinfo {volume} {55}},\ \bibinfo {pages} {4848} (\bibinfo {year}
  {1997}),\ \Eprint{http://arxiv.org/abs/gr-qc/9604020}{gr-qc/9604020}%
  \bibAnnoteFile{NoStop}{1997PhRvD..55.4848S}%
\bibitem{1998PhRvD..58b3005S}%
  \BibitemOpen
  \bibfield{author}{%
  \bibinfo {author} {\bibfnamefont{S.}~\bibnamefont{{Suzuki}}}\ and\ \bibinfo
  {author} {\bibfnamefont{K.-I.}\ \bibnamefont{{Maeda}}},\ }%
  \bibfield{journal}{%
  \Doi{10.1103/PhysRevD.58.023005}{\bibinfo {journal} {\prd}}\ }%
  \textbf{\bibinfo {volume} {58}},\ \bibinfo {eid} {023005} (\bibinfo {year}
  {1998}),\ \Eprint{http://arxiv.org/abs/gr-qc/9712095}{gr-qc/9712095}%
  \bibAnnoteFile{NoStop}{1998PhRvD..58b3005S}%
\bibitem{2003PhRvD..67b4005H}%
  \BibitemOpen
  \bibfield{author}{%
  \bibinfo {author} {\bibfnamefont{M.~D.}\ \bibnamefont{{Hartl}}},\ }%
  \bibfield{journal}{%
  \Doi{10.1103/PhysRevD.67.024005}{\bibinfo {journal} {\prd}}\ }%
  \textbf{\bibinfo {volume} {67}},\ \bibinfo {eid} {024005} (\bibinfo {year}
  {2003}),\ \Eprint{http://arxiv.org/abs/gr-qc/0210042}{gr-qc/0210042}%
  \bibAnnoteFile{NoStop}{2003PhRvD..67b4005H}%
\bibitem{2003PhRvD..67j4023H}%
  \BibitemOpen
  \bibfield{author}{%
  \bibinfo {author} {\bibfnamefont{M.~D.}\ \bibnamefont{{Hartl}}},\ }%
  \bibfield{journal}{%
  \Doi{10.1103/PhysRevD.67.104023}{\bibinfo {journal} {\prd}}\ }%
  \textbf{\bibinfo {volume} {67}},\ \bibinfo {eid} {104023} (\bibinfo {year}
  {2003}),\ \Eprint{http://arxiv.org/abs/gr-qc/0302103}{gr-qc/0302103}%
  \bibAnnoteFile{NoStop}{2003PhRvD..67j4023H}%
\bibitem{LeTiec:2011dp}%
  \BibitemOpen
  \bibfield{author}{%
  \bibinfo {author} {\bibfnamefont{A.}~\bibnamefont{{Le Tiec}}}, \bibinfo
  {author} {\bibfnamefont{E.}~\bibnamefont{{Barausse}}},\ and\ \bibinfo
  {author} {\bibfnamefont{A.}~\bibnamefont{{Buonanno}}},\ }%
  \bibfield{journal}{%
  \bibinfo {journal} {\prl}\ }%
  \textbf{\bibinfo {volume} {108}},\ \bibinfo {eid} {131103} (\bibinfo {year}
  {2012}),\ \Eprint{http://arxiv.org/abs/1111.5609}{arXiv:1111.5609}%
  \bibAnnoteFile{NoStop}{LeTiec:2011dp}%
\bibitem{scinet}%
  \BibitemOpen
  \bibfield{author}{%
  \bibinfo {author} {\bibfnamefont{C.}~\bibnamefont{Loken}}, \bibinfo {author}
  {\bibfnamefont{D.}~\bibnamefont{Gruner}}, \bibinfo {author}
  {\bibfnamefont{L.}~\bibnamefont{Groer}}, \bibinfo {author}
  {\bibfnamefont{R.}~\bibnamefont{Peltier}}, \bibinfo {author}
  {\bibfnamefont{N.}~\bibnamefont{Bunn}}, \bibinfo {author}
  {\bibfnamefont{M.}~\bibnamefont{Craig}}, \bibinfo {author}
  {\bibfnamefont{T.}~\bibnamefont{Henriques}}, \bibinfo {author}
  {\bibfnamefont{J.}~\bibnamefont{Dempsey}}, \bibinfo {author}
  {\bibfnamefont{C.-H.}\ \bibnamefont{Yu}}, \bibinfo {author}
  {\bibfnamefont{J.}~\bibnamefont{Chen}}, \bibinfo {author}
  {\bibfnamefont{L.~J.}\ \bibnamefont{Dursi}}, \bibinfo {author}
  {\bibfnamefont{J.}~\bibnamefont{Chong}}, \bibinfo {author}
  {\bibfnamefont{S.}~\bibnamefont{Northrup}}, \bibinfo {author}
  {\bibfnamefont{J.}~\bibnamefont{Pinto}}, \bibinfo {author}
  {\bibfnamefont{N.}~\bibnamefont{Knecht}},\ and\ \bibinfo {author}
  {\bibfnamefont{R.~V.}\ \bibnamefont{Zon}},\ }%
  \bibfield{journal}{%
  \Doi{10.1088/1742-6596/256/1/012026}{\bibinfo {journal} {J. Phys.: Conf.
  Ser.}}\ }%
  \textbf{\bibinfo {volume} {256}},\ \bibinfo {pages} {012026} (\bibinfo {year}
  {2010})%
  \bibAnnoteFile{NoStop}{scinet}%
\end{thebibliography}
\end{document}